\documentclass[prd,aps,twocolumn,a4paper,showkeys,nofootinbib,floatfix]{revtex4-2}

\usepackage{graphicx,psfrag}
\usepackage{epstopdf}
\usepackage{mathrsfs}
\usepackage{amsmath,amsfonts,amssymb}
\usepackage{multirow}
\usepackage{comment}
\usepackage{tipa}
\usepackage[normalem]{ulem}
\usepackage{color}
\usepackage{hyperref}
\usepackage{enumitem}
\usepackage{tipa}
\usepackage{bm}
\usepackage{algorithm}
\usepackage{algpseudocode}
\usepackage{lipsum}
\usepackage{aas_macros}
\usepackage{array}
\usepackage{soul}
\usepackage{makecell}
\usepackage[table]{xcolor}
\usepackage{pgf}

\definecolor{probRed}{RGB}{250,210,210}
\definecolor{probYellow}{RGB}{255,245,190}
\definecolor{probGreen}{RGB}{200,235,205}

\newcommand{\prob}[1]{%
  \ifdim #1pt<0.25pt
    \pgfmathtruncatemacro{\mix}{max(0,min(100,100-400*(#1)))}%
    \edef\applyprobcolor{%
      \noexpand\cellcolor{probRed!\mix!probYellow}%
    }%
    \applyprobcolor #1%
  \else
    \pgfmathtruncatemacro{\mix}{max(0,min(100,400*(#1-0.25)))}%
    \edef\applyprobcolor{%
      \noexpand\cellcolor{probGreen!\mix!probYellow}%
    }%
    \applyprobcolor #1%
  \fi
}

\makeatletter
\def\switch@array{\let\@array@sw\@array@sw@LaTeX}
\makeatother
\usepackage{multirow}
\usepackage{booktabs}
\usepackage{tabularx}

\newcommand{\be}{\begin{equation}}
\newcommand{\ee}{\end{equation}}
\newcommand{\bea}{\begin{eqnarray}}
\newcommand{\eea}{\end{eqnarray}}
\newcommand{\bel}{\begin{align}}
\newcommand{\eel}{\end{align}}

\hypersetup{
    colorlinks=true,
    linkcolor=red,
    citecolor=blue,
    urlcolor=black
}

\def\rm{\mathrm}
\def\m0{m_\rm{0}}

\def\SBIe{\rm{SBI}_\rm{Early}}
\def\SBIl{\rm{SBI}_\rm{Late}}
\def\truffle{\texttt{TRUFFLE}}

\usepackage{pifont} 

\newcommand{\xmark}{\ding{55}}

\usepackage{color}
\definecolor{cyan}{rgb}{0,0.6,0.6}
\definecolor{lightbrown}{rgb}{0.6,0.6,0}
\definecolor{orange}{rgb}{0.9,0.5,0}
\definecolor{magenta}{rgb}{1,0,1}
\definecolor{purple}{rgb}{0.8,0.4,0.8}
\definecolor{gray}{rgb}{0.8242,0.8242,0.8242}

\newcolumntype{L}[1]{>{\raggedright\arraybackslash}p{\dimexpr#1\textwidth-2\tabcolsep\relax}}
\newcolumntype{C}[1]{>{\centering\arraybackslash}p{\dimexpr#1\textwidth-2\tabcolsep\relax}}
\newcolumntype{R}[1]{>{\raggedleft\arraybackslash}p{\dimexpr#1\textwidth-2\tabcolsep\relax}}

\definecolor{nicegreen}{rgb}{0.1,0.5,0.1}
\definecolor{nicered}{rgb}{0.7,0.1,0.1}

\begin{document}

\title{Low-Latency Alerts of Massive Black Hole Binaries in LISA}

 \author{Rahul \surname{Srinivasan}${}^{1,2,3}$}
 \author{Lorenzo \surname{Copparoni}${}^{1,2,3,6}$}
 \author{Chantal \surname{Pitte}${}^{1,2,3}$}
 \author{Enrico \surname{Barausse}${}^{1,2,3}$}
 \author{Roberto \surname{Trotta}${}^{1,3,4,5}$}

\affiliation{${}^{1}$SISSA, Via Bonomea 265, 34136 Trieste, Italy}
\affiliation{${}^{2}$INFN Sezione di Trieste, Via Valerio 2, 34127 Trieste, Italy}
\affiliation{${}^{3}$IFPU - Institute for Fundamental Physics of the Universe, Via Beirut 2, 34014 Trieste, Italy}
\affiliation{${}^{4}$Department of Physics, Imperial College London, SW7 2AZ London, UK}
\affiliation{${}^{5}$Italian Research Center on High Performance Computing, Big Data and Quantum Computing}
\affiliation{${}^{6}$Univ Toulouse, CNRS, L2IT, Toulouse, France}

\date{\today}

\begin{abstract}
The Laser Interferometer Space Antenna (LISA) will observe massive black hole binaries (MBHBs) for days to months before merger, offering opportunities for advance electromagnetic follow-up. To enable such observations, inference methods must provide rapid, continuously updated, and statistically calibrated posterior estimates, while remaining robust to overlapping sources. We present \truffle\ (Time-evolving Rapid Updates For Follow-up of LISA Events), a simulation-based inference method for MBHBs in LISA data designed to meet these requirements. Our approach uses two networks, selected according to the time to coalescence relative to the end of the analysis window, $\Delta t_c$. An early-warning network, operating at $\Delta t_c \in [4~\mathrm{h}, 2~\mathrm{d}]$, estimates the chirp mass, mass ratio, and coalescence time to support electromagnetic follow-up scheduling. A near-merger network, operating at $\Delta t_c \in [-1~\mathrm{h}, +4~\mathrm{h}]$, infers nine parameters, including the sky position needed for telescope pointing. Both networks process the most recent six days of data using multiband time- and frequency-domain representations, which are summarized by a transformer to condition a normalizing flow. The networks are trained on the full orbit-driven LISA response and amortized over a 30-day range of observation epochs. The near-merger posterior can be corrected through likelihood-based importance sampling, with neural inference taking seconds and likelihood reweighting approximately five minutes on a single GPU. We obtain unbiased posteriors in the presence of overlapping sources, including plunging extreme-mass-ratio inspirals, resolvable Galactic binaries, and additional MBHBs merging up to two hours before the primary merger.

\end{abstract}

\maketitle

\section{Introduction}
\label{sec:intro}

The Laser Interferometer Space Antenna (LISA)~\cite{LISA:2017pwj,LISA:2024hlh} will be sensitive to massive black holes of $10^{4}$--$10^{8}\,M_\odot$ out to redshift $z \sim 20$. Binaries of $10^{5}$--$10^{7}\,M_\odot$ are detected days to weeks before coalescence, and the loudest mergers can reach signal-to-noise ratios (SNRs) above $1000$~\cite{LISA:2024hlh}. LISA opens a mass range that no gravitational-wave (GW) detector has observed. These black holes are three to five orders of magnitude heavier than those in the ground-based catalogs, whose component masses reach $\sim 140\,M_\odot$~\cite{GWTC4}. Semi-analytic models of massive black hole evolution predict from a few to hundreds of mergers per year, depending on the black hole seeds and on the delays between galaxy and black hole mergers~\cite{Barausse:2012fy,Sesana:2014bea,Antonini:2015sza,Klein:2015hvg,Bonetti:2018tpf,Barausse:2020mdt,Barausse:2023yrx}. A fraction of these binaries will merge in gas-rich nuclei, where the circumbinary disc and the mini-discs around each black hole can emit through the final inspiral, and the circumbinary disc can brighten again after the merger~\cite{Bogdanovic2022,Tang2018,dAscoli2018,Gutierrez2022}. A joint GW--electromagnetic (EM) detection would single out the host among the $\gtrsim 10^{2}$ galaxies in a LISA error box~\cite{Lops2023}, turn the binary into a standard siren with a measured redshift~\cite{Tamanini2016}, and show how gas accretes onto the binary and onto the remnant~\cite{mangiagli2022}. Time-domain surveys may even catalogue some of these binaries as periodic quasars long before they enter the LISA band~\cite{XinHaiman2021}.
 
The EM phenomenology is time-critical. Although large model uncertainties exist in the expected emissions from massive black hole mergers, the predicted signatures could include X-ray variability that follows the orbital phase during the inspiral~\cite{Haiman2017,Tang2018,DalCanton2019}, thermal X-ray emission from the mini-discs, which lasts until the binary decouples from the circumbinary disc, close to merger for viscous discs and earlier otherwise~\cite{Dittmann2023}, and which then drops by orders of magnitude, one to two days before merger in the simulations of Ref.~\cite{Krauth2023}, and an afterglow that brightens years later, as the disc migrates inward~\cite{Milosavljevic2005}. The counterpart need not be confined to X-rays. Gas trapped inside the orbit is driven inwards by the inspiral, so that accretion can exceed the Eddington rate just before coalescence~\cite{Armitage:2002uu} and may power a broadband, quasar-like flare that wide-field optical surveys could also detect~\cite{Tamanini2016}. The motion of the binary through the magnetic field of the circumbinary disc also generates a Poynting flux that grows through the inspiral. These emissions include collimated dual jets~\cite{Palenzuela:2010nf} and a brighter, non-collimated component~\cite{Moesta:2011bn}, enhanced by orders of magnitude when the accreting plasma amplifies the field~\cite{Giacomazzo:2012iv}. Part of it may emerge as radio synchrotron emission modulated on the orbital period~\cite{OShaughnessy:2011nwl}, with which wide-field radio instruments could localize a binary identified in advance of merger by its GWs~\cite{Kaplan:2011mz}. Forecasts for LISA standard sirens rely on such counterparts in the optical, radio and soft X-ray bands~\cite{Tamanini2016,mangiagli2022,Laghi:2026ujb}. They predict from a few to about twenty detectable counterparts in four years, depending on obscuration and on the collimation of the radio emission~\cite{mangiagli2022}. The pre-merger signatures evolve on the orbital period, minutes to hours in the final week, and the sharpest change occurs in the last days before merger: a counterpart that is not observed at the right time is lost. 

Observing strategies for these counterparts therefore assume that LISA supplies a sky localization that is updated as the merger approaches~\cite{DalCanton2019}. Data gaps can complicate the cadence of posterior updates and vary the posteriors themselves \cite{Dey:2021dem}. To point telescopes at the merger, an observer needs pre-merger estimates from the GW inspiral of the coalescence time, masses, distance, and sky position. These are not needed all at once. Scheduling an observation needs the merger time and a rough mass estimate days ahead, whereas pointing needs the sky localization, which for most sources is determined only just before merger. The full parameter set is needed only afterwards, to hand the event to the global fit.  This also fixes the operating requirement: inference must run repeatedly, on a sliding analysis window, and refine its answer as data accumulate, rather than once on a final data set. Likelihood-based samplers like Markov chain Monte Carlo (MCMC) and nested sampling give accurate posteriors for MBHBs~\cite{marsat2021,katz2020,Pratten2023,Weaving2024}, but a pre-merger analysis with a stochastic sampler takes days~\cite{cabourndavies2025}, which is too late for EM follow-up.

The idea of a triggered EM search guided by the GW inspiral goes back to the early mission studies~\cite{Kocsis2008,LangHughes2008,LangHughes2009}. Fisher forecasts follow how the parameters sharpen from a month before merger~\cite{mangiagli2020}: the chirp mass is known to tens of per cent a month ahead and to below a per cent in the last hour, while the sky position of a heavy binary reaches $\sim10\,\mathrm{deg}^2$ only in that last hour. Full Bayesian analyses show that the sky posterior could be multimodal, with eight modes set by the symmetries of the LISA response, of which the true position and its reflection through the constellation plane survive longest~\cite{marsat2021}. 

Recent MBHB data analyses for LISA fall into three broad groups: pre-merger search pipelines, rapid inference near-merger, and post-merger deep analysis. 

    \textbf{1)} Pre-merger search pipelines: a stochastic, likelihood-based search run on simulated LISA data~\cite{cornish2022}, matched-filter searches detect signals up to two weeks before merger, in the presence of data gaps~\cite{cabourndavies2025,cabourndavies2026}, and a rapid time--frequency excess-power search that detects binaries from hours to two weeks before merger, processing a ten-day segment in under a second~\cite{nobili2026}. Parameter estimates are then either point values with a detection-significance contour~\cite{nobili2026} or come from a stochastic sampler run after detection, which takes days and is not yet reliable in the last day before merger~\cite{cabourndavies2025}. None of these delivers a calibrated posterior at the cadence of the data. 

    \textbf{2)} Rapid inference near-merger: LISA inference with SBI methods using normalizing flows~\cite{Du2024,cnf2024,spadaro2026} and sequential neural likelihood~\cite{snl2024} deliver posteriors, typically importance-sampled, within a minute. These methods place the coalescence time with a prior of minutes to an hour around the merger, at a single observational epoch and a rigid satellite constellation. A pre-merger analysis for a TianQin-like detector~\cite{zhang2026} localizes a source $15\,\mathrm{min}$ before merger in about a minute per event on a CPU. It is currently designed for a single snapshot, with the merger placed $10$--$20\,\mathrm{min}$ after the end of the data, not importance sampled, and assumes non-spinning binaries.
    
    \textbf{3)} Post-merger deep analysis: Sampler-based pipelines like ~\cite{Geemoo_Deng2025,Katz2025,sharma2024,katz2020} perform sampler-based inference of the MBHBs, which provide accurate posteriors for cataloging and downstream population inference. The methods trade longer compute times for accurate inference.

Existing methods address parts of the low-latency deliverables, but none of the approaches discussed above combines repeated pre-merger posterior updates over a wide coalescence-time prior with the orbit-driven, time-dependent LISA response. The combination is hard for two reasons. The constellation moves, so the data depend on when the analysis window sits on the orbit, as well as on the source parameters; and before merger, the position of the coalescence relative to the window changes with every new analysis. A time shift is therefore not a symmetry of the problem. The device that makes neural parameter estimation inexpensive for ground-based detectors, standardizing the coalescence time of every signal through the time-translation symmetry of the data~\cite{Dax:2021myb,Dax:2021tsq}, is unavailable here, and the coalescence time must be inferred over a wide prior as an ordinary parameter (Sec.~\ref{sec:signal-data}).

The paper is organized as follows. Sec.~\ref{sec:method} describes the pipeline from the data generation to network training and inference. Sec.~\ref{sec:results} presents the validation tests, follows a reference signal from two days before merger to one hour after it, and tests the robustness on data containing other overlapping sources. Sec.~\ref{sec:discussion} discusses alert system deliverables, places the results among existing work, elaborates its limitations. Sec.~\ref{sec:conclusion} provides concluding remarks. The appendices~\ref{app:handedness}--\ref{app:res-sky} discuss network and training specifications, prior-wide inference.

\section{Method}
\label{sec:method}

\begin{figure*}[t]
  \centering
  \includegraphics[width=\textwidth]{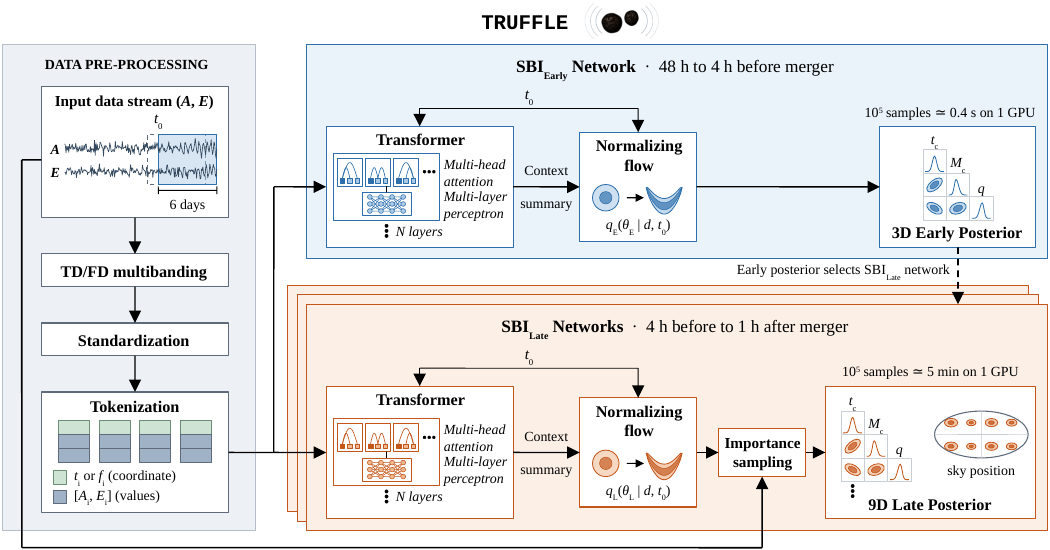}
  \caption{Flowchart of our low-latency LISA inference pipeline. Continuous LISA (A,E) data are processed in six-day windows, multibanded, standardized, and tokenized before entering transformer-based neural posterior estimators conditioned on the observation epoch $t_0$. $\SBIe$ (48~h to 4~h before merger) infers a posterior over chirp mass, mass ratio, and coalescence time, informing the selection of the $\SBIl$ (4 h before to 1 h after merger). The latter infers a nine-dimensional posterior, including sky position, refined through importance sampling using the likelihood computed with the six-day input.}
  \label{fig:flowchart}
\end{figure*}

\subsection{Overview}
\label{sec:overview}
Our inference algorithm, \truffle\ (Time-evolving Rapid Updates For Follow-up of LISA Events), is a simulation-based inference (SBI) \citep{SBIfrontier:Kranmer+2020} method that trains a neural network to provide rapid posterior estimates from real-time data. We illustrate the flow of data through the pipeline in Fig.~\ref{fig:flowchart}. The expensive computation is amortized: a high cost is paid once during training, after which inference for each new data segment is fast. Our alert pipeline can be used from days before merger to about an hour post-merger. The upfront training cost, however, scales poorly with the number of inferred parameters. To limit this cost, we split the algorithm into two regimes, early and late inference, according to the proximity to merger and the number of parameters inferred. Table~\ref{tab:deliverables} lists the three objectives of our low-latency pipeline, the parameters required for each, and when they are needed.

The early inference method, $\SBIe$, provides posteriors for sources that merge 2 days to 4 hours after the observation window. In this regime, we provide rapid inference of the  time to coalescence $\Delta t_c$, chirp mass $\mathcal{M}_c$ and mass ratio $q$ - parameters required to schedule telescopic follow-up observations. For the majority of MBHBs, especially those with $\mathcal{M}_c\gtrsim10^6\,\rm{M_\odot}$, the sky position among other parameters, is not well constrained this early in the inspiral. The late near-merger inference, $\SBIl$, provides posteriors from 4 hours before to 1 hour after the merger. It provides the full multi-dimensional inference of the coalescence time $\Delta t_c$, chirp mass $\mathcal{M}_c$, mass ratio $q$, spins $\chi_1$, $\chi_2$, luminosity distance $D_L$, inclination $\iota$, and most importantly, sky location $\lambda,\,\beta$ in the LISA-frame. The nine-dimensional $\SBIl$ posterior can also be used to seed a slower, full MCMC analysis (Sec.~\ref{sec:hybridMCMC}). 

The pipeline operates on a continuous data stream: an analysis window of fixed length is moved forward along it in real time, and each new position returns an updated posterior within a minute (with eight GPUs, five minutes on one). Each window is compressed into two multiband representations, one in the time domain and one in the frequency domain. Together they keep the sharp timing information near merger and the accumulated phase of the long inspiral at a fraction of the raw data volume. A transformer summarizes both into a single feature vector passed to the normalizing flow which returns the posterior density $q(\theta \mid d, t_0)$. Both transformer and flow are conditioned on the observation start time $t_0$, which fixes the constellation's position on its orbit (Sec.~\ref{sec:datagen}).

\begin{table}[t]
  \caption{Alert deliverables, the parameters inferred, and the information provided, assuming LISA data is available near-merger with little delay.}
  \label{tab:deliverables}
  \begin{ruledtabular}
  \begin{tabular}{lcl}
    Regime & Deliverables & Use \\ \colrule
    $\Delta t_c\sim\mathcal{O}(\rm{1d})$ & $t_c$, $\mathcal{M}_c$, $q$ &
      Scheduling \\
    $\Delta t_c\sim\mathcal{O}(\rm{1h})$ & $\,\,t_c,\,\mathcal{M}_c,\,\lambda,\, \beta\,\,$ &
      Tiling and pointing \\
    $\Delta t_c<0$, post-merger  & full $\theta$ & Hand-off to MCMC \\
  \end{tabular}
  \end{ruledtabular}
\end{table}

Four features distinguish this construction from existing neural parameter estimation for LISA MBHBs. First, the detector response is the full orbit-driven one, with unequal, time-varying arms, and the networks are amortized over a $30$-day range of observation start times rather than tied to one constellation configuration. Second, the coalescence time is inferred as an ordinary parameter over a wide prior. It cannot be removed by simply shifting the signal in time: changing ($t_c$) changes which stage of the inspiral is present in the analysis window and, through LISA’s orbital motion, also changes the detector response. Third, for the $\SBIl$ network, importance sampling against the exact likelihood corrects the network posterior, and its efficiency measures the network's accuracy for each analysis. This requires the network to model the joint density of every parameter the likelihood depends on, not only the alert parameters, with the exception of $\phi$ and $\psi$, which are numerically marginalized inside the likelihood computation. Fourth, we reparameterize the flow so that the network can learn the sky--inclination handedness and avoid assigning posterior mass to spurious modes that appear in the network posterior but are ruled out by the full likelihood (Sec.~\ref{sec:handedness}). A similar strategy may help with other discrete sky degeneracies in future work.

The rest of this section follows the pipeline: simulated data and priors (Secs.~\ref{sec:signal-data}--\ref{sec:datagen}), the multiband representation (Sec.~\ref{sec:multiband}), the network, its training and optimization (Secs.~\ref{sec:architecture}--\ref{sec:handedness}), and importance sampling evaluation (Sec.~\ref{sec:is}). Fig.~\ref{fig:flowchart} illustrates the flow of the method from the input data to the final posteriors.

\subsection{Signal model and detector response}
\label{sec:signal-data}
 
We consider MBHBs in the millihertz band, observed through inspiral, merger, and ringdown. Waveforms are generated with \textsc{IMRPhenomTHM}~\cite{Estelles:2020twz, garcíaquirós2025gpuacceleratedlisaparameterestimation},  as implemented in \textsc{phentax}~\cite{alessandro_santini_2026_19390393}, including the $(2,2)$, $(2,1)$, $(3,3)$ and $(4,4)$ modes and their negative-$m$ counterparts. 
Higher modes make the simulated signals closer to real data~\cite{Pitte:2023ltw, Yi:2025pxe, Yi:2026ucv} and carry much of the information that breaks the sky, inclination and mass-ratio degeneracies near merger. All waveforms start at $f_{\min} = f_{\rm ref} = 3\times10^{-5}\,\mathrm{Hz}$, well before the simulated data begin (Sec.~\ref{sec:datagen}).

The detector response is computed with \textsc{lisax}~\cite{LISAX}, which reads the position of the spacecrafts and the light-travel-times directly from the orbit files generated with \texttt{lisaorbits}~\cite{Bayle}. This means that no rigid-constellation or low-frequency approximation is made at any stage, unlike previous neural parameter estimation for LISA~\cite{spadaro2026}. We use the ESA-trailing orbits and the quasi-orthogonal TDI channels $A$ and $E$ (Sec.~\ref{sec:target}), allowing us to assume independent noises.
We use second-generation TDI and neglect the $T$ channel as it carries little signal information~\cite{Prince2002}.

Ground-based neural inference differs in complexity from LISA inference in two ways. First, ground-based detectors move negligibly during a signal, so the detector response can be treated as time-independent. Second, the merger-time prior is typically much narrower: networks can be trained over a fraction of a second around a known trigger~\cite{Dax:2021tsq}. These simplifications do not apply here. LISA's constellation and antenna pattern rotate over the year, parameterized by observation time $t_0$, and the relative merger time $\Delta t_c$ is inferred over a wide prior, spanning hours to days. The analysis window has a fixed duration $T_{\rm obs} = 2^{19}\,\mathrm{s} \simeq 6.07\,\mathrm{d}$ and a cadence $\Delta t = 32\,\mathrm{s}$, set by the highest signal frequency within our priors and window so that no band-limited information is lost (see Sec.~\ref{sec:multiband}).
 
The noise is stationary and Gaussian, drawn independently for $A$ and $E$ from a fixed one-sided power spectral density (PSD) from SCiRD~\cite{LISA2018ScienceReq} that includes the Galactic-binary confusion foreground approximation for one year of data. Neglecting their small cross-correlation is a reasonable approximation for near-equal arms over a window of a few days. Data gaps, glitches are not modeled (Sec.~\ref{sec:limitations}). The training data contain one MBHB per realization. Other overlapping resolvable sources are excluded from the training but introduced in robustness tests in Sec.~\ref{sec:res-robustness}. These assumptions simplify the present analysis but are not intrinsic to the SBI framework: more realistic noise, instrumental artifacts, and additional sources can be incorporated directly into the simulations used for training.

\subsection{Simulation Priors and Network Inferences}
\label{sec:priors}

\begin{table}[t]
\begin{ruledtabular}
\setlength{\tabcolsep}{3pt}
\footnotesize
\begin{tabular}{llccc}
Symbol & Parameter & Simulation & \multicolumn{2}{c}{Inference} \\
 &  & prior & $\SBIl$ & $\SBIe$ \\
 \noalign{\vskip 2pt}
 \colrule
 \noalign{\vskip 2pt}
$\log_{10}\mathcal{M}_c$ & Chirp mass (det.)   & $[5.0, 7.0]$       &            & \checkmark \\
                         &                     & $[6.5, 7.0]$       & \checkmark &            \\
$q$                      & Mass ratio          & $[1, 10]$          &            & \checkmark \\
                         &                     & $[1, 4]$           & \checkmark &            \\
$\Delta t_c$                    & Coalescence time    & $[4\,\mathrm{h}, 2\,\mathrm{d}]$ &  & \checkmark \\
                         &                     & $[-1, 4]\,\mathrm{h}$ & \checkmark &         \\
$\chi_1$                 & Primary spin        & $[-0.9, 0.9]$      & \checkmark & \xmark   \\
$\chi_2$                 & Secondary spin      & $[-0.9, 0.9]$      & \checkmark & \xmark   \\
$D_L$                    & Distance (Gpc)      & $[20, 100]$        & \checkmark & \xmark   \\
$\cos\iota$              & Inclination         & $[-1, 1]$          & \checkmark & \xmark   \\
$\lambda$                & Longitude           & $[0, 2\pi)$        & \checkmark & \xmark   \\
$\sin\beta$              & Latitude            & $[-1, 1]$          & \checkmark & \xmark   \\
$\phi$                   & Reference phase     & $[0, 2\pi)$        & \xmark   & \xmark   \\
$\psi$                   & Polarization        & $[0, \pi)$         & \xmark   & \xmark   \\
\noalign{\vskip 2pt}
\cline{1-5}
\noalign{\vskip 2pt}
$t_0$                    & Obs.\ start time    & $[70, 100]\,\mathrm{d}$ & --    & --         \\
\end{tabular}
\end{ruledtabular}
\caption{Simulator parameters and their uniform priors. A check mark marks a parameter the network infers, a crossmark one it does not infer but still varies in training, i.e.\ marginalizes over. Entries without a mark apply to the other network only. Spins are aligned with the orbital angular momentum, $q = m_1/m_2$, and $\phi$ is the phase at the reference frequency. $t_0$ is neither inferred nor marginalized: it is supplied to the network (Sec.~\ref{sec:architecture}). Sky angles are sampled and inferred in the LISA frame at merger.}
\label{tab:priors}
\end{table}

Table~\ref{tab:priors} lists the twelve simulator parameters. Sky position, coalescence time and polarization are sampled in the LISA (L) frame at merger. In the L-frame, the antenna pattern of a source depends mainly on its angles and only weakly on $t_0$, whereas in a barycentric frame the pattern of a fixed sky position rotates with the constellation over the year.
The L-frame therefore keeps the map the network must learn as nearly independent of $t_0$ as possible. 

The coalescence time needs a definition, because different modes peak at different times. We adopt the \textsc{IMRphenomTHM} convention, and we refer to the coalescence time as the peak time for the $(2,2)$-mode amplitude, given by the \texttt{epoch}, which does not depend on which modes are included. Working in the L-frame, we define $\Delta t_c$ as the arrival time at the constellation barycenter, relative to the end of the analysis window, i.e., $\Delta t_c \equiv t_{c} - t_{\rm end}$.

Positive $\Delta t_c$ places the merger in the future, while a negative $\Delta t_c$ places it inside the window. The two networks are trained on adjacent intervals of this variable: $\SBIe$ on $[+4\,\mathrm{h}, +2\,\mathrm{d}]$, where the data are pure inspiral, and $\SBIl$ on $[-1\,\mathrm{h}, +4\,\mathrm{h}]$, whose lower bound lets the window run up to an hour past coalescence so that merger and ringdown are captured in full. The two intervals meet at $+4\,\mathrm{h}$ and cover the last two days before merger without a gap (Fig.~\ref{fig:countdown}). The mass priors differ by design. $\SBIe$ spans $\log_{10}(\mathcal{M}_c/M_\odot) \in [5.0, 7.0]$ and $q \in [1, 10]$ and returns an early mass estimate that selects which $\SBIl$ network to run. $\SBIl$ is envisaged as a set of independently pre-trained networks, each on a chirp-mass band of $\sim 0.5~\mbox{dex}$ and mass ratio band of 4, so that the compression can be tailored to the band (Sec.~\ref{sec:multiband}). This keeps the input short, $153$ tokens against $553$ for $\SBIe$, and frees capacity for the harder nine-parameter inference. Adjacent bands overlap, e.g.\ $[6.05, 6.55]$ and $[6.5, 7.0]$, so that every source lies well inside at least one band. When the early estimate is ambiguous, both neighbors can be run, since one evaluation takes seconds. Here, based on the $\SBIe$ posteriors of a \emph{reference} signal (see Fig.~\ref{fig:corner_evolution}), we train and demonstrate $\SBIl$ over a single band, $\log_{10}\mathcal{M}_c \in [6.5, 7.0]$ with $q \in [1, 4]$, and extending its priors to $\log_{10}\mathcal{M}_c \le 6.5,\,q \geq 4$ are left for future work.
 
$\SBIe$ infers $\log_{10}\mathcal{M}_c$, $q$ and $\Delta t_c$, which are the parameters that observers need to schedule follow-up, while the other eight parameters are varied in training and thus marginalized over. The required parameters are, in general, measurable a day or more before merger.

The observation start time $t_0$ is known at analysis time and is supplied to the network (Secs.~\ref{sec:datagen} and~\ref{sec:architecture}), from which it can determine the constellation's position and orientation for a given orbit file.

\subsection{Data generation and loading}
\label{sec:datagen}
\label{sec:sliding}

\emph{Training and validation sets.} Rather than simulating every analysis window separately, we evaluate the forward model once for each source-parameter draw. The waveform is generated at the lowest value of the $\Delta t_c$ prior, $4\,\mathrm{h}$ before merger for $\SBIe$ and $1\,\mathrm{h}$ after merger for $\SBIl$. Its angles are transformed from the L frame at merger to the equatorial frame, and the signal is propagated through the orbit-driven response and TDI to produce a clean full-resolution $A/E$ time series. The merger is placed at a fixed sample of this stream, and only the noiseless signal is stored.

The analysis window and its value of $\Delta t_c$ are selected at training time. Noise is also generated on the fly, and the multiband compression is applied only after the window has been chosen. A single stored signal can therefore provide many independent training examples with different noise realizations and window positions. The effective number of training examples is set by the number of epochs drawn during optimization rather than by the number of stored waveform simulations.

The training sets were generated on $112$ CPU cores per node. The $\SBIl$ set contains $1.7\times10^{7}$ sources and required $14$ node-hours, corresponding to $35\,\mathrm{min}$ of wall-clock time. The $\SBIe$ set contains $5.0\times10^{7}$ sources and required $56$ node-hours, or about one hour of wall-clock time across two batches. This corresponds to about $0.3$--$0.5$ core-seconds per simulated source, including the full-resolution time series and both multiband views.
 
Training examples are cut from these streams with a sliding window. At each epoch, a source receives one colored-noise realization over its full length, from which $k = 8$ windows ($k = 4$ for $\SBIe$) are cut at values of $\Delta t_c$ drawn from the prior. The windows are successive views of one event through one noise realization, simulating the continuously accumulating satellite data. By generating the full signal once and then slicing random observation windows inside it, we can inexpensively (in compute and storage) obtain many $\Delta t_c$ samples per simulation. This procedure is equivalent to drawing $\Delta t_c$ and the noise jointly with the other parameters. Because multiple examples of the same signal at different $\Delta t_c$ can appear within the same training batch, examples within a batch are not fully independent. For the validation and test examples, the window positions and the noise realizations are fixed, so that the monitored loss is comparable across epochs. One optimizer step uses $16\,384$ examples, $2048$ sources $\times\ 8$ windows for $\SBIl$ and $4096$ sources $\times\ 4$ windows for $\SBIe$, with all windows of a source in the same step.

\emph{Amortization over $t_0$.} Amortizing over $t_0$ adds a dependence the network must learn on top of the source physics. For $\SBIl$, we split the training into two stages (Sec.~\ref{sec:training}), one at a fixed $t_0$ and one over the $t_0$ prior. In stage~1, every stream starts at a fixed  $t_0$. The network is trained on $1.7\times10^{7}$ sources and, over $\simeq 670$ epochs, sees $\simeq 2.6\times10^{8}$ noise realizations and $\simeq 2.1\times10^{9}$ windows. In stage~2 it is fine-tuned on $1.7\times10^{7}$ new sources whose streams start at $512$ different times spread over the $t_0$ prior, over $\simeq 100$ epochs it sees $\simeq 5\times10^{7}$ noise realizations and $\simeq 4\times10^{8}$ windows, so amortization costs as many new simulations as the fixed-$t_0$ stage but a seventh of the training epochs. $\SBIe$, whose nine-parameter marginalization is not required and whose inspiral-only data depend less on the constellation orientation, is trained in a single stage directly over the $t_0$ prior, on $5.0\times10^{7}$ sources whose streams start at $1536$ different times. 

The simulator processes one stream start time per batch, so each training file holds streams with one start time. The two networks receive $t_0$ in different conventions. $\SBIl$ is given the start of the data stream, one value for all windows of a source, which differs from the start of the analysis window by at most $5\,\mathrm{h}$. Over such an interval the constellation turns by a fraction of a degree, and shifting $t_0$ by $0.19\,\mathrm{d}$ leaves the posterior unchanged. $\SBIe$, whose windows slide over $44\,\mathrm{h}$, is given the start of each analysis window, to which a uniform jitter of $\pm 6\,\mathrm{h}$ is added during training so that the network does not tie its output to the simulated grid.

\subsection{Data compression for network input}
\label{sec:multiband}
\label{sec:target}

The simulations are passed both in time and frequency domains. By using this redundancy in our data representation, we improve the efficiency of the network in learning the posterior properties across parameter space. Specifically, some features, such as the time of merger, are easier to learn in the time domain, while others, such as mass, are easier to learn in the frequency domain.~\cite{Alvey:2023npw}. However, note that in this implementation, the two representations are not equivalent. To avoid spectral leakage when moving the data to the frequency domain, we apply a Planck window to taper the data. While this effect is negligible in the $\SBIe$ prior, it can result in a significant loss of information for events in the $\SBIl$ mass prior. We set the sampling cadence $\Delta t = 32\,\mathrm{s}$ to accommodate the lightest sources in our priors. 

\begin{figure*}
    \centering
    \includegraphics[width=\linewidth]{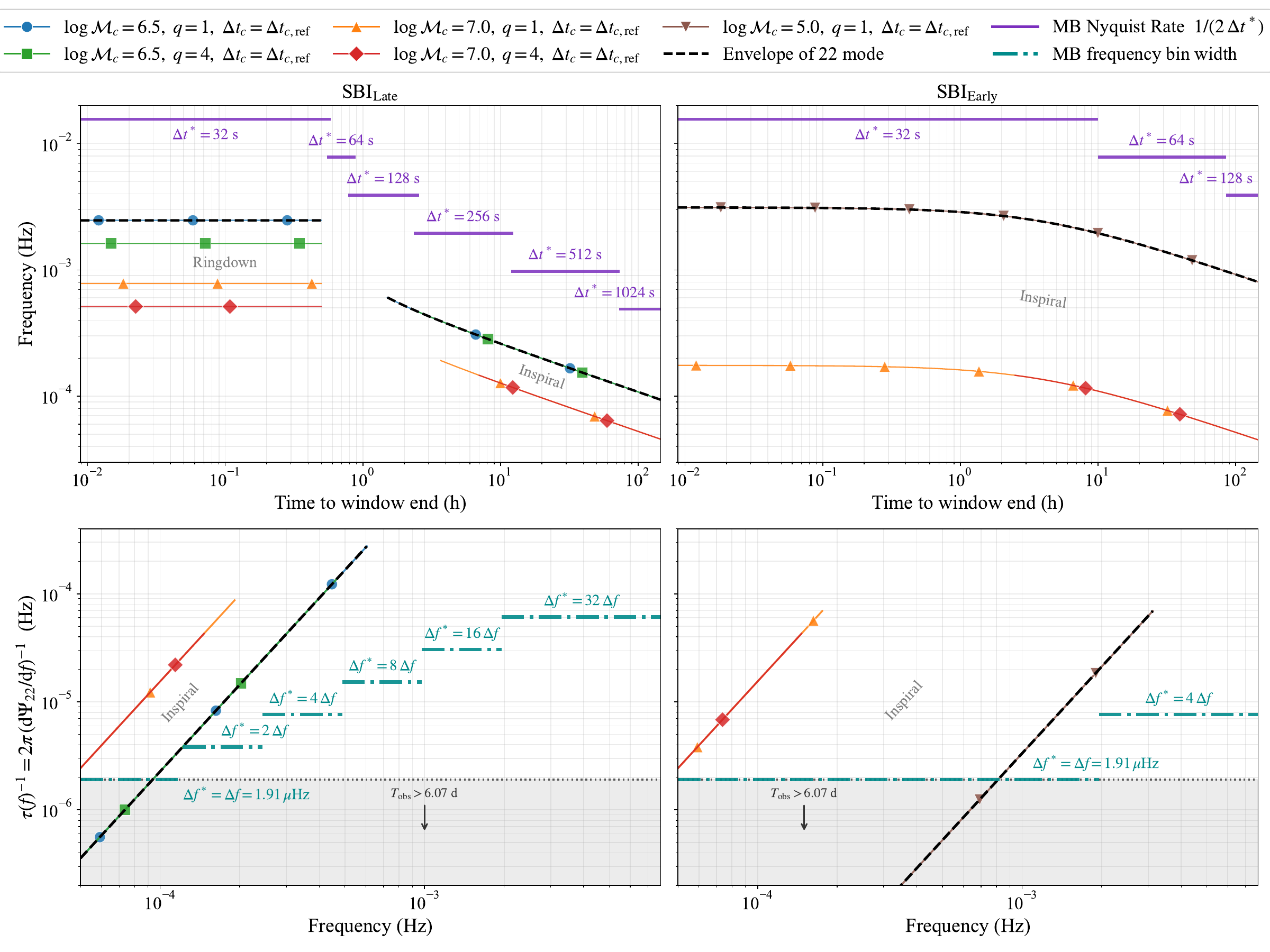}
    \caption{Multiband plans of $\SBIl$ (left) and $\SBIe$ (right). Top: leading-order $(2,2)$ frequency against time before the end of the analysis window, for sources at the corners of each prior with the merger at the time $\Delta t_c^{\rm mb}$, together with their upper envelope and the Nyquist frequency $1/(2\Delta t^*)$ of each time-domain band. Bottom: inverse time to coalescence $\tau(f)^{-1}$ against frequency for the same sources, together with the block-averaged frequency resolution $\Delta f^*$ of each frequency-domain band; a band preserves the signal where $\Delta f^*$ lies below the curves.}
    \label{fig:MB}
\end{figure*}

The transformer (Sec.~\ref{sec:architecture}) receives a compressed representation of the time- and frequency-domain views, as its cost grows quadratically with sequence length. Most of the information for the sources in the networks' prior accumulates the majority of the SNR within the latest $\simeq6\,$d, setting the duration of our observation window. The information is also distributed very unevenly in time: the signal stays at low frequency for most of the window and sweeps upward only in its final hours. We set the compression from the leading-order frequency evolution of the $(2,2)$ mode and check with the $(2,2,0)$ ringdown frequency that the finest resolution retained covers the end of the signal.

Each network's compression plan is computed once, from the lightest binary of its prior ($\mathcal{M}_{c,\rm ref} = 10^{6.5}\,M_\odot$ for $\SBIl$, $10^{5}\,M_\odot$ for $\SBIe$), which reaches every frequency earliest in the window and is therefore the conservative choice. We set the merger at $\Delta t_c^{\rm ref} = -30\,\mathrm{min}$ for $\SBIl$, and the lower prior edge at $+4\,\mathrm{h}$ for $\SBIe$. The plan is applied unchanged to every window, so the compression is a single linear operator independent of the data. Fig.~\ref{fig:MB} and Tables~\ref{tab:bands-td} and~\ref{tab:bands-fd} give the resulting schedules.

Although the $\Delta t_c$ prior for $\SBIl$ extends down to $-60\,\mathrm{min}$, we set the multiband scheduling based on $\Delta t_c^\rm{mb}=-30$~min as this is the typical limit to which we expect to use the pipeline and the posteriors show little to no refinement beyond $-30$~min. Moreover, for mergers with $\Delta t_c<\Delta t_c^\rm{mb}$, merger and early ringdown fall in the $64$- or $128$-s band, whose Nyquist frequencies ($7.8$ and $3.9\,\mathrm{mHz}$) still exceed the $(2,2)$ frequency through ringdown for every source in the prior ($\lesssim 3.6\,\mathrm{mHz}$). For lighter sources in the prior the higher ringdown harmonics can be filtered out of the time-domain (TD) representation, but they are still present in the frequency-domain (FD) representation and in the importance sampling stage.
 
The two views have opposite resolution profiles, fine near merger and fine at low frequency, so each retains information the other discards. See Sec.~\ref{app:multibanding} for a detailed discussion on how we compute the TD and FD resolutions.

For $\SBIl$, information lost in the compression costs importance-sampling efficiency rather than accuracy (provided sufficient proposal support), since the likelihood of Sec.~\ref{sec:is} uses the full-resolution window. For $\SBIe$, which is not importance-sampled, it can only broaden the posterior. The calibration tests of Sec.~\ref{sec:res-calibration} however show that the compression does not bias either network's posterior.

\begin{table}[t]
\caption{Time-domain multiband plan. $\Delta t^*$ is the sampling interval of a band after anti-aliased decimation. Intervals are times before the end of the analysis window. Adjacent bands overlap slightly before their edges are trimmed.}
\begin{ruledtabular}
\begin{tabular}{ccc}
\noalign{\vskip 2pt}
$\Delta t^*$ (s) & $\SBIl$ interval (h) & $\SBIe$ interval (h)\\ \colrule
\noalign{\vskip 2pt}
1024 & 72.53 -- 145.64 & -- \\
512  & 11.95 -- \phantom{0}73.39 & -- \\
256  & \phantom{0}2.35 -- \phantom{0}12.37 & -- \\
128  & \phantom{0}0.78 -- \phantom{00}2.56 & 84.87 -- 145.64 \\
64   & \phantom{0}0.55 -- \phantom{00}0.89 & 10.01 -- \phantom{0}84.98 \\
32   & \phantom{0}0.00 -- \phantom{00}0.59 & \phantom{0}0.00 -- \phantom{0}10.04 \\
\end{tabular}
\end{ruledtabular}
\label{tab:bands-td}
\end{table}

\begin{table}[t]
\caption{Frequency-domain multiband plan. $\Delta f^*$ is the resolution of a band after block-averaging and $\Delta f = 1/T_{\rm obs} = 1.91\,\mu\mathrm{Hz}$ the native resolution, common to both networks. $\SBIe$, whose prior extends to $\mathcal{M}_c = 10^{5}\,M_\odot$, keeps the native resolution up to $1953\,\mu\mathrm{Hz}$, because its light sources sweep through that range too rapidly for block-averaging.}

\begin{ruledtabular}
\begin{tabular}{ccc}
$\Delta f^*/\Delta f$ & $\SBIl$ interval ($\mu$Hz) & $\SBIe$ interval ($\mu$Hz) \\ \colrule
\noalign{\vskip 2pt}
 1 & $50$ -- $122$ & $50$ -- $1953$ \\
 2 & $122$ -- $244$ & -- \\
 4 & $244$ -- $488$ & $1953$ -- $7813$ \\
 8 & $488$ -- $977$ & -- \\
16 & $977$ -- $1953$ & -- \\
32 & $1953$ -- $7813$ & -- \\
\end{tabular}
\end{ruledtabular}
\label{tab:bands-fd}
\end{table}

The compression reduces each window, per channel, to $N$ real time-domain samples and $L$ complex frequency-domain bins, being $N = 946$, $L = 261$ for $\SBIl$ and $N = 7049$, $L = 1765$ for $\SBIe$. Both views are cut into patches of $P$ consecutive samples or bins, $P = 8$ for $\SBIl$ and $P = 16$ for $\SBIe$, with the $A$ and $E$ channels grouped inside each patch, so that every token holds both channels at the same times or frequencies. Writing $x_{[i]} \equiv (x_{Pi}, \dots, x_{Pi+P-1})$ for the $i$-th patch of a stream $x$, time-domain token $j$ is
\begin{equation}
  \mathbf{u}^{\rm TD}_j = \bigl[\,A_{[j]},\, E_{[j]},\, \tau_j / T_{\rm obs}\,\bigr] \in \mathbb{R}^{2P+1},
  \label{eq:tok-td}
\end{equation}
and frequency-domain token $k$ is
\begin{equation}
\begin{aligned}
  \mathbf{u}^{\rm FD}_k = &\bigl[\,\mathrm{Re}\,\tilde{A}_{[k]}, \mathrm{Im}\,\tilde{A}_{[k]},
                            \mathrm{Re}\,\tilde{E}_{[k]}, \mathrm{Im}\,\tilde{E}_{[k]}, \\
                          &f_k / f_{\rm Nyq}, \log_{10}(f_k + f_1)\,\bigr] \in \mathbb{R}^{4P+2},
\label{eq:tok-fd}
\end{aligned}
\end{equation}
where $A$, $E$ ($\tilde{A}$, $\tilde{E}$) are the multibanded time- (frequency-) domain channels, $\tau_j$ and $f_k$ are the patch-centre time and frequency on the non-uniform grid, $\tau_j$ is measured from the window start and normalized by $T_{\rm obs}$, $f_{\rm Nyq} = 1/(2\Delta t)$, and $f_1 = 51.5\,\mu\mathrm{Hz}$ is the lowest bin frequency. Each frequency token carries its centre frequency both linearly and logarithmically, the log scale spreads the low-frequency inspiral bins evenly per decade, while the linear scale keeps resolution at the high-frequency end. Streams whose length is not a multiple of $P$ are zero-padded. This gives $119$ time-domain and $33$ frequency-domain tokens for $\SBIl$, and $441$ and $111$ for $\SBIe$. The start time $t_0$ is not a data token as it enters through the class token and the flow context (Sec.~\ref{sec:architecture}).
 
$A$ and $E$ record the same wavefront through two antenna patterns, so much of the information on sky position, polarization and distance lies in their relative amplitude and phase. Pairing the channels inside a token lets the first embedding layer form this comparison directly instead of leaving it to attention between per-channel tokens. The shorter sequence is a secondary benefit. Separate channel tokens would double the count from $152$ to $304$ for $\SBIl$ and $552$ to $1104$ for $\SBIe$. The explicit coordinate tells the network where a patch sits on the non-uniform grid, which the learned position embedding, indexed by token slot only, would otherwise have to infer.
 
Before tokenization, every input entry is standardized. Each time sample, and the real and imaginary parts of each Fourier bin, has its own mean subtracted and is divided by its own standard deviation, both estimated once from $8192$ noisy training windows ($1024$ random training sources $\times\ 8$ windows, generated as in (Sec.~\ref{sec:training}). Since this bin-wise standardization is based on the empirical distribution of the training data, it already captures the frequency-dependent noise scale. We do not additionally whiten the data using the noise PSD. A per-entry scale is needed because the spread of the data varies strongly along the window. The $\SBIl$ time-domain standard deviation spans three orders of magnitude and peaks in the final hours, which holds the late inspiral and the merger for $\Delta t_c < 0$. For the $\SBIe$ it still varies by a factor of $\sim 20$. A single global scale would leave most of the stream numerically negligible. In the frequency domain, the real and imaginary parts of a bin share one standard deviation, so the phase is not distorted. The source parameters are standardized at the flow input by a fixed affine map, then each is mapped to $[0, 1]$ over its prior range and z-scored with the mean and standard deviation of $512$ training draws. In $\SBIl$ the handedness fold (Sec.~\ref{sec:handedness}) acts between these two steps.

\subsection{Network architecture}
\label{sec:architecture}
 
Each network has two components: a transformer encoder~\cite{Attention-is-all-you-need_2017} that compresses the time- and frequency-domain tokens of Sec.~\ref{sec:target} into a fixed-length summary, and a conditional normalizing flow~\cite{SplineFlows_2019} that maps this summary to a posterior density. Both components are trained jointly (Sec.~\ref{sec:training}). The full layer-by-layer specification is given in Tables~\ref{table:network} and~\ref{table:transformer} of Appendix~\ref{app:network}.

\emph{Token embedding.} The time- and frequency-domain tokens are embedded separately to a width of $512$ and then concatenated into a single sequence with a learned class token. This gives $153$ tokens for $\SBIl$ and $553$ for $\SBIe$. Positional embeddings encode both the order of the patches and whether they come from the time- or frequency-domain view. Combining the two views in one sequence allows the transformer to jointly use the sharp timing information of the time-domain representation and the accumulated phase information of the frequency-domain representation.

\emph{Observation start time.} The start time $t_0$ (stream start for $\SBIl$, window start for $\SBIe$. Sec.~\ref{sec:datagen}) determines where the LISA constellation is on its orbit and therefore the detector response. We encode its orbital phase using $\sin(2\pi t_0)$ and $\cos(2\pi t_0)$, with $t_0$ in years. Since the LISA orbit is only approximately periodic, we also include the second harmonic, $\sin(4\pi t_0)$ and $\cos(4\pi t_0)$, to allow the network to capture deviations from this simple periodic dependence. These features enter the network in two places. All four are embedded and added to the class token, making the detector geometry available throughout the transformer, while $\sin(2\pi t_0)$ and $\cos(2\pi t_0)$ are also appended directly to the summary passed to the flow.

\emph{Encoder.} The combined sequence is processed by six transformer blocks, each with eight attention heads. Attention is applied over the full sequence, allowing information to be exchanged between all time- and frequency-domain patches. The resulting sequence is compressed to a $128$-dimensional summary before being passed to the flow.
 
\emph{Context summary.} The output at the class-token position summarizes the window: the class token attends to every data token in each block, and the other outputs are discarded. For $\SBIe$, whose input sequence is much longer, we additionally use attention pooling with eight learned query vectors. Each query attends over the full token sequence, producing eight summaries that are combined and added to the class-token summary. The resulting representation is mapped to $128$ features, followed by a layer normalization and a two-layer perceptron, giving the summary $\mathbf{z}$. With the start-time columns, this forms a $130$-dimensional flow context. In $\SBIl$, a linear map of the context to one logit, the handedness head of Sec.~\ref{sec:handedness}, predicts $\mathrm{sgn}(\sin\beta\cos\iota)$ for the auxiliary loss of Eq.~\eqref{eq:handaux}.
 
\emph{Flow.} The posterior $q(\theta \mid d, t_0)$ is modeled with a conditional rational-quadratic neural spline flow~\cite{SplineFlows_2019}. The flow takes the $130$-dimensional network summary as context and maps the inferred parameters to a standard normal base distribution. For $\SBIl$, the parameters first pass through the handedness fold of Sec.~\ref{sec:handedness}. The two networks use the same flow construction. We use a normalizing flow because it provides both posterior samples and the density $q(\theta \mid d, t_0)$ required for importance sampling (Sec.~\ref{sec:is}). The full flow specification is also given in Table~\ref{table:network}.

\emph{Size.} $\SBIl$ has $15.0$M trainable parameters: $13.6$M in the tokenizer and encoder, of which $0.27$M embed the start time $t_0$, $1.38$M in the flow, and $131$ in the handedness head. $\SBIe$ has $16.5$M: $13.8$M in the tokenizer and encoder, $1.6$M in the attention-pool readout and $1.08$M in the flow.

\subsection{Training: a two-stage curriculum}
\label{sec:training}

The encoder and flow are trained jointly from random initialization by minimizing the expected negative log-posterior, $-\log q(\theta \mid d, t_0)$, with the handedness loss of Eq.~\eqref{eq:handaux} added for $\SBIl$.

$\SBIl$ is trained in two stages. In the first stage, the network learns the source physics, sliding-window structure and handedness at a fixed constellation configuration. In the second, this pretrained network is extended across a month of observation start times, so that inference is amortized over the changing detector response. $\SBIe$ is trained in a single stage directly on streams whose start times span the full $t_0$ prior (Sec.~\ref{sec:datagen}). It does not use the handedness loss because it does not infer the sky position or inclination.
 
\emph{Batches.} Training runs on NVIDIA A100 GPUs with synchronous data parallelism. Each optimizer step averages the gradient over $16\,384$ examples: eight windows for each of $2048$ sources for $\SBIl$ and four windows for each of $4096$ sources for $\SBIe$ (Sec.~\ref{sec:sliding}). Validation windows are cut from the held-out files with window positions and noise realizations frozen, so the validation loss is comparable across epochs.
 
\emph{Optimization.} AdamW~\cite{Loshchilov2019} with weight decay $10^{-5}$ and gradient-norm clipping at $5$. In each stage the learning rate rises linearly during a short warm-up~\cite{Goyal2017} and then follows one cosine cycle~\cite{Loshchilov2017} to a floor, updated at every step. No dropout is used: with $\sim 10^{7}$ distinct sources and a new noise realization and new window positions each time a source is drawn, no training window is ever seen twice, so memorization is not a concern.
 
\emph{Stage 1: one observation start time.} All training sources share $t_0 \simeq 86\,\mathrm{d}$, fixing the constellation configuration so that the network first learns the source parameters and sliding-window structure. The training set contains $1.7\times10^{7}$ sources, of which $3.9\times10^{5}$ sources ($3.1\times10^{6}$ observations) are sampled per epoch. The handedness fold and auxiliary loss of Sec.~\ref{sec:handedness} are included in this stage.

\emph{Stage 2: amortization over the start-time prior.} We then train on a new set of $1.7\times10^7$ sources spanning $512\,t_0$ samples, and hence constellation configurations. The validation set contains independent sources and $t_0$ values, testing generalization to unseen configurations. The stage-1 weights are retained, while the new $t_0$ inputs of Sec.~\ref{sec:architecture} are initialized to have zero contribution. The network therefore starts stage~2 from the stage-1 solution and only has to learn how the posterior changes with constellation configuration. This stage converges in about one seventh of the stage-1 epochs. We retain the stage-1 input standardization and restart the optimizer with a new warm-up and cosine cycle.

\subsection{Sky--inclination handedness}
\label{sec:handedness}

\begin{figure}[t]
  \centering
  \includegraphics[width=\columnwidth]{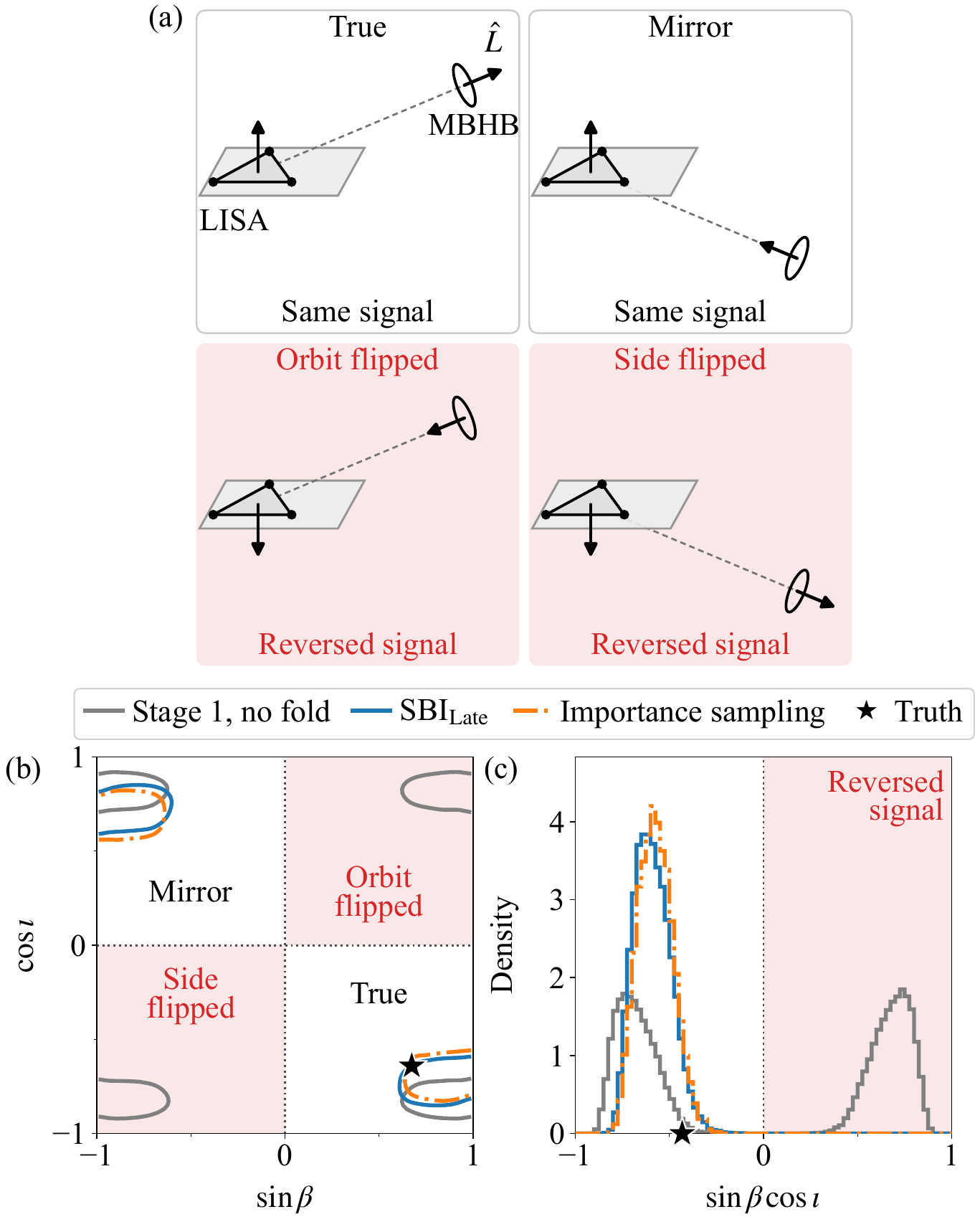}
  \caption{Sky--inclination handedness. (a) The four sign combinations of $(\sin\beta, \cos\iota)$ of an example signal, seen from the side of LISA's plane (gray sheet): the true one and its mirror image through the plane (top), and the single flips of the orbit or of the side of the plane (bottom, red). The source is drawn with its orbital angular momentum $\hat{L}$. LISA (triangle) is also shown with its angular momentum represented by the arrow. (b) Joint posterior $2\sigma$ contours of $(\sin\beta, \cos\iota)$ for the example signal in the window ending $1.5\,\mathrm{h}$ before merger, from the stage-1 network without the fold (gray), $\SBIl$ (blue), and importance sampling (orange, dash-dot). The star marks the truth, the dotted lines the quadrant boundaries, and the red quadrants the reversed sense. (c) Distribution of $\sin\beta\cos\iota$ for the same three sample sets. The handedness loss term corresponds to the sign of $\sin\beta\cos\iota$.}
  \label{fig:handedness}
\end{figure}

This subsection concerns $\SBIl$ only. In the L-frame, reflecting the source about the constellation plane ($\sin\beta \to -\sin\beta$) while flipping the inclination ($\cos\iota \to -\cos\iota$) leaves the signal nearly unchanged. Fig.~\ref{fig:handedness} illustrates the real and false modes of the $\sin\beta -\cos\iota$ joint posterior. The real posterior modes correspond to the true configuration and its mirror image, at $(\sin\beta^*, \cos\iota^*)$ and $(-\sin\beta^*, -\cos\iota^*)$, respectively. Neither sign is measurable alone, but their product, which we refer to as the handedness $h = \mathrm{sgn}(\sin\beta\,\cos\iota)$, is. It fixes the sense in which the polarization rotates in the constellation plane, and with it the sign of the relative phase between $A$ and $E$. Flipping only one of the two signs reverses that phase, and the likelihood down-weights the single-flip configurations $(\sin\beta, -\cos\iota)$ and $(-\sin\beta, \cos\iota)$, shown in panel (a), shaded.

A network trained as described so far does not learn $h$ with our limited training budget. Every such network we trained, across different network sizes, places approximately equal mass on all four sign combinations (panel~(b), gray; see Appendix~\ref{app:handedness}). Half of each posterior therefore lies in modes excluded by the data, which also caps the importance-sampling efficiency at $\epsilon \le 0.5$ for any source with $\sin\beta \neq 0$.

The handedness information is nevertheless present in the transformer. A linear probe on the class-token output of the final transformer block recovers $h$ with $99.5\%$ accuracy on noisy windows, whereas the same probe on the context passed to the flow recovers the correct modes $50\%$ of the time. We suspect that optimization becomes trapped near the symmetric solution. Learning the correct posterior requires two things to happen together: the transformer summary must retain $h$, and the flow must use it to model the joint sign structure of $\cos\iota$ and $\sin\beta$. Neither change alone improves the loss. The intervention below is designed to break this coupling.

We remove the local minima in the loss landscape with two changes. First, the flow models a folded parameter vector, its first transform replaces $\cos\iota$ by $\cos\iota\,\mathrm{sgn}(\sin\beta)$,
\begin{equation}
  F:\ (\cos\iota,\ \sin\beta) \mapsto \bigl(\cos\iota\,\mathrm{sgn}(\sin\beta),\ \sin\beta\bigr),
  \label{eq:handfold}
\end{equation}
with $\mathrm{sgn}(0) \equiv +1$ and all other parameters unchanged. $F$ is its own inverse and $|\det J_F| = 1$ on each half of parameter space, so $\log q(\theta \mid d, t_0) = \log \tilde{q}(F(\theta) \mid d, t_0)$ exactly, where $\tilde{q}$ is the density modeled by the flow. The rest of the pipeline, importance sampling included, still sees $\theta$ in its original coordinates. After the fold, the handedness is encoded in the sign of a single coordinate, $\cos\iota,\mathrm{sgn}(\sin\beta)$. The flow can therefore learn this dependence directly.

Second, during training, we add a linear head that predicts the handedness $h$ from the flow context. Its binary cross-entropy $\mathcal{L}_{\rm BCE}$ is added to the posterior loss,
\begin{equation}
\begin{aligned}
\mathcal{L} ={}&
-\log q(\theta \mid d, t_0) \\
&+ w\,\mathcal{L}_{\rm BCE}\left(
\sigma\left(\texttt{L}_{130\to1}(\mathbf{c})\right),
\mathbf{1}[h>0]
\right),
\end{aligned}
\label{eq:handaux}
\end{equation}
where $\mathbf{c}$ is the $130$-dimensional flow context, $\texttt{L}_{130\to1}$ is a linear classification head, $\sigma$ is the logistic sigmoid function, and $w=5$ is an empirically chosen weight. The binary handedness labels, $\mathbf{1}[h>0]$, are computed directly from the injected source parameters. The linear classification head is used only during training and discarded at inference.

This auxiliary loss encourages the context to retain handedness information, while the folded parameterization allows the flow to use it. With both changes, the network assigns posterior mass only to the true configuration and its mirror image, and the importance-sampling efficiency approximately doubles. Appendix~\ref{app:handedness} shows where the handedness information is lost in the original network and how these two changes recover it.

\subsection{Importance sampling}
\label{sec:is}

The flow provides a rapid approximation $q(\theta \mid d, t_0)$ to the posterior, but the network approximation need not be exact. We therefore use importance sampling with the likelihood, similar to Ref.~\cite{Dax2023}. The flow posterior acts as a proposal distribution: reweighting its samples corrects deviations from the likelihood-based posterior, while the sampling efficiency provides a diagnostic of the proposal quality.

\emph{Weights.} For samples $\theta_i \sim q(\theta \mid d, t_0)$, $i = 1, \dots, N$, the importance weights are
\begin{equation}
  w_i = \frac{\mathcal{L}(d \mid \theta_i)\,\pi(\theta_i)}{q(\theta_i \mid d, t_0)}
  \,\propto \frac{\mathcal{L}(d \mid \theta_i)}{q(\theta_i \mid d, t_0)},
  \label{eq:isweights}
\end{equation}
where the second expression follows because the prior $\pi$ is uniform over the training bounds. We obtain likelihood-based posterior samples by resampling the flow samples with probability proportional to $w_i$.

\emph{Likelihood.} $\mathcal{L}$ is evaluated on the full-resolution analysis window, not on its multiband compression: the $16\,384$ samples per channel at $\Delta t = 32\,\mathrm{s}$ that the network's input was built from.
We use the exact Gaussian likelihood in the time domain, $\ln\mathcal{L} = -\frac{1}{2}\langle d,d\rangle +\langle d, h \rangle - \tfrac{1}{2}\langle h, h \rangle$ up to a constant, with $\langle a, b \rangle = \sum_{c \in \{A, E\}} \mathbf{a}_c^{\mathsf T} \mathbf{C}_c^{-1} \mathbf{b}_c$, where $\mathbf{C}_c$ is the noise covariance of channel $c$ built from the same PSD as the network training simulations, Galactic confusion included. Working in the time domain avoids the periodicity assumed by a frequency-domain likelihood on a finite window, which matters here because the merger can lie close to the end of window, where it would be discarded due to the tapering procedure. 
The templates use the same waveform model, mode content, orbits, and second-generation TDI as the simulations.

\emph{Marginalization over $\phi$ and $\psi$.} The two angles not inferred by the network are integrated out inside the likelihood. Their dependence factorizes: the phase $\phi$ multiplies each mode $(\ell, m)$ by $e^{i m \phi}$, and the polarization angle rotates the two polarizations by $2\psi$. 
Since the LISA response acts linearly for a fixed sky position we can write the template as:
\begin{equation}
    \begin{split}
    h(\mathbf{\theta},\,\phi,\,\psi) = \sum_{\ell,m} h_{\ell m}(\mathbf{\theta},0,0)\cos(2\psi+m\phi) +\\
    h_{\ell m}(\mathbf{\theta},0,\pi/4)\sin(2\psi+m\phi).
    \label{eq:mode_decomposition}
\end{split}
\end{equation}
The mode-by-mode inner products are therefore computed once per sample for the fixed values of phase and polarization, and the likelihood is then evaluated cheaply on a $1024 \times 512$ grid over $\phi \in [0, 2\pi]$ and $\psi \in [0, \pi]$ and integrated with the trapezoidal rule. 
We validated this integration method against a nested sampling run with all the parameters fixed with the exception of phase and polarization.
This marginalization allows the flow to focus only on the relevant nine parameters and yet be reweighted by the likelihood, which depends on 11 parameters.

For each analysis window we draw $N = 100\,000$ samples from the flow, compute their weights, and report the effective sample size $\mathrm{ESS} = \bigl(\sum_i w_i\bigr)^2 / \sum_i w_i^2$ and the sampling efficiency $\epsilon = \mathrm{ESS}/N$. The likelihood is vectorized on a GPU, where reweighting one window costs about five minutes on one GPU ($282\,\mathrm{s}$ median for $N = 10^5$; Sec.~\ref{sec:res-cost}), compared with $\approx 0.4\,\mathrm{s}$ for drawing the $10^5$ samples from the flow.

\begin{figure}[t]
    \centering
    \includegraphics[width=\columnwidth]{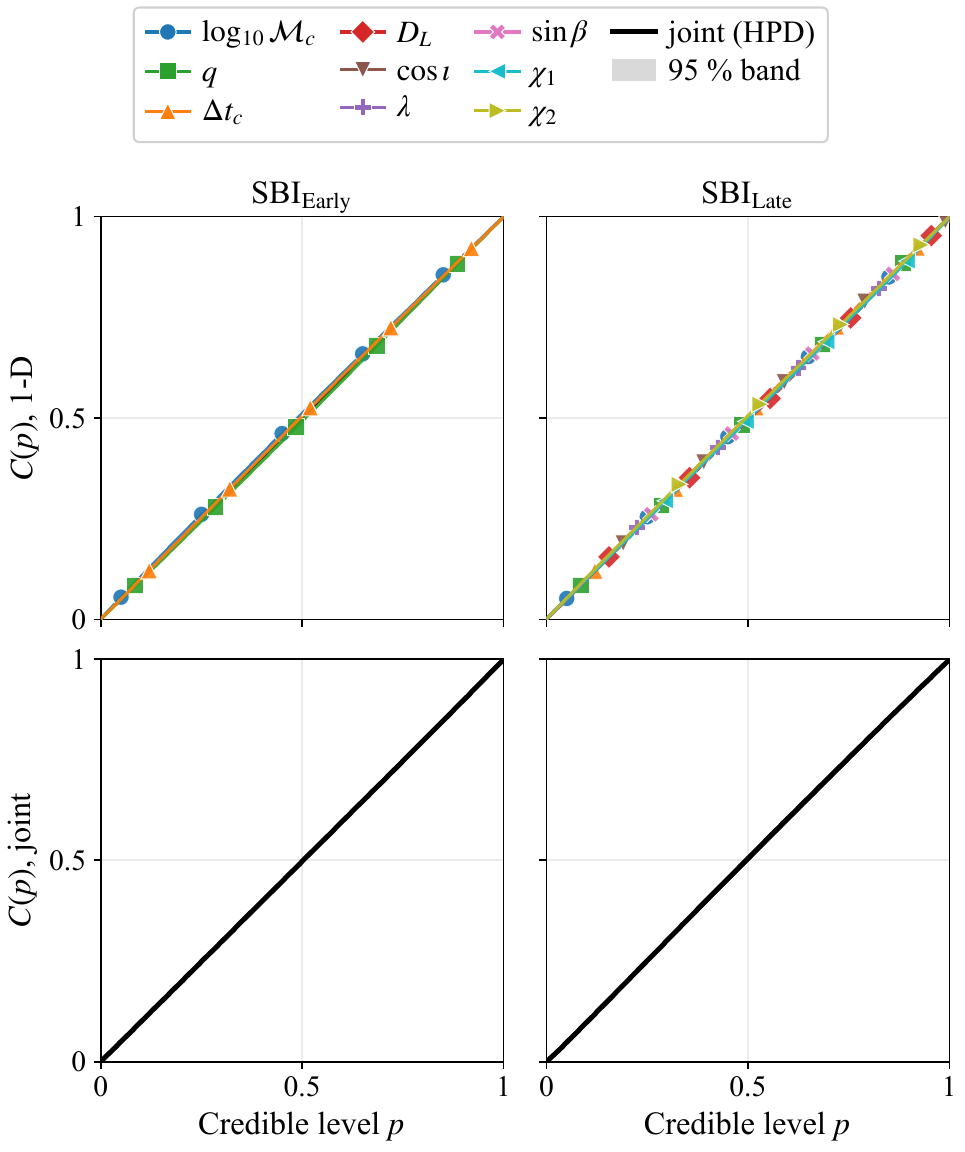}
    \caption{Calibration (P--P) of $\SBIe$ (left) and $\SBIl$ (right) on $1.11\times10^{6}$ independent test signals each, drawn from the network's prior, with $1024$ posterior draws per signal. Top: for each parameter, the fraction $C(p)$ of signals whose true value lies below the $p$ quantile of its marginal posterior. Bottom: the fraction of signals whose true parameters lie inside the $p$ highest-posterior-density region of the joint posterior over all inferred parameters ($\log_{10}\mathcal{M}_c$, $q$, $\Delta t_c$ for $\SBIe$ and all nine for $\SBIl$). A calibrated posterior follows the diagonal.}
    \label{fig:pp}
\end{figure}

\section{Results}
\label{sec:results}

We first check that both networks are calibrated over their priors (Sec.~\ref{sec:res-calibration}), then follow a \emph{reference} signal through its last two days (Sec.~\ref{sec:res-countdown}), compare the network posteriors with the likelihood (Sec.~\ref{sec:res-likelihood}), and test $\SBIl$ on data containing additional sources (Sec.~\ref{sec:res-robustness}). Additional studies of $\SBIl$ across the source population and as a function of sky position are presented in Appendix~\ref{app:res-population},~\ref{app:res-sky}. Corner-plot contours and sky areas $\Omega_{86.5\%}$ are quoted at $86.5\%$ ($2\sigma$) confidence, while one-dimensional widths $\Delta_{68\%}$ correspond to $68\%$ ($1\sigma$) confidence. Results refer to the network posterior unless labeled importance sampled (IS).

Throughout the manuscript, we track the algorithm’s performance using a \emph{reference} signal (see \emph{Test sets} in Sec~\ref{sec:res-evolution}): one noise realization of a source, analyzed in windows that slide towards and past its merger, each window's posterior being the alert the pipeline would issue at that moment. The \emph{reference} is located at a relatively unfavorable sky location: high L-frame latitude where sky localization is expected to be less precise than for an otherwise comparable source near the equator. This source provides a balance between high SNR but relatively low sky resolution. For completeness, we also explicitly compare results with the same source at a more-favorable, equatorial sky-location. The equatorial source has the same parameters as the \emph{reference} but with $\sin\beta = 0$.

\subsection{Calibration}
\label{sec:res-calibration}
 
Fig.~\ref{fig:pp} shows the probability-probability (P--P) plot of the nominal vs empirical credible level, where the ideal curve is the $45^\rm{o}$ diagonal. The top and bottom panels show the PP plot for the one-dimensional marginal and joint distributions, respectively. A diagonal PP plot for both the marginal and joint distributions is a necessary condition for an on-average (over source parameters) calibrated network. We show that both networks are calibrated over their priors. On $1.1\times10^{6}$ signals each, the largest deviation of a one-dimensional coverage curve from the diagonal is $1.2\,\%$ for $\SBIe$ ($\log_{10}\mathcal{M}_c$) and $0.8\,\%$ for $\SBIl$ ($D_L$). The joint coverage curves, which also test correlations and, for $\SBIl$, the multimodal sky--inclination structure (Sec.~\ref{sec:handedness}), deviate by at most $0.4\,\%$ and $0.6\,\%$. With $10^{6}$ signals, these deviations are statistically resolved but small.
 
Calibration is an average over the prior, a network that extracts only part of the information can be calibrated and still be wider than the likelihood posterior for a given signal. Secs.~\ref{sec:res-countdown} and~\ref{sec:res-likelihood} examine individual signals.

\subsection{One event through the pipeline}
\label{sec:res-evolution}

\begin{figure*}[t]
    \centering
    \includegraphics[width=.9\textwidth]{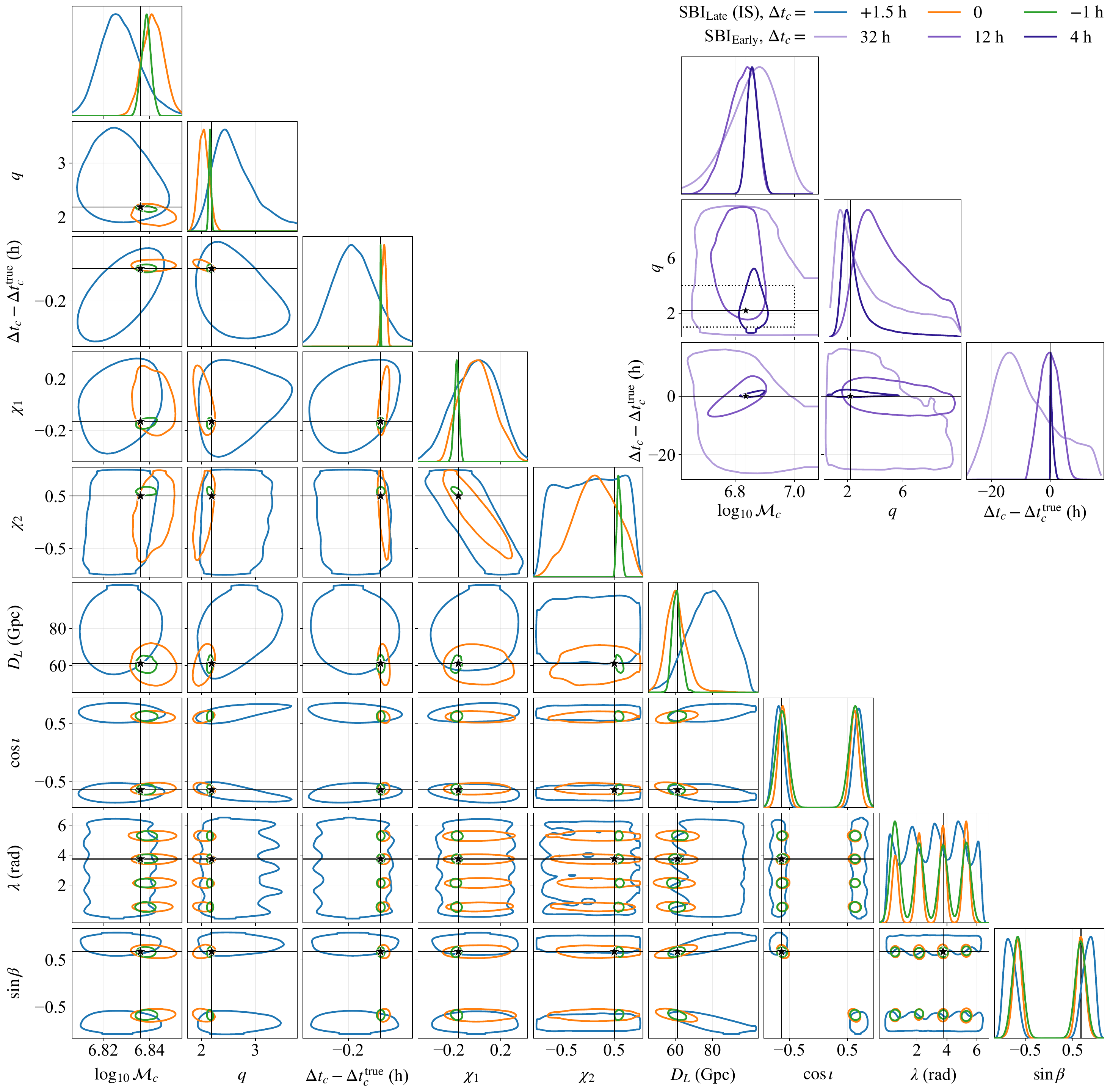}
    \caption{The \emph{reference} signal through both networks on one simulated data stream. Main corner and sky inset: importance-sampled $\SBIl$ posterior in windows ending $1.5\,\mathrm{h}$ before merger ($\Delta t_c = +1.5\,\mathrm{h}$, blue), at merger ($\Delta t_c = 0\,\mathrm{h}$, orange) and $1\,\mathrm{h}$ after it ($\Delta t_c = -1\,\mathrm{h}$, green). The likelihood-reweighted, resampling efficiencies are $27\%$, $2\%$ and $0.2\%$, respectively. Upper right: $\SBIe$ network posterior of $\log_{10}\mathcal{M}_c$, $q$ and $\Delta t_c$ in windows ending $32$, $12$ and $4\,\mathrm{h}$ before merger (light to dark purple). The black dotted box within ($\log_{10}\mathcal{M}_c, q$) marks the $\SBIl$ prior. Contours enclose $2\sigma$ ($86.5\,\%$) of the highest probability density and the one-dimensional marginals are scaled to unit peak for ease of comparison. The coalescence time is shown as $t_c - t_c^{\rm true}$. Stars and solid black lines mark the true values.}
    \label{fig:corner_evolution}
\end{figure*}

\begin{figure*}[t]
\centering
\includegraphics[width=\textwidth]{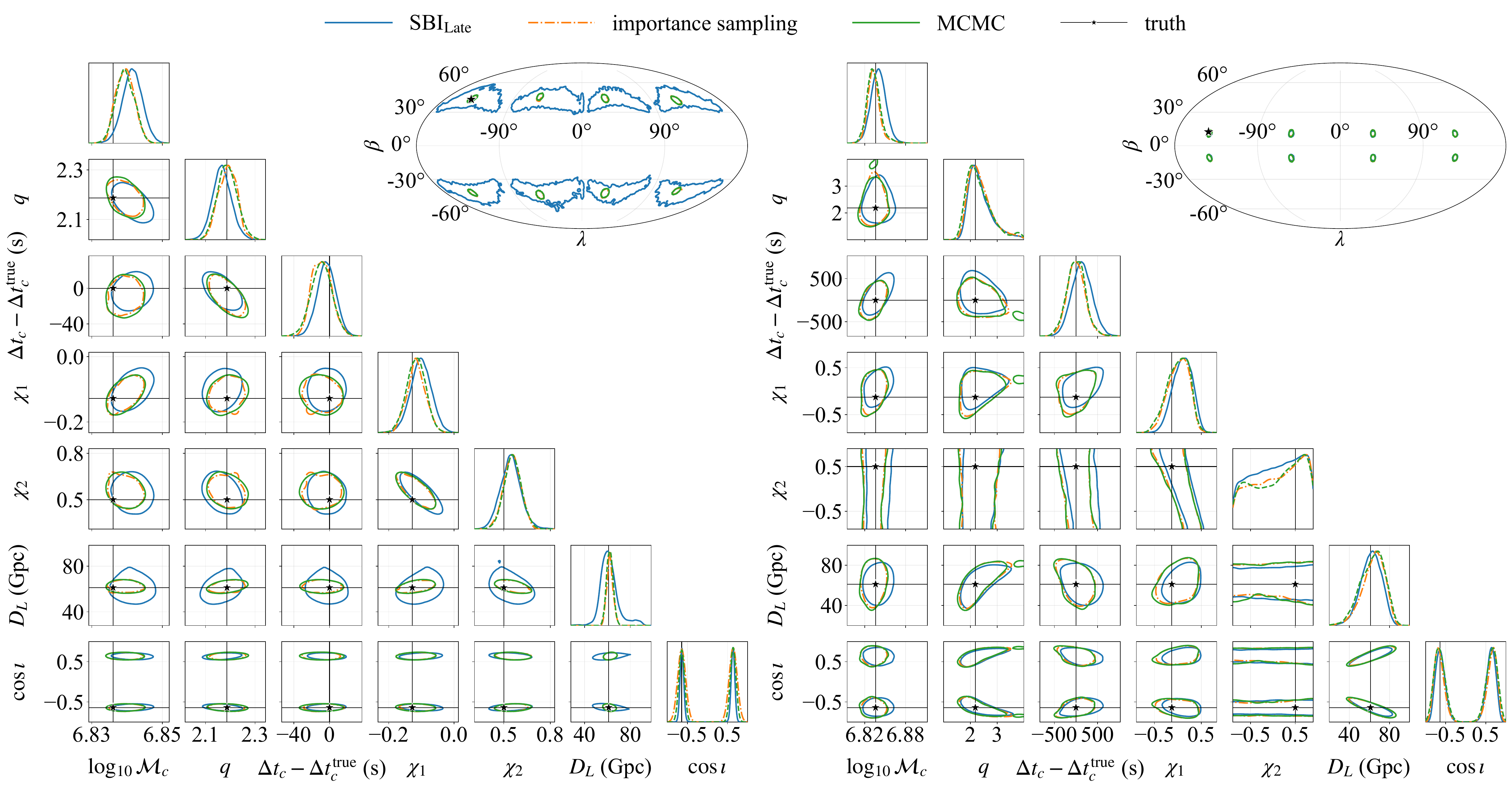}
\caption{$\SBIl$ (blue), importance-sampled (orange, dash-dot) and parallel-tempered MCMC (green) posteriors for the \emph{reference} signal $30\,\mathrm{min}$ \textit{after} merger (left) and the same source moved near the constellation plane ($\sin\beta = 0.21$) and $30\,\mathrm{min}$ \textit{before} merger (right). We show the sky map in the inset, although we infer all parameters jointly. Contours enclose the $2\sigma$ ($86.5\,\%$) interval. The diagonal 1-dimensional marginals are scaled to unit peak. Stars and black lines mark the true values. The IS efficiency (effective sample size per 100 network samples) is 0.34\% on the left and $16.8\,\%$ on the right.}
\label{fig:mcmc}
\end{figure*}

The test sets follow one source as its merger approaches, as an alert pipeline would. We fix a \emph{reference} source within the training priors ($\log_{10}\mathcal{M}_c = 6.84$, $q = 2.19$, $\chi_1 = -0.13$, $\chi_2 = 0.50$, $D_L = 61.1\,\mathrm{Gpc}$, $\cos\iota = -0.64$, $\lambda = 3.75$, $\sin\beta = 0.67$, $\phi = 0.82$, $\psi = 2.31$ with an SNR of $21$ in the window ending $4\,\mathrm{h}$ before merger and $297$ in the window ending $30\,\mathrm{min}$ after it). Its full inspiral--merger--ringdown stream is simulated once, at $t_0 = 85.74\,\mathrm{d}$, and a single noise realization is added. The windows are cut on a descending grid of merger times. For $\SBIl$ the grid runs from $\Delta t_c = +4\,\mathrm{h}$ to $-1\,\mathrm{h}$ in $10$-min steps, refined to $1$~min between $+30$ and $-10$~min, resulting in $67$ windows. For $\SBIe$ it runs from $+2\,\mathrm{d}$ to $+4\,\mathrm{h}$, split into 265 windows at 10 minute intervals. Consecutive windows share signal and noise, as successive analyses of real data would. 

Fig.~\ref{fig:corner_evolution} follows the \emph{reference} signal through both networks on one data stream: three $\SBIe$ windows with $\Delta t_c=+32,\,+12,\,4$~h, and three $\SBIl$ windows with ($\Delta t_c=+1.5,\,0,\,-1$~h), the latter importance resampled. The early posterior is prior-dominated at $32\,\mathrm{h}$ (SNR $\simeq 5$): the mass ratio fills its prior, the merger time has a $2\sigma$ width of $21\,\mathrm{h}$ with its median offset by $\simeq10\,\mathrm{h}$, and only the chirp mass begins to concentrate, with a long tail to its lower edge ($\simeq0.5\,\mathrm{dex}$ at $2\sigma$). By $12\,\mathrm{h}$ (SNR $\simeq10$) the chirp mass is known to $\simeq0.1\,\mathrm{dex}$ and the merger time to $6.5\,\mathrm{h}$, and at $4\,\mathrm{h}$ (SNR $\simeq21$) to $0.04\,\mathrm{dex}$ and $\simeq1\,\mathrm{h}$, while the mass ratio remains poorly constrained (width of $2.9$). 

The $\SBIe$ estimate of $\log_{10}\mathcal{M}_c, \,q$ selects which $\SBIl$ network to run (see Sec.~\ref{sec:priors}). As the early posteriors lie predominantly within a single $\SBIl$ network prior ($\log_{10}\mathcal{M}_c \in [6.5, 7]$, $q \in [1,4]$), that network alone is evaluated for the \emph{reference} example.

Before the merger $\Delta t_c=+1.5$~h, $\SBIl$'s posteriors for $\log_{10}\mathcal{M}_c$, $\Delta t_c$, and $\cos\iota$ are well constrained with respect to their prior range ($\Delta t_c$ prior spans 5h), with narrow bimodality in $\cos\iota$. However, the mass ratio, spins, distance, and sky position are broad. Once the merger is in the window $\Delta t_c\le0$, all parameters narrow and the sky posterior separates into eight compact modes, four in longitude, $\pi/2$ apart, each mirrored through the constellation plane ($\beta \to -\beta$). Note that the $\cos\iota$-$\sin\beta$ posteriors contain none of the reflected wrong-handedness modes (here, off diagonal), as the fold of Sec.~\ref{sec:handedness} intends. The network itself places less than $0.03\%$ of its draws on them in every analysis window, where a network without the fold placed half (Appendix~\ref{app:handedness}). The IS efficiency falls from $27\,\%$ to $2\,\%$ and $0.2\,\%$ across the three windows indicating that once the merger is observed, the likelihood narrows faster than the network posterior, implying that the network requires further training with a denser dataset. This is one of the limitations of the current network (see Sec.~\ref{sec:limitations}).

\subsection{Comparison with the likelihood}
\label{sec:res-likelihood}

\emph{Comparison with MCMC.} Fig.~\ref{fig:mcmc} compares network, IS and parallel-tempered MCMC posteriors from the same data. On the left, we show the comparison for the \emph{reference} signal $30\,\mathrm{min}$ after merger. Both IS and MCMC agree in every parameter and in the sky. The network posterior has the right location and correlations but is wider, most visibly in the sky, where it covers eight broad patches around the eight compact modes and $D_L$, with a tail to $\sim80\,\mathrm{Gpc}$. The sky position alone can change the sky posterior and the IS efficiency drastically. The right panel shows the reference source moved closer to the constellation plane, with the analysis window ending 30 minutes before merger. Despite the lower accumulated SNR and the absence of the merger from the window, its sky posterior is confined to eight substantially more compact modes, which the network reproduces in agreement with MCMC. The IS efficiency is 16.8\%, compared with 0.34\% for the reference configuration in the left panel. We examine the dependence of sky localization on source geometry in detail in Sec.\ref{sec:res-evolution} and Appendix\ref{app:res-sky}.

\begin{figure}[t]
  \centering
  \includegraphics[width=\columnwidth]{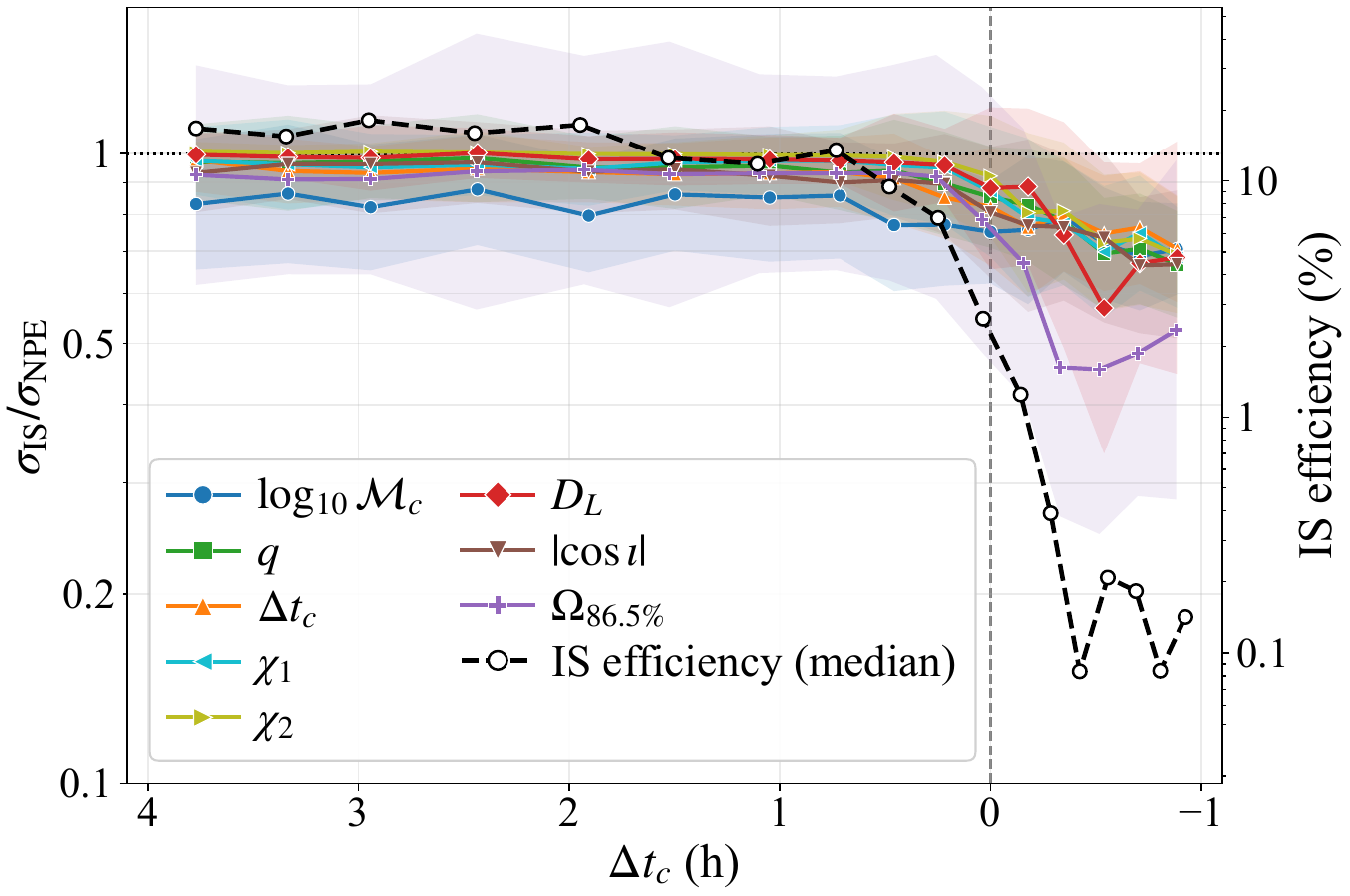}
  \caption{Importance-sampling check of the $\SBIl$ posteriors against time to merger, for $\sim1000$ validation sources stratified in $\Delta t_c$. Left axis: median ratio of the IS to the network posterior standard deviation, $\sigma_\mathrm{IS}/\sigma_\mathrm{NPE}$, per parameter, in $\Delta t_c$ bins of $50$ sources with $\mathrm{ESS} \ge 30$. $\Omega_{86.5\%}$ is the $86.5\,\%\,\mathrm{C.L.}$ sky area on the sky map. Bands show the source-to-source scatter, the median multiplied and divided by $e^{s}$, with $s$ the standard deviation of the log ratio in the bin. Right axis (black dashed, open circles): median IS efficiency, the effective sample size as a percentage of the network draws.}
  \label{fig:is_vs_npe}
\end{figure}

\emph{Network vs IS posteriors.} In Fig.~\ref{fig:is_vs_npe}, we importance sample the $\SBIl$ posteriors of $\sim1000$ validation signals drawn from the prior and plot the ratio of the standard deviation of the IS and network posteriors, stratified in $\Delta t_c$. We also show the IS efficiency along the secondary y-axis. Before merger, where the median IS efficiency is $7$--$18\,\%$, the IS uncertainty widths are within $\simeq 5\,\%$ of the network widths for most parameters, except $\log_{10}\mathcal{M}_c$, for which they are $15$--$25\,\%$ narrower, about $20\,\%$ on average. After merger, the network is, on average, $1.25$--$1.5$ times wider than the IS posterior and about twice as wide in sky area, while the median efficiency drops to $0.5\,\%$ and then $0.1\,\%$. Overall, the network closely matches the likelihood before merger and becomes systematically broader after merger.

The lower post-merger efficiency coincides with a substantial narrowing of the likelihood posterior relative to the training prior. Likely due to the substantial increase in SNR upon including the merger, the lower IS efficiency for higher SNR is seen in other SBI methods \citep{spadaro2026}. The sharp post-merger posteriors makes its structure more difficult for an amortized network to resolve with a fixed training budget. Even modest differences between the network and likelihood posteriors can then produce highly unequal importance weights. Increasing the density of training examples in these regions may improve efficiency, although alternative explanations, including limitations of the input compression and network architecture, remain to be investigated. We recommend reweighting a larger number of network samples for post-merger analyses, targeting an effective sample size of $\mathcal{O}(1000)$.

\subsection{Evolution of the posterior with time to merger}
\label{sec:res-countdown}
 
\begin{figure*}[t]
\centering
\includegraphics[width=\textwidth]{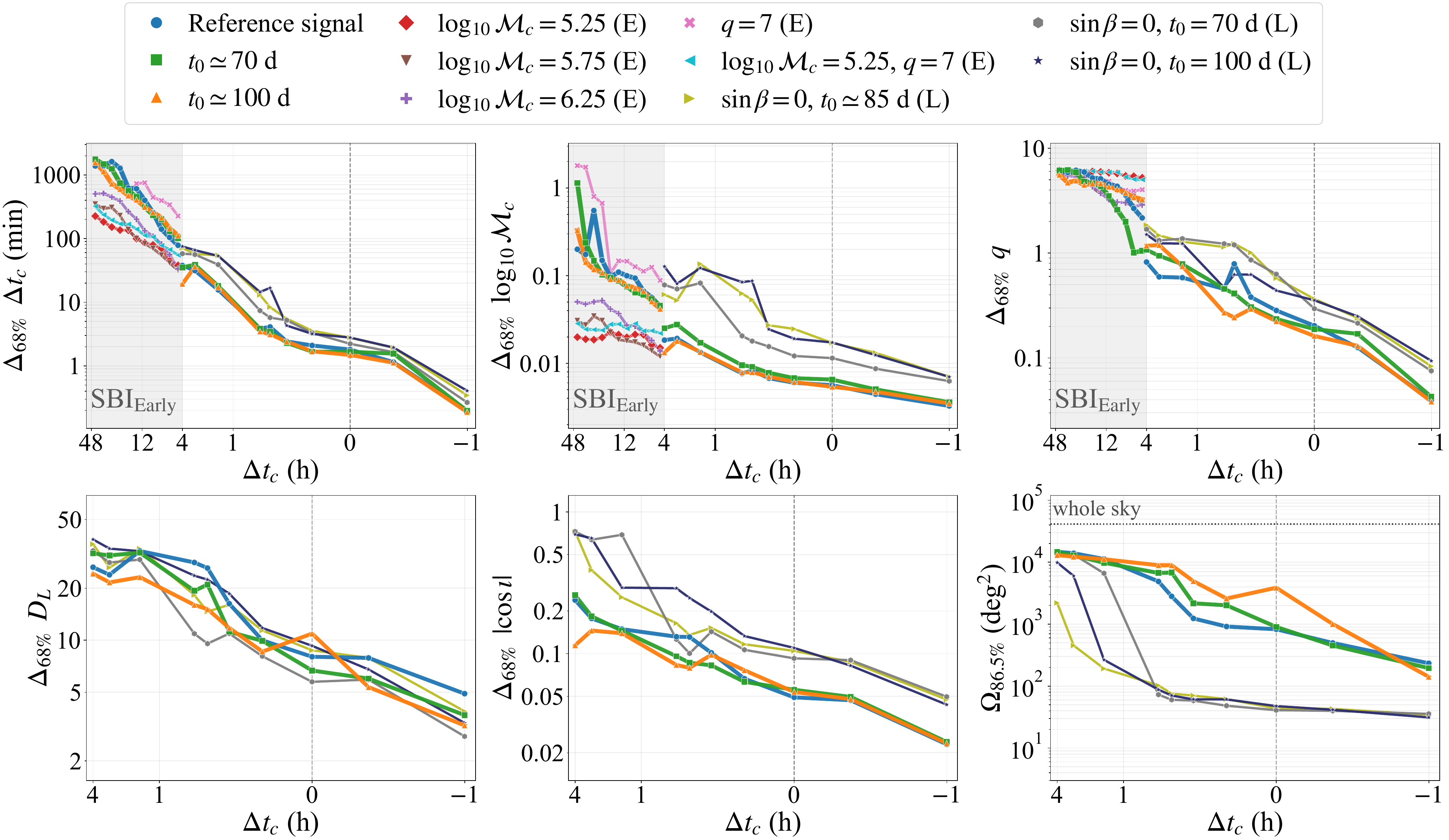}
\caption{Posterior precision during the final two days before merger for the parameters most relevant to an alert. We show the coalescence time, chirp mass, mass ratio, luminosity distance, absolute cosine of inclination, and sky area. The first three are inferred by both $\SBIe$ and $\SBIl$, while the remaining quantities are inferred only by $\SBIl$. Each curve follows one source through successive analysis windows and shows the $68\%$ credible-interval width $\Delta_{68\%}$ of the highest probability density as a function of $\Delta t_c$. $\SBIe$ covers $48$ to $4\,\mathrm{h}$ before merger, shown by the shaded region, and is plotted using the median width in bins of $\Delta t_c$. $\SBIl$ covers $4\,\mathrm{h}$ before to $1\,\mathrm{h}$ after merger and uses importance-sampled posteriors. The dashed vertical line marks the merger. The sources are the \emph{reference} signal and the parameter variants listed in the legend. Labels (E) and (L) mark sources analyzed only by $\SBIe$ or $\SBIl$ because they lie outside the prior of the other network.}
\label{fig:countdown}
\end{figure*}

Figure~\ref{fig:countdown} shows the precision of the alert parameters as the merger approaches. For the \emph{reference} signal, the two networks analyze the same data stream across their respective time ranges. The same noise realization is used across the sources common to $\SBIe$ and $\SBIl$, so the transition at $4~\mathrm{h}$, where the inference switches from $\SBIe$ to $\SBIl$, compares the two networks on the same underlying event.

We repeat this analysis for several controlled variations of the \emph{reference} signal. To test the dependence on observation epoch, the \emph{reference} signal is also placed near the two edges of the $t_0$ prior at $73.9$ and $99.5\,\mathrm{d}$. For $\SBIl$, we additionally place the source in the constellation plane, $\sin\beta=0$, at $t_0=70$, $85.7$, and $100\,\mathrm{d}$. For $\SBIe$, which covers the wider mass prior, we consider $\log_{10}\mathcal{M}_c=6.25$, $5.75$, and $5.25$, together with a $q=7$ variant and a combined $\log{10}\mathcal{M}_c=5.25$, $q=7$ case.

We focus here on quantities needed for early warning and electromagnetic follow-up. The spins become more relevant near and after merger, where the network posterior can instead be used to initialize likelihood-based inference (Sec.~\ref{sec:hybridMCMC} and Fig.~\ref{fig:population}).

The $\SBIe$ curves show the network posterior, while the $\SBIl$ curves use importance sampling. The transition therefore compares the early network posterior with the likelihood-corrected late posterior. Because the extrinsic posteriors are strongly multimodal, we show $|\cos\iota|$ and the $86.5\%\,\mathrm{C.L.}$ sky area $\Omega_{86.5\%}$ rather than marginal widths in $\lambda$ and $\sin\beta$. The latter is quoted at $86.5\%~\mathrm{C.L}$ throughout the manuscript because it is a two-dimensional credible region.

The top row of Fig.~\ref{fig:countdown} shows the three parameters inferred by both networks. For the \emph{reference} signal, the uncertainty in coalescence time falls from about $17\,\mathrm{h}$ two days before merger to $\sim12\,\mathrm{s}$ one hour after it. The chirp-mass uncertainty also falls by more than two orders of magnitude and reaches below $0.01\,\mathrm{dex}$ near merger. The mass ratio improves later. It remains only weakly constrained through most of the early-inspiral window and begins to narrow substantially during the final $\sim12\,\mathrm{h}$. Most of the improvement occurs in the last few hours as higher harmonics become stronger. The curves obtained at different observation epochs largely overlap, showing that the $t_0$ amortization remains stable across the tested range.

\emph{Early warning.} The \emph{reference} signal accumulates SNR slowly. The SNR is approximately $4$, $6$, $11$, and $21$ in windows ending $48$, $24$, $12$, and $4\,\mathrm{h}$ before merger. The chirp mass becomes well constrained between about $20$ and $12\,\mathrm{h}$ before merger. Four hours before merger, the coalescence-time width is $\approx65\,\mathrm{min}$ and the chirp-mass width is $0.04\,\mathrm{dex}$, about $9\%$ in $\mathcal{M}_c$. At the start of the $\SBIl$ range, the corresponding widths are about $40\,\mathrm{min}$ and $0.02\,\mathrm{dex}$. The two networks are trained independently and use different priors and compressions, so an exact match at the transition is not expected. For an alert stream, $\SBIe$ provides the earlier estimate needed for scheduling, while the more precise $\SBIl$ posterior becomes available closer to merger when sky localization also becomes useful.

The time at which a signal becomes informative depends strongly on the source mass. The lighter variants with $\log_{10}\mathcal{M}_c\le6.25$ are already well inside the sensitive band two days before merger. At that time their chirp masses are already measured to $0.02$--$0.05\,\mathrm{dex}$ and their coalescence times to about $3$--$9\,\mathrm{h}$. Four hours before merger, these improve to $\approx0.010\,\mathrm{dex}$ and $25\,\mathrm{min}$. Their mass ratios remain much less constrained. A larger mass ratio can further delay informative early warning inference. For the $q=7$ variant of the \emph{reference} source, the SNR is only $13$ at $4\,\mathrm{h}$ and the coalescence-time width remains about $3.8\,\mathrm{h}$.

\emph{Near and after merger.} Once the merger enters the window, the common parameters continue to improve rapidly. At $\Delta t_c=0$, the widths in coalescence time, chirp mass, and mass ratio are approximately $1.8\,\mathrm{min}$, $0.006\,\mathrm{dex}$, and $0.2$. One hour after merger, they reach about $12\,\mathrm{s}$, $0.003\,\mathrm{dex}$, and $0.04$. The in-plane source has broader posteriors in these parameters because it is fainter than the \emph{reference} signal. Its SNR is $146$ compared with $297$ for the \emph{reference} signal at $\Delta t_c=-0.5\,\mathrm{h}$.

The lower row shows the quantities inferred only by $\SBIl$. Distance and inclination improve steadily as the merger approaches. Their precision is again worse for the in-plane source because of its lower SNR. The sky area behaves oppositely. For the \emph{reference} signal, the sky posterior remains at $\sim10^{4}\,\mathrm{deg}^2$ until close to merger. It falls to about $800\,\mathrm{deg}^2$ at merger and to about $230\,\mathrm{deg}^2$ one hour later. The posterior remains distributed over eight modes. By contrast, an in-plane source is localized much earlier. Its sky area is about $200\,\mathrm{deg}^2$ at $1.5\,\mathrm{h}$ before merger and falls below $100\,\mathrm{deg}^2$ from about $30\,\mathrm{min}$ before merger onward, depending on the observation epoch. At late times, the four remaining sky patches approach the smoothing floor of the map.

This difference is geometric rather than a consequence of SNR. In the constellation plane, the relative amplitudes in the $A$ and $E$ channels constrain the longitude throughout the inspiral, up to the $\pi/2$ degeneracy in longitude. The two sky modes related by reflection through the constellation plane also coincide. A generic source retains all eight modes until the merger provides additional directional information. The in-plane source can therefore have a substantially smaller sky area despite having a lower SNR. The geometry of these modes and their dependence on inclination and latitude are studied in detail in Appendix~\ref{app:res-sky}.

At one hour after merger, the inclination widths reach $\Delta_{68\%}(|\cos\iota|)\simeq0.02$ for the \emph{reference} signal and $\simeq0.05$ for the in-plane source. The corresponding distance widths are about $3$--$5\,\mathrm{Gpc}$ and $3$--$4\,\mathrm{Gpc}$. Spin evolution is discussed separately in Appendix~\ref{app:res-population}. For precision beyond the network posterior near merger, the network can instead be used to initialize likelihood-based inference as discussed in Sec.~\ref{sec:hybridMCMC}.

We characterize how the precision of each $\SBIl$ parameter varies across the prior using $\sim10^6$ signals in Appendix~\ref{app:res-population}. The corresponding trends in sky localization are examined in detail in Appendix~\ref{app:res-sky}.

\begin{figure*}[t]
  \centering
  \includegraphics[width=\textwidth]{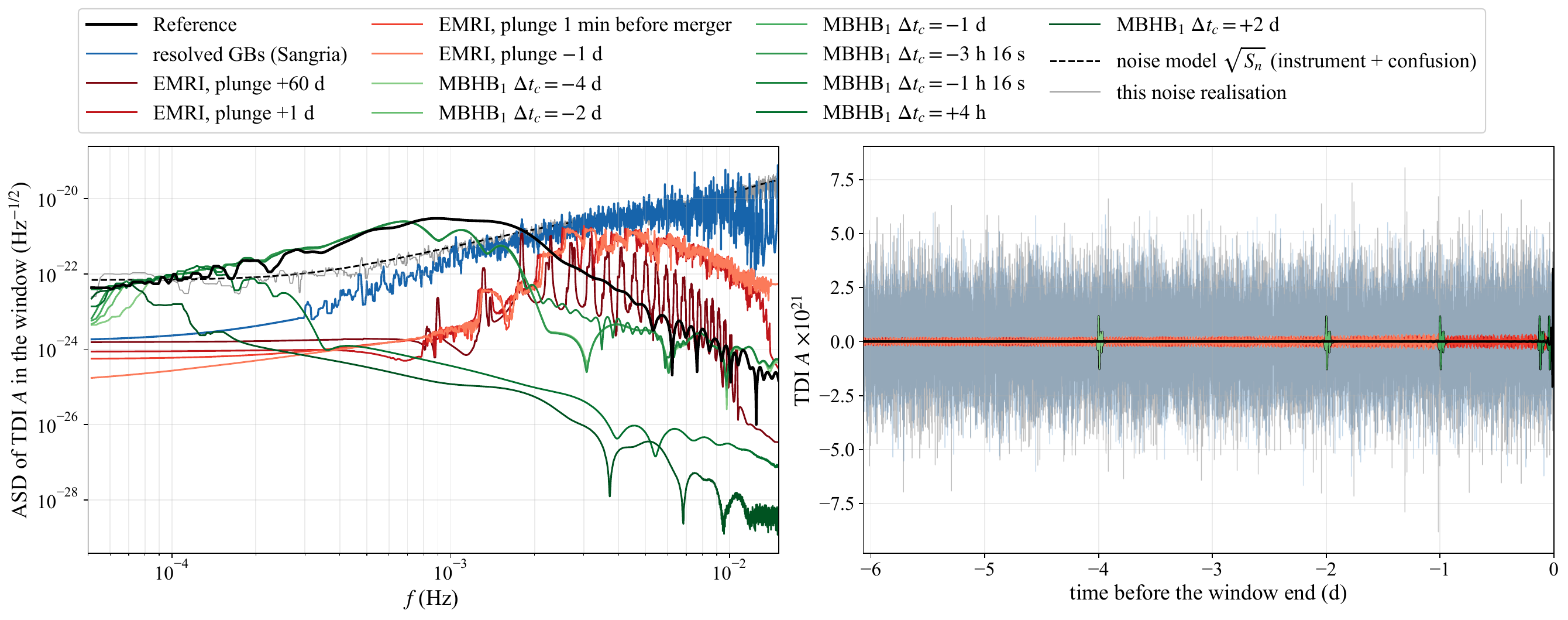}
  \caption{Signals injected in the robustness tests, within the 6.07-d window of the \emph{reference} signal ending 30\,min after its merger ($\Delta t_c=-30$\,min). Left: amplitude spectral density of TDI $A$ for each signal in the window --- the \emph{reference} signal, the resolvable Galactic binaries of the Sangria population (GBs), EMRIs and MBHBs at different merger times --- against the noise model $\sqrt{S_n}$ (instrument and Galactic confusion). Right: TDI $A$ over the window for all signals.}
  \label{fig:robustness-data}
\end{figure*}

\subsection{Robustness to other sources}
\label{sec:res-robustness}

\begin{figure*}[t]
  \centering
  \includegraphics[width=\textwidth]{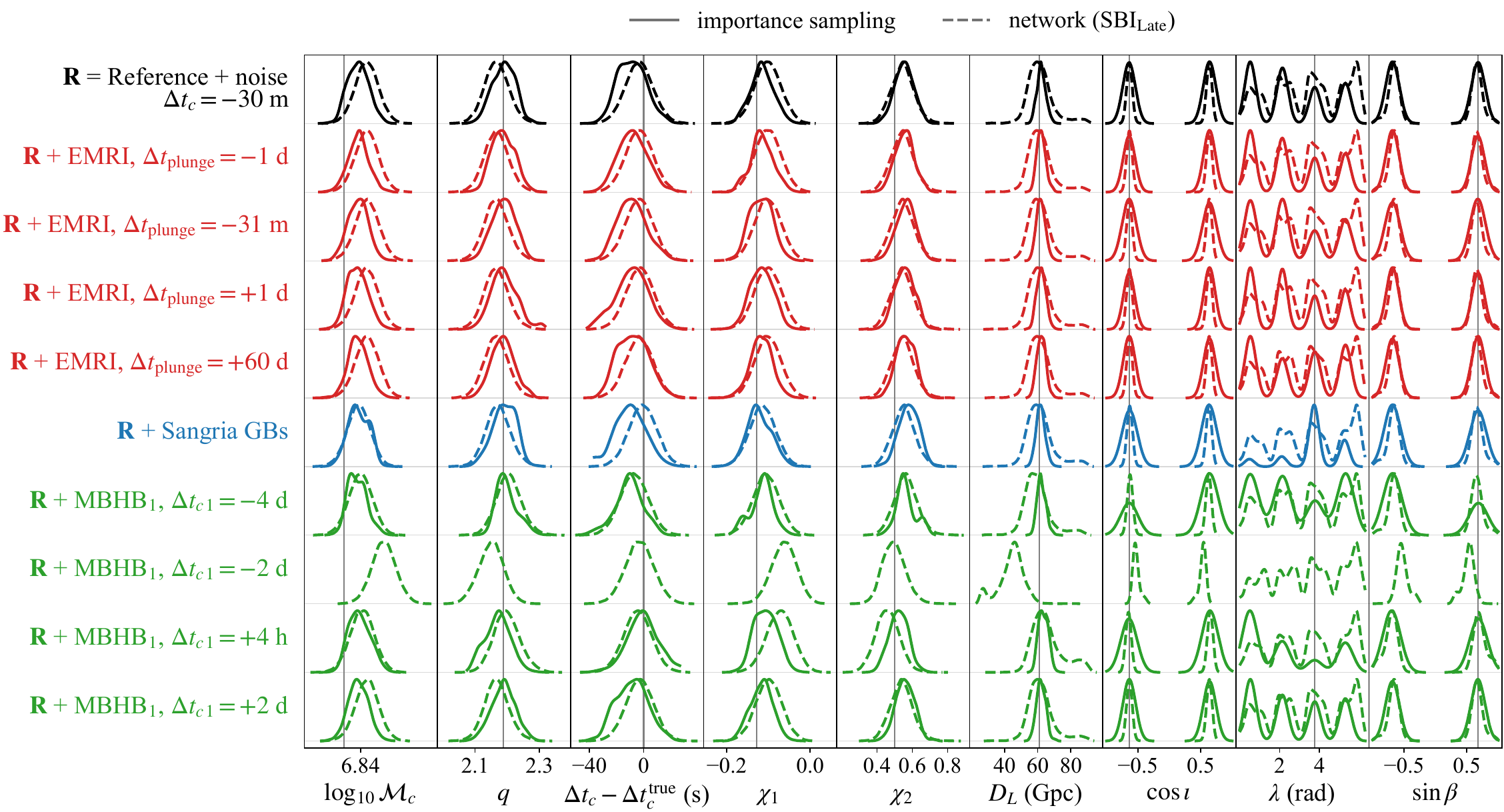}
  \caption{One-dimensional $\SBIl$ posteriors of the \emph{reference} signal, in the window ending $30\,\mathrm{min}$ after its merger, without (black) and with (colored) an added signal: an EMRI plunging $1\,\mathrm{d}$ before the window end, $1\,\mathrm{min}$ before the merger of the \emph{reference} signal, or $1\,\mathrm{d}$ or $60\,\mathrm{d}$ after the window end; the resolvable Galactic binaries of the Sangria population; or a second MBHB (MBHB$_1$) merging $4\,\mathrm{d}$ or $2\,\mathrm{d}$ before, or $4\,\mathrm{h}$ or $2\,\mathrm{d}$ after the window end. Signed times are relative to the end of the window. Dashed curves are network posteriors and solid curves are importance-sampled ones, drawn only where $\mathrm{ESS} \ge 50$. For MBHB$_1$ at $-2\,\mathrm{d}$ importance sampling fails ($\mathrm{ESS} = 1$). All rows share one noise realization. Vertical gray lines mark true values. }
  \label{fig:robustness-ridge}
\end{figure*}

\begin{figure*}[t]
  \centering
  \includegraphics[width=\textwidth]{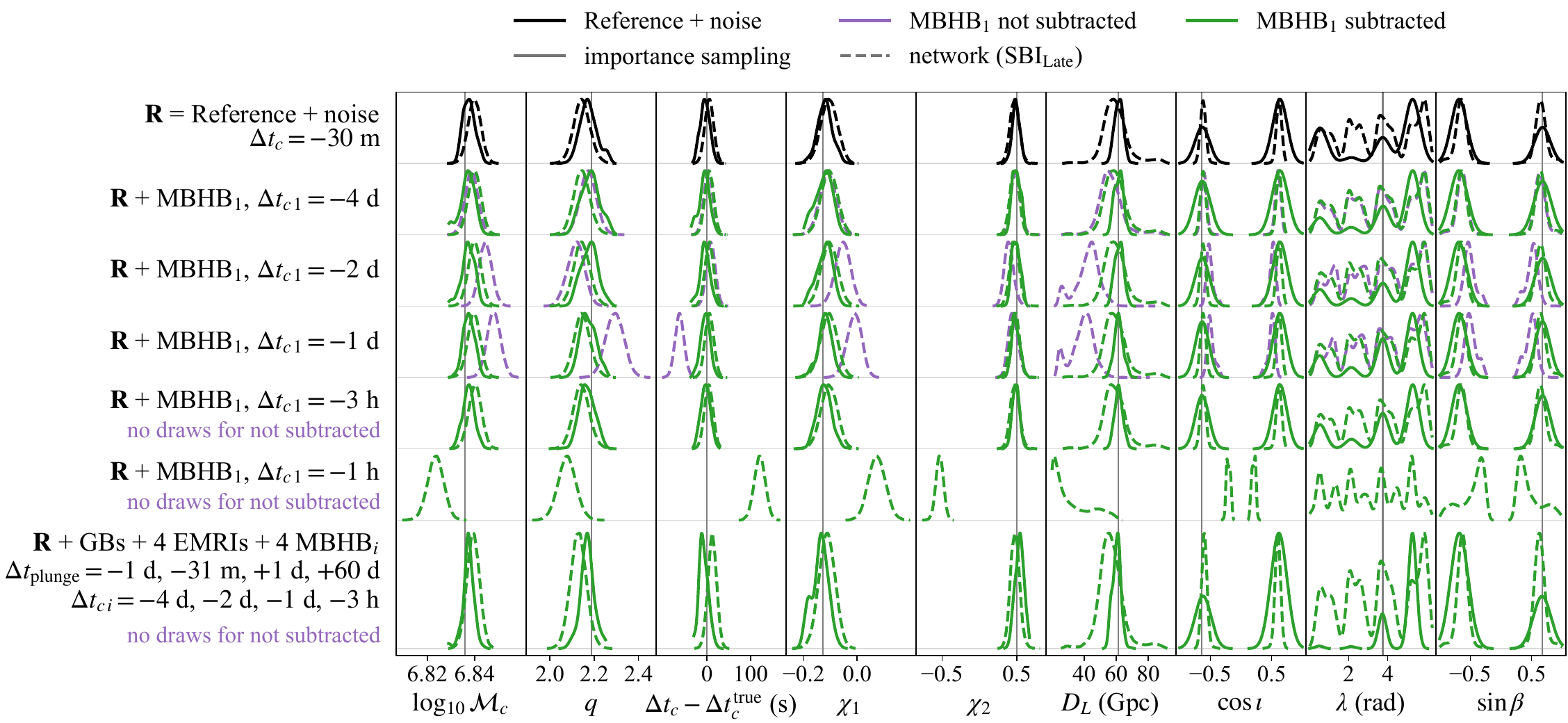}
  \caption{As Fig.\ref{fig:robustness-ridge}, but comparing the $\SBIl$ network posteriors of the \emph{reference} signal before (dashed) and after (solid) subtracting the waveform corresponding to the maximum a posteriori estimate of MBHB$_1$. In the last row, the data contain the resolvable Sangria Galactic binaries, the four EMRIs of Fig.\ref{fig:robustness-ridge}, and four copies of MBHB$_1$ merging $4\,\mathrm{d}$, $2\,\mathrm{d}$, $1\,\mathrm{d}$, and $\simeq3\,\mathrm{h}$ before the window end. The four MBHB contaminants are subtracted sequentially (Fig.~\ref{fig:robustness-all}); the Galactic binaries and EMRIs are left unchanged. No network draws fall within the prior for unsubtracted MBHB$_1$ mergers at approximately $-3$ and $-1\,\mathrm{h}$, or when all contaminants are present without subtraction, so the corresponding curves are omitted. After subtracting MBHB$_1$ at $-1\,\mathrm{h}$, importance sampling fails ($\mathrm{ESS}=1$), and the network posterior is constructed from the $49\%$ of draws within the prior.}
  \label{fig:robustness-ridge-sub}
\end{figure*}

The networks are trained on a single MBHB in stationary Gaussian noise (Sec.~\ref{sec:signal-data}), whereas LISA data will contain overlapping signals from multiple sources. We test $\SBIl$ on the window of the \emph{reference} signal ending $30\,\mathrm{min}$ after its merger (SNR $297$), adding signals the network has never seen. Each added signal goes through the same response and TDI as the \emph{reference} signal, and we keep the noise realization fixed, so any change in the posterior is due to the added signal. Fig.~\ref{fig:robustness-data} shows the different components of the data in frequency and time-domain. The \emph{reference} data stream only contains the noise realization (grey) generated with the sum of the instrumental and unresolvable Galactic binaries (GB) confusion PSD, and the \emph{reference} waveform (black). Contaminant signals are added onto the \emph{reference} data one at a time. We explore the effect of adding the resolvable GBs (blue), EMRIs at different plunge times (reds) and an additional MBHB (MBHB$_1$, greens) at different time-of-coalescence $t_c$. In \emph{All contaminants at once}, we add the GBs, EMRIs and 4 MBHB$_1$ together for a combined inference. The plunge times and time-to-coalescence are stated, as usual, in relation to the end of the observing window. We specify the population and source parameters of the contaminant sources below: 

\begin{itemize}
    \item \emph{Resolvable Galactic binaries:} the $9\,980$ double white dwarfs of the Sangria Galaxy~\cite{Sangria} with a 4-year SNR above $7$ and $f < 15\,\mathrm{mHz}$, simulated through our own response and TDI with the pipelines of Refs.~\cite{Srinivasan2025, Korol+2022}. Each GB is a quasi-monochromatic signal from its own sky position. Their summed SNR in the window is $269$. The unresolved remainder is already part of the training PSD.
    \item \emph{An EMRI:} equatorial orbit (primary mass $10^{6.25}\,M_\odot$, secondary mass $10^{1.75}\,M_\odot$ or mass ratio $q =  3.2\times10^4$, luminosity distance $D_L = 0.7\,\mathrm{Gpc}$, primary spin $a=0.3$, initial eccentricity $e=0.2$), Kerr eccentric equatorial waveform of FastEMRIWaveforms~\cite{Chua2021,Katz2021,ChapmanBird2025}), plunging $60\,\mathrm{d}$ or $1\,\mathrm{d}$ after the window end, and $1\,\mathrm{d}$ or $31\,\mathrm{min}$ before window end. The latter is a plunge just $1\,\rm{min}$ before the merger of the \emph{reference} signal. The window SNR of the EMRIs range from $9$ to $24$.
    \item \emph{Secondary MBHB}, MBHB$_1$, that merges approximately $4\,\rm{d}$, $2\,\rm{d}$, $1\,\rm{d}$, $3\,\rm{h}$, or $1\,\rm{h}$ before the end of observation window, as well as $4\,\rm{h}$ or $2\,\rm{d}$ after it. MBHB$_1$ is sampled once from the $\SBIl$ prior ($\log_{10}\mathcal{M}_c = 6.80$, $q = 3.85$, $D_L = 45.3\,\mathrm{Gpc}$). Its window SNR is $\simeq260$ if it merges inside the window and $23$ and $5$ if it merges $4\,\mathrm{h}$ and $2\,\mathrm{d}$ after the end of the observation window.
\end{itemize}

Figure~\ref{fig:robustness-ridge} compares the one-dimensional posteriors of the \emph{reference} signal with and without contamination. We show the network posterior in each case, together with the importance-sampled (IS) posterior when the sampling efficiency exceeds $0.05\%$. To quantify the effect of contamination, we report shifts in posterior medians in units of $\sigma_{\rm ref}$, the standard deviation of the corresponding uncontaminated posterior (network or IS). The uncontaminated reference posterior is obtained using the same noise realization as the contaminated cases.

\emph{EMRIs} leave the posterior unchanged. Network widths change by at most $1\,\%$ and medians by less than $0.1\,\sigma_\rm{ref}$; the IS medians agree with the uncontaminated IS posterior to within $0.2\,\sigma_\rm{ref}$, comparable to the Monte Carlo noise of IS at ESS $\sim10^{2}$. The EMRI power lies mostly at $3$--$10\,\mathrm{mHz}$, above the band of the \emph{reference} signal ($0.3$--$2\,\mathrm{mHz}$; Fig.~\ref{fig:robustness-data}). For the $+1\,\mathrm{d}$ plunge, the IS widths of $t_c$ and $q$ are $14$--$18\,\%$ larger, from a tail towards earlier $t_c$ that the network rarely proposes; this is consistent with IS sampling noise in that region. These results are in agreement with Bayesian studies~\cite{Agrawal:2026ohy}.
 
\emph{Resolvable Galactic binaries} also leave the posterior unchanged. The network medians move by at most $0.7\,\sigma_\rm{ref}$ ($\log_{10}\mathcal{M}_c$) and its widths by at most $5\,\%$; after importance sampling ($\mathrm{ESS} = 126$) the medians and widths of the unimodal parameters are within $0.5\,\sigma_\rm{ref}$ and $20\,\%$ of the uncontaminated ones. These binaries add at most noise-level power in the $0.3$--$2\,\mathrm{mHz}$ band of the \emph{reference} signal (Fig.~\ref{fig:robustness-data}). This does not hold for every Galaxy model: a realization with $6.5$ times as many resolvable binaries (window SNR $1200$, the fiducial population of Ref.~\cite{Srinivasan2025}, which follows Ref.~\cite{Korol+2022}) biases the network posterior by $7$--$11\,\sigma_\rm{ref}$ in $q$ and $t_c$, and importance sampling then fails ($\mathrm{ESS} = 2$) rather than correcting it.

\emph{A secondary MBHB} can bias the posteriors based strongly on its merger time with respect to that of the \emph{reference}. When the second MBHB merges well outside the observation window ($+4\,\mathrm{h}$, $+2\,\mathrm{d}$), the network medians are shifted by at most $1.3\,\sigma_\rm{ref}$ ($\chi_2$, at $+4\,\mathrm{h}$) and the IS medians by at most $0.8\,\sigma_\rm{ref}$. However, when the MBHB merges inside the observation window it biases the \emph{reference} signal more as its merger gets closer to the \emph{reference} one.

For a secondary merger $4\,\mathrm{d}$ before the window end, network posterior medians shift by at most $1.1\,\sigma_{\rm ref}$. Moving the merger to $2\,\mathrm{d}$ before the window end increases the shifts to $1.4$ and $1.9\,\sigma_{\rm ref}$ in $\mathcal{M}_c$ and $D_L$, respectively, while a merger $1\,\mathrm{d}$ before the window end shifts $t_c$ by $6.7\,\sigma_{\rm ref}$. Importance sampling fails to resolve these posteriors reliably ($\mathrm{ESS}\leq18$), so we report the network shifts. For secondary mergers at $\Delta t_{c,1}=-3\,\mathrm{h}\,16\,\mathrm{s}$ and $-1\,\mathrm{h},16,\mathrm{s}$ (hereafter approximately $-3$ and $-1\,\mathrm{h}$), no network draws fall within the prior, so no posterior is shown. This failure is readily detectable and indicates that the network cannot produce a usable posterior for these strongly contaminated data.

Facilitated by the fact that the MBHB posterior is mostly unaffected by a secondary signal merging outside the observation window, we extend the range of merger separations that the network handles by inferring the additional MBHB from the data using the network. We estimate the parameters of the secondary MBHB using $\SBIl$ (on a window ending $30\,\mathrm{min}$ after its merger), subtract the waveform corresponding to the maximum a posteriori estimate, and then infer the parameters of the \emph{reference} signal from the residual data. 

The results of the subtraction method for different secondary MBHB merger times relative to the end of the observation window ($\Delta t_{c\,1}$) are reported in Fig.~\ref{fig:robustness-ridge-sub}. For secondary mergers approximately $4\,\mathrm{d}$, $2\,\mathrm{d}$, $1\,\mathrm{d}$, and $3\,\mathrm{h}$ before the window end, subtraction leaves residual SNRs of $6$–$9$. The resulting reference-signal posteriors closely agree with those obtained without contamination: network medians differ by less than $0.15\,\sigma_{\rm ref}$ and posterior widths by less than $2\%$, while IS medians remain within $0.25\,\sigma_{\rm ref}$. 

At $\Delta t_{c\, 1} \simeq -1\,\mathrm{h}$ the subtraction fails, and so does IS of the residual. The \emph{reference} signal (SNR $167$ in the window of MBHB$_1$) biases the inference of MBHB$_1$ itself and leaves a residual of SNR $143$. In this case the posteriors of the two signals are correlated and a sequential subtraction fails, mergers this close need joint inference~\cite{Samajdar:2021egv, Pizzati:2021apa}.

\begin{figure*}[t]
  \centering
  \includegraphics[width=\textwidth]{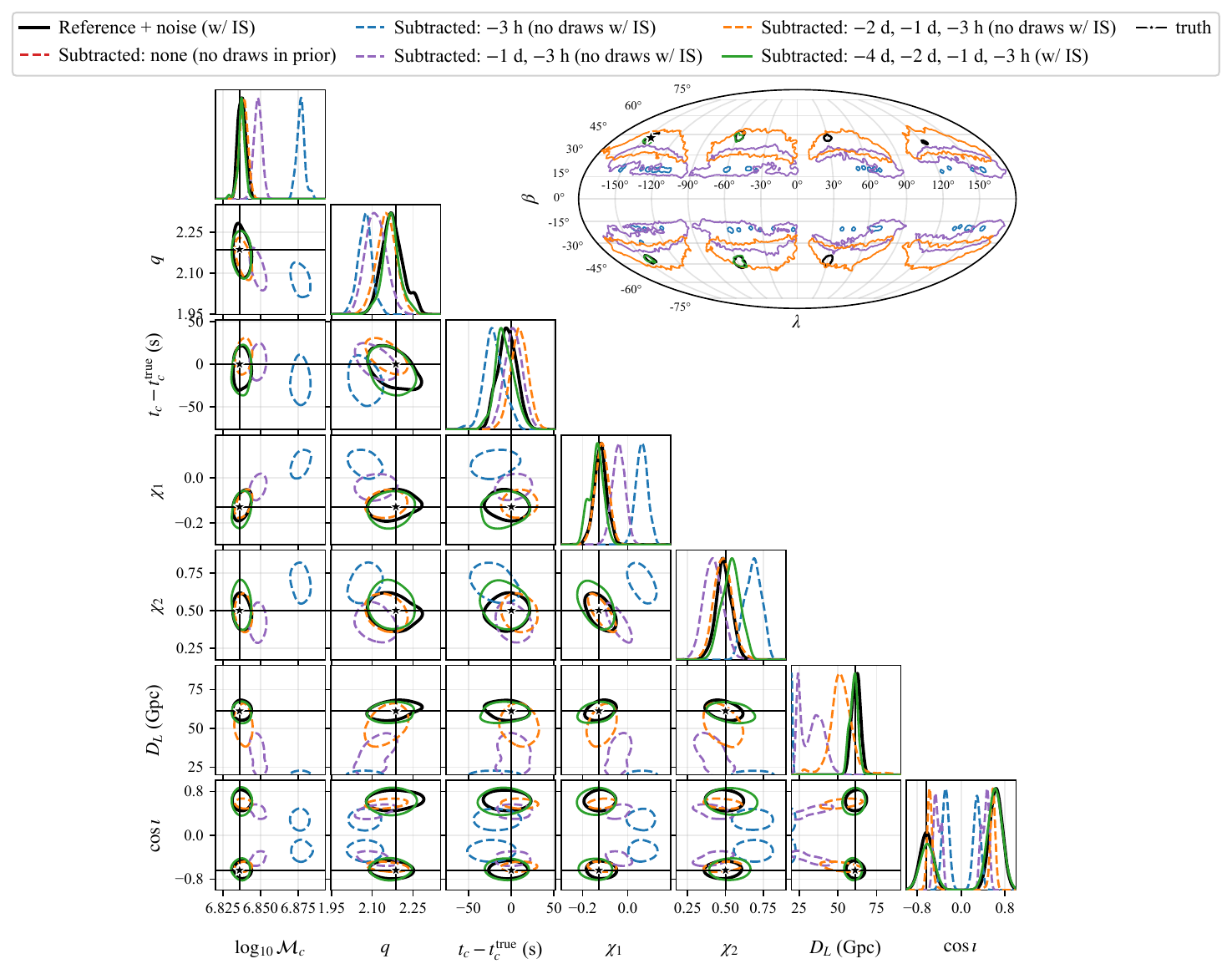}
  \caption{Robustness of the $\SBIl$ posterior of the \emph{reference} with all contaminants present at once: four EMRIs (summed SNR $40$), plunging 60 d or 1 d after the end of the \emph{reference} window, 1 d before it, or 1 min before the merger of \emph{reference}, the resolvable GB population from Sangria (SNR $269$), and 4 MBHBs merging 4 d, 2 d, 1 d, 3 h before the window end, each sequentially subtracted with its corresponding MAP waveform. Each successive contaminant MBHB is subtracted by the MAP waveform estimated from a window ending $30\,\mathrm{min}$ after its merger, and with the earlier MBHB contaminants already removed. Black: \emph{reference} alone in the same noise realization and importance-sampled (IS). Dashed lines show reference inference after different numbers of subtracted contaminants; their IS has $\mathrm{ESS} \le 7$ and is not shown. Solid green shows posterior after IS and after all four MBHB contaminants are subtracted (residual SNR $17$).}
  \label{fig:robustness-all}
\end{figure*}

\emph{All contaminants at once.} 
The merger rate of MBHBs can span several orders of magnitude~\cite{Barausse:2012fy,Sesana:2014bea,Antonini:2015sza,Klein:2015hvg,Bonetti:2018tpf,Barausse:2020mdt,Barausse:2023yrx, 2024MNRAS.533.1907B}, with the more populous cases displaying merger rates $\mathcal{O}(1000)~\mbox{yr}^{-1}$.  In those configurations multiple events may merge within a single analysis window. To test this plausible scenario, we analyze the inference of the \emph{reference} signal (merging 30~m before the window end), but in the presence of noise, the $9980$ resolvable Sangria GBs of total $\rm{SNR = 269}$ (with respect to the instrumental and confusion noise), four EMRIs (at the different plunge times; total $\rm{SNR=40}$) and four additional MBHBs, merging $\approx \rm{4\,d, 2\,d, 1\,d, 3\,h}$ before the window end of total $\rm{SNR = 250}$. 

In Fig.~\ref{fig:robustness-all}, we show the posteriors of the \emph{reference} in the absence (black) and presence (colored) of all contaminants. Different colors represent different numbers of sequential subtractions, each using the maximum a posteriori (MAP) waveform of the network posterior of the respective MBHB$_1$. Passing the data directly to the network without any subtraction (red) produces no samples in the prior and hence no contours. Sequentially subtracting the loudest sources produces posteriors that have progressively lower biases, until the subtraction of all MBHB$_1$ produces unbiased posteriors (green, also shown in Fig.~\ref{fig:robustness-ridge-sub}, last row). In the full chain each subtraction is estimated on data from which the earlier copies have already been removed, and each leaves a residual of SNR $5$--$10$, so that after the four subtractions $\mathrm{SNR} = 17$ of MBHB$_1$ remains in the window. The IS posterior of these data agrees with the uncontaminated one: medians within $0.25\,\sigma_\rm{ref}$ and widths within $6\,\%$, except $t_c$ ($37\,\%$ wider, within sampling noise). In three separate partial runs, with only the $3\,\mathrm{h}$ copy, the $1\,\mathrm{d}$ and $3\,\mathrm{h}$ copies, or the $2\,\mathrm{d}$, $1\,\mathrm{d}$ and $3\,\mathrm{h}$ copies subtracted, $460$, $370$ or $261$ of MBHB$_1$ SNR remain in the window, the chirp mass is biased by $17$, $5$ and $0.5\,\sigma_\rm{ref}$ and the distance by $16$, $10$ and $3\,\sigma_\rm{ref}$, and importance sampling fails ($\mathrm{ESS} \le 7$).

\subsection{Hybrid $\SBIl-\rm{MCMC}$ inference}
\label{sec:hybridMCMC}
We investigate whether the likelihood-corrected $\SBIl$ posterior can accelerate a subsequent MCMC analysis. As a proof of concept, we compare initialization from the IS posterior with initialization from a broad box centered on the injected parameters, with half-widths of $0.075$ in $\log_{10}\mathcal{M}_c$ and $5\,\mathrm{h}$ in coalescence time. The latter provides a controlled reference initialization rather than a fully blind search. We apply both approaches to a longer, $24\,\mathrm{d}$ data segment and compare the likelihood evaluations required to reach the posterior. We split the MCMC run in an initial search phase using only Stretch proposals and a proposal which jumps between the 8 degenerate modes in the sky. The search stage ends when maximum log-likelihood does not change significantly for more than 2500 steps. The resulting posterior is used to initialize an MCMC run using the same setup as the one used as comparison with $\SBIl$ in Sec.~\ref{sec:res-likelihood}, using a mixture of single component adaptive metropolis, differential evolution, stretch and sky jump proposals. The $\SBIl$ initialized run immediately converges to the correct maximum likelihood point, and requires about $5~\mbox{h}$ to converge to the correct posterior. The run initialized on  the prior requires about five hour to fall to the maximum likelihood point, after which the sampling MCMC starts. Compared to the $\SBIl$ initialized run, the prior initialized run requires about twice as many likelihood computations.
 
\subsection{Computational cost}
\label{sec:res-cost}

For a single analysis window, the network produces $10^5$ posterior samples in about $0.15\,\mathrm{s}$ on one NVIDIA A100 GPU, while the multiband compression adds less than $0.01\,\mathrm{s}$. Importance sampling dominates the latency. On one A100, reweighting $10^5$ network draws takes $282\,\mathrm{s}$. When the draws are split over eight A100 GPUs, the full pipeline from the input data window to the final importance weights and effective sample size takes $68\,\mathrm{s}$. Of this, about $9\,\mathrm{s}$ is spent building the likelihood, $28\,\mathrm{s}$ compiling it, and $30\,\mathrm{s}$ evaluating the likelihood. The remaining overhead is below one second. The current implementation rebuilds and recompiles the likelihood for every window, so part of this $\sim37\,\mathrm{s}$ overhead could be reduced in a persistent optimized implementation. A likelihood-corrected posterior is therefore available in about $1\,\mathrm{min}$ on eight GPUs, or about $5\,\mathrm{min}$ on one GPU.

For comparison, the parallel-tempered MCMC runs of Fig.~\ref{fig:mcmc}, initialized at the true parameters, took about $11\,\mathrm{h}$ on one A100 for the \emph{reference} signal $30\,\mathrm{min}$ after merger and $18\,\mathrm{h}$ for the near-equatorial source $30\,\mathrm{min}$ before merger. These runs used fixed iteration budgets rather than a convergence-based stopping criterion.
 
\section{Discussion}
\label{sec:discussion}

\subsection{What an alert stream delivers}
\label{sec:disc-alerts}

Scheduling information becomes useful hours to days before merger, while accurate pointing generally arrives only close to merger. As shown in Fig.~\ref{fig:countdown}, four hours before merger the coalescence time of the \emph{reference} signal is constrained to about an hour and its chirp mass to roughly $10\,\%$. A lighter source reaches about $20\,\mathrm{min}$ precision in coalescence time and $2$--$3\,\%$ in chirp mass at the same stage, with useful estimates already available two days earlier. These uncertainties are sufficient to schedule observations around the expected merger time.

Sky localization improves much later. The figure shows that, for a generic orientation, the sky area remains at $\sim 10^{4}\,\mathrm{deg}^2$ until close to merger and reaches a few hundred square degrees only after it. Favorable geometries can localize earlier: the in-plane copy of the \emph{reference} signal reaches $\sim 200\,\mathrm{deg}^2$ about $1.5\,\mathrm{h}$ before merger and $\sim 10^{2}\,\mathrm{deg}^2$ about half an hour before merger (Sec.~\ref{sec:disc-sky}).
 
The \emph{reference} signal is deliberately challenging rather than representative of the best targets for EM follow-up. At $D_L = 61\,\mathrm{Gpc}$, corresponding to $z \approx 6.2$, it reaches an SNR of only $21$ four hours before merger, and any counterpart at this distance would be difficult to detect. More promising targets are closer sources and systems with favorable orientations, such as the equatorial variants discussed in Sec.~\ref{sec:res-countdown}.

For scale, Ref.~\cite{mangiagli2022} consider fields of view of $9.6\,\mathrm{deg}^2$ for the Rubin Observatory, $\sim10\,\mathrm{deg}^2$ for SKA, and $0.4\,\mathrm{deg}^2$ for the Athena Wide Field Imager. The $310\,\mathrm{deg}^2$ localization of the \emph{reference} signal $30\,\mathrm{min}$ after merger would therefore require roughly $30$ Rubin or SKA pointings. The in-plane source would require about $10$ pointings half an hour before merger and about $20$ at $1.5\,\mathrm{h}$ before merger, whereas a generic pre-merger localization of $\sim10^{4}\,\mathrm{deg}^2$ would require $\mathcal{O}(10^3)$. These estimates are only indicative: the $310\,\mathrm{deg}^2$ area is distributed across eight modes, which could in practice be targeted separately or distributed across multiple observatories.

\subsection{Accuracy of the network posteriors}
\label{sec:disc-accuracy}

The results of Sec.~\ref{sec:res-likelihood} show that the network posteriors are generally close to the likelihood before merger, but become systematically broader once the merger and ringdown enter the analysis window. This is also where the posterior becomes most concentrated relative to the training prior and the importance-sampling efficiency decreases.

Importance sampling therefore provides both a correction and a useful reliability diagnostic. When its efficiency is sufficient, the reweighted samples recover the likelihood posterior; a low efficiency instead flags cases where the network posterior does not match it closely enough. In an alert pipeline, the network posterior could be issued immediately and replaced by the importance-sampled posterior about a minute later on eight GPUs, or about five minutes later on one GPU (Sec.~\ref{sec:res-cost}), provided that the effective sample size exceeds a chosen threshold. We use $\mathrm{ESS} \ge 50$ in Sec.~\ref{sec:res-robustness}. Otherwise, the ESS is too low for a reliable correction and the uncorrected network estimate is flagged as provisional with additional likelihood analysis necessary. Finally, using the network posterior to initialize a likelihood-based sampler (Sec.~\ref{sec:hybridMCMC}) could recover the full likelihood precision while retaining the speed of the initial alert.

\subsection{Sky localization in practice}
\label{sec:disc-sky}

The eight modes of the sky posterior arise from the LISA response rather than from the network. There are four modes in longitude and their reflections through the constellation plane. The same structure appears in MCMC analyses of the data (Fig.\ref{fig:mcmc}) and in earlier likelihood-based studies\cite{marsat2021}. For EM follow-up, the relevant question is not only the total sky area but also how that area is distributed among the modes. For the \emph{reference} signal, the $310\,\mathrm{deg}^2$ localization $30\,\mathrm{min}$ after merger consists of eight patches of $\approx 40\,\mathrm{deg}^2$ each. These patches can be targeted individually, whereas the same total area distributed in a continuous band would be much less useful. This is particularly relevant for coordinated follow-up, where observatories at different locations on Earth can target different modes. Since the mode weights are calibrated, the patches can also be observed in order of posterior probability (Appendix~\ref{app:res-sky}).

The localization strongly depends on inclination and latitude. As $\cos\iota \to 0$, the handedness that distinguishes the reflected modes becomes unmeasurable and the posterior spreads over $5\times10^{3}$--$1.6\times10^{4},\mathrm{deg}^2$. Polar sources show a similar loss of localization. Face-on and in-plane sources therefore carry most of the localization power. During the final hour before merger, one third of the importance-sampled signals reach $\Omega_{86.5\%} < 100\,\mathrm{deg}^2$. Three quarters of these are in-plane with $|\sin\beta| < 0.26$ or face-on with $|\cos\iota| > 0.76$, even though these geometries make up only $43\%$ of the population (Appendix~\ref{app:res-population}).

Lighter systems with $\mathcal{M}_c \lesssim 10^{6}~M_\odot$ may break more of these degeneracies. They merge at higher frequencies, where the gravitational wavelength becomes comparable to the LISA arm length and the long-wavelength approximation becomes less accurate. Finite light-travel-time effects across the arms then introduce additional frequency-dependent structure into the detector response, which can help distinguish otherwise degenerate sky positions. Our population already shows stronger breaking of the $\pi/2$ longitude degeneracy as the ringdown frequency increases. Lighter systems also remain in band for weeks to months, allowing more orbital modulation of the detector response to accumulate. The limitations of the present mass range and observation duration, and the extensions needed to study this regime, are discussed in Sec.~\ref{sec:limitations}.
 
\subsection{Other sources and the global fit}
\label{sec:disc-global}

The robustness tests of Sec.\ref{sec:res-robustness} suggest that $\SBIl$ could operate alongside a global-fit pipeline~\cite{Littenberg2023,Strub2024,Katz2025} without requiring every overlapping source to be modeled jointly. In the configurations tested, EMRIs and a Sangria-like population of resolvable Galactic binaries have little effect on the reference posterior. A Galaxy model with $6.5$ times as many resolvable binaries, with a combined window SNR of $1200$, instead biases the posterior by up to $11\,\sigma_{\rm ref}$. Such a high-multiplicity foreground could be mitigated if at least some of Galactic binaries have already been cataloged and subtracted by the global fit.

A secondary MBHB can instead introduce substantial biases when its merger lies within the analysis window, but sequential inference and subtraction recover the reference posterior for the tested merger separations down to approximately $2.5\,\mathrm{h}$. Closer mergers remain problematic and will likely require joint inference. The present results are limited to the tested source configurations and do not establish robustness across the full LISA source population.

This subtraction strategy currently requires the contaminating MBHB to lie within the $\SBIl$ prior, since the network must first obtain a posterior estimate for the source being removed. More general contaminant populations and wider mass ranges remain to be tested and are discussed in Sec.~\ref{sec:limitations}. In the opposite direction, the network posterior can provide the global fit with a rapid proposal for each MBHB, together with its importance-sampling efficiency as a measure of proposal quality.
 
\subsection{Comparison with existing work}
\label{sec:positioning}

\begin{table*}[t]
  \caption{
  Comparison with other fast parameter estimation and low-latency pipelines for space-based GW interferometers. The table reports the prior for the chirp mass and time of coalescence assumed in the different works, which physical parameters are resolved by the analysis, whether the pipeline is likelihood-calibrated through importance sampling (IS), and the reported run time. }
  \label{tab:results-comparison}
  \centering
  \begin{ruledtabular}
  
\renewcommand{\arraystretch}{1.25}
  \begin{tabular}{lcccccccccccccr}
 &  &  & \multicolumn{2}{c}{Search/inference ranges} & \multicolumn{8}{c}{Deliverables} &  &   \\
\cmidrule(lr){4-5}\cmidrule(lr){6-13}

Method & Detector & Response & $\mathcal{M}_c/M_\odot$ & $\Delta t_c$ & $\mathcal{M}_c$ & $t_c$ & $q$ & $\lambda,\,\beta$ & $D_L$ & $\iota$ & $\chi_{1,2}$ & $\phi$,$\psi$ & IS  & Updates \\ \colrule
        Davies \emph{et al.} Search & LISA & TDI 1.5 & $[3,87]\times10^5$ & $[14~\mbox{d},0.5~\mbox{d}]$ & \checkmark & \checkmark & \xmark & \xmark & \xmark & \xmark & \xmark & \xmark & no & 1\,h (CPUs)\\
        
        \hspace{1.75cm} PE~\cite{cabourndavies2025}\footnotemark[1] & LISA & TDI 1.5 & $[8,100]\times10^5$ & $2\,\mathrm{h}$ range & \checkmark & \checkmark & \checkmark & \checkmark & \checkmark & \checkmark & \checkmark & \checkmark & no & $1$--$5\,\mathrm{d}$ (CPUs) \\

        Deng \emph{et al.}~\cite{Deng:2025qhx}\footnotemark[2] & LISA & Frozen low-f & $[0.1,100]\times10^5$ & $[20~\mbox{d},-1~\mbox{d}]$ & \checkmark & \checkmark & \checkmark & \xmark & \xmark & \xmark & \checkmark & \xmark & no & $\sim1\,\mathrm{m}$ (CPU)\\ 
        
        Nobili \emph{et al.}~\cite{nobili2026}\footnotemark[3] & LISA & Frozen low-f & $[2,50]\times10^5$ & $[14~\mbox{d},2~\mbox{d}]$ & \checkmark & \checkmark & \xmark & \xmark & \xmark & \xmark & \xmark & \xmark & no & $<1\,\mathrm{s}$ (CPU)\\ 
        
        \textsc{dingo}-LISA~\cite{spadaro2026} \footnotemark[4]& LISA & Fixed-$t_0$ low-f & $[0.8,2]\times 10^7 $ & $30~$m range & \checkmark & \checkmark & \checkmark & \checkmark & \checkmark & \checkmark & \checkmark & \checkmark & yes & $20\,\mathrm{s}$ (GPU) \\ 
        
        Zhang \emph{et al.}~\cite{zhang2026}\footnotemark[5] & TianQin & TDI 1 & $[0.2, 1]\times10^6$ & $[10~\mbox{m},20\mbox{m}]$ & \checkmark & \checkmark & \checkmark & \checkmark & \checkmark & \xmark & \xmark & \xmark & no & $ 10~\mbox{s}$ (GPU) \\ 
        
        This work $\SBIe$ & LISA & TDI 2 & $[1, 100]\times 10^5$ & $[2~\mbox{d}, 4~\mbox{h}]$ & \checkmark & \checkmark & \checkmark & \xmark & \xmark & \xmark & \xmark & \xmark & no & $<1\,\mathrm{s}$ (GPU) \\ 
        
        \hspace{1.45cm}  $\SBIl$ \footnotemark[6]& LISA & TDI 2 & $[3.2, 10]\times 10^6$ & $[4~\mbox{h}, -1~\mbox{h}]$ & \checkmark & \checkmark & \checkmark & \checkmark & \checkmark & \checkmark & \checkmark & \xmark & yes & $< 1\,\mathrm{m}$ (GPUs)\\

  \end{tabular}
  \end{ruledtabular}
  \footnotetext[1]{The search providing parameter estimates through template bank match filter. PE uses trigger-dependent mass priors.}
  \footnotetext[2]{Search pipeline using the $\mathcal{F}$-statistic method, extending the searchused in the \texttt{gee-moo} global-fit pipeline for the Sangria dataset~\cite{Geemoo_Deng2025}.}
  \footnotetext[3]{Time-frequency excess-power search without explicit detector-response evaluation. Minimum-false-alarm point estimates of $t_c$ and $\mathcal{M}_c$.}
  \footnotetext[4]{SBI pipeline that uses the low-frequency response approximation at fixed observation epoch.}
  \footnotetext[5]{Nonspinning waveforms. Infers the symmetric mass ratio $\eta$ rather than $q$. $\iota$, polarization, and coalescence phase are marginalized.}
  \footnotetext[6]{IS of $10^5$ network draws takes $\le$1~min on eight A100, network draws its $10^5$ samples in $\approx 0.4\,\mathrm{s}$ (see Sec.~\ref{sec:res-cost}).}
\end{table*}

Table~\ref{tab:results-comparison} places this work in the context of other studies of MBHBs in space-based detectors. Some provide point estimates based on false-alarm statistics and others provide posteriors. Our contribution is posterior inference calibrated by IS which can be updated continuously as an alert stream, while most of the individual ingredients have appeared separately in earlier work. The rows differ in detector model, mass range, distance, SNR, and in how uncertainties are defined, so the comparison is intended to provide an overview rather than a ranking.

Time-resolved information on MBHBs before merger is not new. Fisher forecasts have followed parameter constraints from a month before merger~\cite{mangiagli2020}, while likelihood-based analyses have examined several pre-merger lead times for zero-noise signals~\cite{marsat2021}. An excess-power search detects LISA MBHBs up to $17\,\mathrm{d}$ before merger, with a measured false-alarm rate and no training~\cite{nobili2026}. Its typical chirp-mass errors range from a few to tens of per cent, with one more difficult source performing substantially worse. It therefore detects sources much earlier than our first alert, $48\,\mathrm{h}$ before merger, while providing comparable chirp-mass precision for some sources in the early-warning regime. Neural posterior estimation has also been applied to LISA MBHBs close to merger around a reference time~\cite{spadaro2026}, and to TianQin MBHBs $15\,\mathrm{min}$ before merger, with comparison to PTMCMC~\cite{zhang2026}.

Marsat et al.\cite{marsat2021} find that for sources with redshifted total mass $2\times10^{6}~M_\odot$ and SNR $666$--$945$, only the true sky mode and its reflection remain once the merger enters the data. The other six modes disappear in the final minutes before merger. Our \emph{reference} signal has a detector-frame total mass of $1.7\times10^{7},M_\odot$ and SNR $297$, yet retains all eight modes $30\,\mathrm{min}$ after merger. Marsat et al. show that the additional modes begin to lose likelihood near $2\,\mathrm{mHz}$ as the frequency dependence of the response becomes important, with the effect strengthened by the rapid accumulation of SNR close to merger. They also note that the balance between the time and frequency dependence of the response changes with the mass of the system. The \emph{reference} signal radiates mostly below $2\,\mathrm{mHz}$ (Fig.\ref{fig:robustness-data}) and has a lower SNR, which is consistent with the persistence of all eight modes in our case.

Unlike the other analyses in Table~\ref{tab:results-comparison}, ours provides rapid calibrated posteriors continuously from two days before merger to one hour after merger using an orbit-driven LISA response with unequal, time-varying arms and second-generation TDI. The coalescence time is inferred over a prior spanning hours to days rather than being restricted around a known trigger, and the network is amortized over $30\,\mathrm{d}$ of observation start epochs. We demonstrate, as proof-of-concept, the robustness of \truffle\ inference in the presence of Galactic binaries, EMRIs, or a second MBHB.
 
\subsection{Limitations and outlook}
\label{sec:limitations}

The present networks are demonstrated in a restricted setting.

\emph{Data model.} We assume stationary Gaussian noise with a Galactic confusion foreground included in the noise spectrum, and treat the $A$ and $E$ channels as uncorrelated. For the high-SNR MBHBs considered here, we expect these approximations to be less important than for weaker sources, but this has not been tested explicitly. We also do not include data gaps or glitches. Injection and recovery use the same waveform model, which includes higher modes but not precession. Precession may itself help break some of the sky degeneracies discussed in Sec.~\ref{sec:disc-sky}. Future training should include waveform variation so that the network can marginalize over waveform systematics.

\emph{Prior and observation duration.} $\SBIl$ currently covers only $\log_{10}\mathcal{M}_c \in [6.5,7]$ with $q \le 4$, and must ultimately be extended to the full $\SBIe$ mass range. Lighter systems with $\log_{10}\mathcal{M}_c \lesssim 6$ remain in band for longer and will require longer observing windows, likely of order $7$--$30\,\mathrm{d}$. The optimal duration will depend on the trade-off between accumulated SNR and the amount of data that must be compressed. Both networks are also amortized over only $30\,\mathrm{d}$ in observation start time, rather than over the full year of LISA orbital motion. For the heavy systems studied here, the $6$-day observation window samples less than $2\%$ of the LISA orbit, so only a small fraction of the available orbital modulation is used. Lighter systems will require longer observation windows to exploit this additional localization information. If the likelihood breaks a sky degeneracy that the network fails to learn, the auxiliary-label strategy of Sec.~\ref{sec:handedness} could also be extended to the corresponding sky mode.

\emph{Other sources.} The robustness tests of Sec.~\ref{sec:res-robustness} cover a limited set of contaminating populations. For MBHB$_1$ and the EMRIs, we use one set of intrinsic parameters and shift the same sources in time to vary their overlap with the target. We also test only the current $\SBIl$ mass band. Lighter MBHBs overlap more strongly in frequency with Galactic binaries and EMRIs, while an additional MBHB with a different mass may overlap less strongly with the target. These cases remain to be tested. A broader extension would train on random contaminants spanning different source classes, numbers and masses, both inside and outside the current prior. This could be initialized from the present network through transfer learning rather than trained again from random weights.

\emph{Importance sampling.} The networks were developed primarily for pre-merger alerts. Once the merger enters the window, the likelihood posterior can become substantially narrower than the network posterior for some orientations and sky locations, reducing the importance-sampling efficiency from $\simeq 15\%$ before merger to as low as $\simeq 0.1\%$ after it (Fig.\ref{fig:is_vs_npe}). The reweighted posteriors nevertheless agree with the MCMC reference when a sufficient effective sample size is obtained (Fig.\ref{fig:mcmc}). Additional sources can reduce the efficiency further (Sec.~\ref{sec:res-robustness}). Denser training simulations could improve the network approximation, although at increased training cost. Alternatives include concentrating training simulations in regions where the posterior becomes narrow and exploring more efficient representations for longer signals. This may become more important for lighter systems, whose chirp mass and sky position can be measured more precisely.

The next steps, in order of priority, are
(i) benchmarking against externally generated mock LISA data with realistic source content, including residuals from a global-fit subtraction of Galactic binaries, such as data released by the LISA Distributed Data Processing Center~\cite{Mojito};
(ii) joint inference of MBHBs that merge within about an hour of one another;
(iii) data gaps and non-stationary noise;
(iv) wider priors, including additional $\SBIl$ mass bands with a tested transition from $\SBIe$, $q \le 10$, and amortization over the full year;
(v) network-initialized likelihood sampling (Sec.~\ref{sec:disc-accuracy}) and integration with the global fit.

\section{Conclusion}
\label{sec:conclusion}

We demonstrate \truffle, a simulation-based inference pipeline for low-latency parameter estimation of massive black hole binaries in LISA data. Two neural posterior estimators cover the evolution from two days before merger to one hour after it. $\SBIe$ infers the chirp mass, mass ratio, and coalescence time from $48$ to $4\,\mathrm{h}$ before merger. $\SBIl$ infers nine parameters, including the sky position, from $4\,\mathrm{h}$ before to $1\,\mathrm{h}$ after merger. Both networks are trained on sliding windows of continuous data streams, with the orbit-driven LISA response (unequal, time-varying arms and second-generation TDI), and are amortized over a $30$-day range of observation epochs. Both are calibrated over their training priors, as verified on $1.1\times10^{6}$ independent test signals each.

The pipeline meets the requirements for scheduling EM follow-up and for an early characterization of the binary. For the \emph{reference} source, with detector-frame chirp mass $\mathcal{M}_c\simeq7\times10^{6}\,M_\odot$, four hours before merger the coalescence time is known to about an hour and the chirp mass to about $10\%$. At merger, these improve to about $2\,\mathrm{min}$ and $1\%$. One hour later they reach $12\,\mathrm{s}$ and $0.8\%$. Lighter systems become informative earlier: a source with $\mathcal{M}_c\simeq2\times10^{5}\,M_\odot$ is timed to about $3\,\mathrm{h}$ two days before merger and to about $20\,\mathrm{min}$ four hours before it.

Accurate sky localization arrives later and depends strongly on the source orientation. For the \emph{reference} signal, the sky posterior remains broad before merger; thirty minutes after merger it covers $310\,\mathrm{deg}^2$, spread over eight compact modes set by the symmetries of the LISA response. Sources near the constellation plane or close to face-on are localized earlier and more precisely. The network reproduces the sky--inclination handedness only with a reparameterization of the flow and an auxiliary training loss, an example of how known structure in the likelihood can be built into neural posterior estimation.

Importance sampling provides both a likelihood-based correction and a per-event diagnostic. Before merger, the network posterior is generally close to the likelihood posterior. Once merger and ringdown enter the window, the likelihood posterior can become substantially narrower and the importance-sampling efficiency drops. Where the effective sample size is sufficient, the reweighted posterior agrees with parallel-tempered MCMC on the same data. The network posterior can therefore be issued immediately and replaced a few minutes later by a likelihood-corrected one, or used as a proposal for a conventional sampler when higher precision is required.

The robustness tests give a first indication of how \truffle\ could operate without the LISA global fit. For the heavy MBHBs studied here, the tested EMRIs and a Sangria-like population of resolvable Galactic binaries have little effect on the inferred posterior.  Louder signals, such as an additional MBHB or very dense Galactic binary foreground can spoil the parameter inference. The latter may be solved either by training the network over several GB realizations or by subtracting the GBs found in the global fit run. For systems merging at least three hours apart we are able to correctly infer the properties of both systems by subtracting the MAP of the early merging source found by the network from the data stream. This does not completely solve the multiple sources issue, since MBHBs merging within about an hour of each other still require joint inference. As SBI has been shown to be able to address trans-dimensional Bayesian problems ~\cite{Houba:2025dnr}, the joint inference will be a natural extension of our pipeline in future work.

The present study is restricted in several ways. We assume stationary Gaussian noise and use one waveform family without precession; the near-merger network covers a limited mass range; both networks are amortized over $30\,\mathrm{d}$ of observation epochs rather than the full LISA orbit; and after merger the near-merger network remains broader than the likelihood posterior for some sources. The next step is to test the full data sequence on externally generated LISA data with realistic source content, including waveform mismatch, non-stationary noise, and data gaps. Further steps include extending $\SBIl$ to lighter systems and longer observing windows, and jointly treating closely spaced MBHB mergers.

The main result is that the two parts of the low-latency problem, early warning and near-merger localization, can be addressed within one framework: the networks provide a first posterior within seconds, and importance sampling checks and refines it against the likelihood near merger.

\textbf{Data Availability:  }
The code used to reproduce the results presented in this paper and to construct the neural network pipeline may be made available upon reasonable request to the authors.

\begin{acknowledgments}
The authors thank Riccardo Buscicchio, Massimo Dotti, Francesco Nobili, Malvina Bellotti, Antoine Petiteau, Stephen Green, and Maximillian Dax for useful discussions. RS and LC thank Café Lovat for ample, timely cappuccinos.
We acknowledge support from the European Union’s Horizon ERC Synergy Grant ``Making Sense of the Unexpected in the Gravitational-Wave Sky'' (Grant No. GWSky-101167314, to E.B., R.S., L.C. and C.P.). 
R.T. acknowledges co-funding from Next Generation EU, in the context of the National Recovery and Resilience Plan, Investment PE1 Project FAIR ``Future Artificial Intelligence Research''. This resource was co-financed by the Next Generation EU [DM 1555 del 11.10.22]. R.T. is partially supported by the Fondazione ICSC, Spoke 3 ``Astrophysics and Cosmos Observations'', Piano Nazionale di Ripresa e Resilienza Project ID CN00000013 ``Italian Research Center on High-Performance Computing, Big Data and Quantum Computing'' funded by MUR Missione 4 Componente 2 Investimento 1.4: Potenziamento strutture di ricerca e creazione di ``campioni nazionali di R\&S (M4C2-19)'' - Next Generation EU (NGEU). 
This work has been supported by the Agenzia Spaziale Italiana (ASI), Project n. 2024-36-HH.0, ``Attività per la fase B2/C della missione LISA''. Computational work has been made possible through SISSA-CINECA and CINECA-INFN agreements providing access to resources on LEONARDO at CINECA.MB.
\end{acknowledgments}

\vskip 0.5cm

\appendix

\section{Handedness: training and verification}
\label{app:handedness}

This appendix documents how the handedness defect described in Sec.~\ref{sec:handedness} was identified and how the fold and auxiliary loss were developed and validated.

\emph{The defect.}
Table~\ref{tab:hand-modes} shows the posterior mass of the \emph{reference} signal in each of the four sign combinations of $(\sin\beta,\cos\iota)$. Across all analysis windows, from $\Delta t_c=+4\,\mathrm{h}$ to $-1\,\mathrm{h}$, the unfolded network distributes its posterior mass approximately equally among the four combinations. By contrast, importance sampling retains only the true combination and its reflection through the constellation plane, with approximately half the posterior mass in each. This defect persists across all networks trained without the fold, including transformer widths of $256$ and $512$, suggesting that it is not caused by insufficient model capacity. The input itself contains the necessary information: the sign of $\mathrm{Im}(\tilde A\tilde E^*)$, computed from the compressed frequency-domain input, agrees with $-h$ in $93\%$ of noisy training windows and $97$--$100\%$ of clean, full-resolution windows.

\begin{table}[t]
\caption{Posterior mass of the \emph{reference} signal across the four sign combinations of $(\sin\beta,\cos\iota)$, estimated from $10^5$ stage-1 network draws per window, before (unfolded) and after (folded) the procedure of Sec.~\ref{sec:handedness}. T denotes the true combination, R its exact reflection degeneracy, and $\mathrm{F}_c$ and $\mathrm{F}_b$ the likelihood-disfavored single flips. Cell colors indicate posterior mass, from red ($0$) through yellow ($0.25$) to green ($0.5$). Importance sampling assigns approximately equal mass to T and R, with negligible mass in the other combinations. The production network yields $\mathrm{F}_c+\mathrm{F}_b\leq3\times10^{-4}$ across all $67$ near-merger windows.}
\label{tab:hand-modes}
\begin{ruledtabular}
\begin{tabular}{r|cccc|cccc}
 & \multicolumn{4}{c|}{Unfolded}
 & \multicolumn{4}{c}{Folded} \\

$\Delta t_c$ [h]
 & \textcolor{green!60!black}{\textbf{T}}
 & \textcolor{green!60!black}{\textbf{R}}
 & \textcolor{red!75!black}{$\mathrm{\mathbf{F_c}}$}
 & \textcolor{red!75!black}{$\mathrm{\mathbf{F_b}}$}
 & \textcolor{green!60!black}{\textbf{T}}
 & \textcolor{green!60!black}{\textbf{R}}
 & \textcolor{red!75!black}{$\mathrm{\mathbf{F_c}}$}
 & \textcolor{red!75!black}{$\mathrm{\mathbf{F_b}}$} \\
 \hline

$+4.00$
 & \prob{0.252} & \prob{0.248} & \prob{0.250} & \prob{0.250}
 & \prob{0.504} & \prob{0.496} & \prob{0.000} & \prob{0.000} \\

$+2.50$
 & \prob{0.251} & \prob{0.250} & \prob{0.247} & \prob{0.253}
 & \prob{0.499} & \prob{0.501} & \prob{0.000} & \prob{0.000} \\

$+0.84$
 & \prob{0.253} & \prob{0.249} & \prob{0.247} & \prob{0.251}
 & \prob{0.501} & \prob{0.499} & \prob{0.000} & \prob{0.000} \\

$+0.38$
 & \prob{0.251} & \prob{0.249} & \prob{0.250} & \prob{0.250}
 & \prob{0.503} & \prob{0.496} & \prob{0.000} & \prob{0.000} \\

$+0.07$
 & \prob{0.248} & \prob{0.249} & \prob{0.250} & \prob{0.253}
 & \prob{0.501} & \prob{0.498} & \prob{0.001} & \prob{0.001} \\

$-0.10$
 & \prob{0.250} & \prob{0.252} & \prob{0.247} & \prob{0.252}
 & \prob{0.500} & \prob{0.500} & \prob{0.000} & \prob{0.000} \\

$-1.00$
 & \prob{0.260} & \prob{0.233} & \prob{0.271} & \prob{0.236}
 & \prob{0.523} & \prob{0.475} & \prob{0.001} & \prob{0.001} \\

\end{tabular}
\end{ruledtabular}
\end{table}
 
\emph{Where the information is lost.}
To identify where handedness information is lost, we train linear probes to predict $h$ from intermediate activations of the unfolded network, using $15\,360$ validation windows. The information remains accessible throughout the transformer: a probe on the class-token output of the final transformer block achieves $99.5\%$ accuracy, while probes on the class token or mean token representation of each block achieve $95$--$100\%$. The degradation occurs after the transformer. Accuracy falls to $58\%$ on the 128-feature readout and approaches chance level on the flow context, reaching $49\%$ with a linear probe and $53\%$ with a two-layer probe, compared with a majority-class baseline of $52\%$. By comparison, control probes for $\log_{10}\mathcal{M}_c$ and $t_c$ on the same flow context achieve approximately $95\%$ accuracy. These results indicate that the transformer encodes handedness information, but the readout largely discards it before passing the context to the flow.

We also investigate whether the flow's coupling masks could contribute to the defect. Although $\cos\iota$ and $\sin\beta$ always lie on the same side of the alternating mask in each coupling layer, the same spline-flow architecture successfully learns the corresponding sign dependence in toy problems. Thus, the coupling masks do not prevent the flow from representing this dependence.

\emph{Training with the fold and auxiliary loss.}
We introduce the fold and auxiliary head into the trained, unfolded network and continue training. The combined posterior mass in the two single-flip configurations decreases from $0.49$ to $0.05$ by epoch $12$ ($2.3\times10^3$ steps) and falls below $0.005$ by epoch $30$. The auxiliary head reaches $100\%$ accuracy on windows with well-defined handedness by epoch $12$. The validation loss initially increases by $0.5$ during the transition, recovers within five epochs, and subsequently decreases by $1$~nat.

\emph{Importance-sampling efficiency.}
Eliminating two of four equally weighted modes should approximately double the IS efficiency, provided the proposal otherwise remains unchanged. We compare the folded network with its unfolded predecessor using identical analysis windows and IS settings. Across eight windows spanning $\Delta t_c=+4\,\mathrm{h}$ to $-1\,\mathrm{h}$, the effective sample size increases in every case, by a median factor of $2.0$ ($1.6$--$2.1$ in five windows, with larger improvements in the remaining three). At $\Delta t_c=+4$ and $+2.5\,\mathrm{h}$, the IS efficiency increases from $15$--$19\%$ to $32$--$35\%$.

\section{Multibanding}
\label{app:multibanding}

We use the frequency evolution of the $(2,2)$ mode and $(2,2,0)$ ringdown frequency to set the multibanding resolution. At leading order the inspiral $(2,2)$ mode evolves as~\cite{Maggiore,PeterMatthews}
\begin{equation}
\begin{aligned}
  f_{22}(\tau) &= \frac{1}{\pi} \left(\frac{5}{256\,\tau}\right)^{3/8} \mathcal{M}_c^{-5/8},\\
  \tau(f_{22}) &= \frac{5}{256}\,\mathcal{M}_c^{-5/3}\,(\pi f_{22})^{-8/3},
\end{aligned}
\label{eq:foft}
\end{equation}
in geometric units, where $\tau$ is the time to merger at which the mode reaches $f_{22}$. The ringdown sets the highest frequency the mode reaches~\cite{Berti:2005ys},
\begin{equation}
f_{220}=\frac{1.5251-1.1568\,(1-a_f)^{0.1292}}{2\pi\,M_f},
\label{eq:fring}
\end{equation}
with $\eta=q/(1+q)^2$, remnant spin $a_f=\sqrt{12}\,\eta-2.9\,\eta^2$~\cite{Buonanno2008} and remnant mass $M_f=\mathcal{M}_c\,\eta^{-3/5}(1-0.192\,\eta)$. At fixed chirp mass $f_{220}$ is largest for equal masses, so the extreme case is the lightest equal-mass binary of each prior. For $\SBIl$ ($\mathcal{M}_c = 10^{6.5}\,M_\odot$) Eq.~\eqref{eq:fring} gives $f_{220} \simeq 2.5\,\mathrm{mHz}$ for non-spinning components, at most $\simeq 3.6\,\mathrm{mHz}$ for rapidly spinning remnants, and about twice that for $(4,4,0)$, all below the $15.6\,\mathrm{mHz}$ Nyquist frequency of the $32$-s cadence. $\SBIe$ never sees the merger: its windows end at least $4\,\mathrm{h}$ before coalescence, where even its lightest sources ($\mathcal{M}_c = 10^{5}\,M_\odot$) have reached only $f_{22} \simeq 3.1\,\mathrm{mHz}$ by Eq.~\eqref{eq:foft}, and $\simeq 6.3\,\mathrm{mHz}$ in $(4,4)$.

The time-domain (TD) view divides the window into contiguous bands and decimates each by a factor that decreases towards the end of the window. A band ends where the lightest source's $f_{22}$ reaches a quarter of the band's Nyquist frequency $1/(2\Delta t^*)$, so within every band the Nyquist frequency is at least four times $f_{22}$; this leaves room for the $(3,3)$ and $(4,4)$ modes, at $1.5$ and $2$ times $f_{22}$, and raises the sampling rate well before aliasing sets in. Each band is decimated with an anti-aliasing polyphase filter; adjacent bands overlap by a few coarse samples, and the edge samples of each band are discarded before the bands are joined. The result is a stream of $946$ samples per channel for $\SBIl$ (six bands, $\Delta t^* = 1024$ to $32\,\mathrm{s}$) and $7049$ for $\SBIe$ (three bands, $128$ to $32\,\mathrm{s}$), on a non-uniform time axis whose sample times enter the network as token coordinates (Sec.~\ref{sec:target}). The finest band is at the $32$-s analysis cadence, not at LISA's native rate.
 
The frequency-domain (FD) view reads the same band boundaries as frequency edges and averages contiguous complex bins within each band, keeping the native resolution $\Delta f = 1/T_{\rm obs}$ at low frequency, where the inspiral is long-lived. Averaging preserves the signal only if its phase changes little across the block, i.e.\ if the block width $\Delta f^*$ stays below $\tau(f)^{-1}$ (Fig.~\ref{fig:MB}, bottom). For $\SBIl$ the block sizes are the TD decimation factors in reverse, $1$ to $32$. The lighter sources of $\SBIe$ sweep too fast through the middle band for blocks of two: at leading order these would keep only $68\%$ of the $(2,2)$ power of a $10^{5}\,M_\odot$ source, against $99\%$ without averaging, so $\SBIe$ averages only above $1.95\,\mathrm{mHz}$, in blocks of four. This gives $261$ bins for $\SBIl$ and $1765$ for $\SBIe$, both between $5\times10^{-5}$ and $7.8\times10^{-3}\,\mathrm{Hz}$: the lower edge is the floor of the noise model, the upper edge, $1/(4\,\Delta t)$, lies above the highest signal frequencies of both networks, so higher-mode power is retained. A symmetric Planck taper with $\epsilon = 10^{-3}$ is applied before the Fourier transform to suppress leakage from the window edges; the TD stream is left untapered so that a merger near the end of the window is not attenuated.
 
Counting real and imaginary parts separately, the two views replace the $16384$ raw samples per channel with $946 + 2\times261 = 1468$ numbers for $\SBIl$, a factor of $11$, and $7049 + 2\times1765 = 10\,579$ for $\SBIe$, a factor of $1.5$. The compression is weaker for $\SBIe$ because its lightest sources already reach mHz frequencies days before merger. This is one reason for a separate, mass-targeted network near merger: the narrow mass prior of $\SBIl$ allows a stronger compression for the harder nine-parameter inference.

\section{Network Specifications}
\label{app:network}

Table~\ref{table:network} summarizes the complete architecture of $\SBIl$ and $\SBIe$. The two networks share the same transformer width, number of encoder blocks and flow construction, but differ in their input resolution, sequence length and number of inferred parameters. $\SBIe$ additionally uses the attention-pool readout introduced in Sec.\ref{sec:training}, while $\SBIl$ includes the handedness fold and auxiliary classification head described in Sec.\ref{sec:handedness}.

Table~\ref{table:transformer} expands the transformer encoder used by both networks. Each encoder contains six identical pre-layer-normalization blocks, with eight-head self-attention followed by a two-layer feed-forward network. The parameter count of the encoder is independent of the number of tokens and is therefore the same for $\SBIl$ and $\SBIe$, despite their different sequence lengths.

\begin{table*}[ht]
\centering
\small\setlength{\tabcolsep}{4pt}
\begin{tabular}{l c c >{\raggedright\arraybackslash}p{6.cm} r r}
\hline
\textbf{Layer} & \textbf{Input shape} & \textbf{Output shape} & \textbf{Details} & \multicolumn{2}{c}{\textbf{Params}}\\
& & & & \textbf{$\SBIl$} & \textbf{$\SBIe$} \\
\hline
\noalign{\vskip 2pt}
\multicolumn{6}{l}{\emph{\textbf{Tokenizer}}} \\
\noalign{\vskip 2pt}
TD patching & $(B, 2, N)$ & $(B, n_{\rm TD}, 2P{+}1)$ & $\mathbf{u}^{\rm TD}$ (Equation~\ref{eq:tok-td}) & -- & -- \\
\noalign{\vskip 2pt}
FD patching & $(B, 2, 2L)$ & $(B, n_{\rm FD}, 4P{+}2)$ & $\mathbf{u}^{\rm FD}$ (Equation~\ref{eq:tok-fd}) & -- & -- \\
\noalign{\vskip 2pt}
TD embedding & $(B, n_{\rm TD}, 2P{+}1)$ & $(B, n_{\rm TD}, 512)$ & $\texttt{L}_{2P+1\!\to\!512}$, \texttt{GELU}, $\texttt{L}_{512\!\to\!512}$, \texttt{LN} & $272\,896$ & $281\,088$ \\
\noalign{\vskip 2pt}
FD embedding & $(B, n_{\rm FD}, 4P{+}2)$ & $(B, n_{\rm FD}, 512)$ & $\texttt{L}_{4P+2\!\to\!512}$, \texttt{GELU}, $\texttt{L}_{512\!\to\!512}$, \texttt{LN} & $281\,600$ & $297\,984$ \\
\noalign{\vskip 2pt}
$t_0$ embedding & $(B, 4)$ & $(B, 512)$ & $[\sin, \cos](2\pi t_0)$, $[\sin, \cos](4\pi t_0)$; $\texttt{L}_{4\!\to\!512}$, \texttt{GELU}, $\texttt{L}_{512\!\to\!512}$; added to the class token & $265\,216$ & $265\,216$ \\
\noalign{\vskip 2pt}
Sequence assembly & -- & $(B, n_{\rm tok}, 512)$ & $[\,\texttt{CLS} \mid \mathrm{TD} \mid \mathrm{FD}\,]$; learned class token & $512$ & $512$ \\
\noalign{\vskip 2pt}
Position embedding & $(B, n_{\rm tok}, 512)$ & $(B, n_{\rm tok}, 512)$ & Learned vector per token slot (position and view), added & $78\,336$ & $283\,136$ \\
\noalign{\vskip 2pt}
\hline
\noalign{\vskip 2pt}
\multicolumn{6}{l}{\emph{\textbf{Encoder}}} \\
\noalign{\vskip 2pt}
$6 \times$Transformers & $(B, n_{\rm tok}, 512)$ & $(B, n_{\rm tok}, 512)$ & Table~\ref{table:transformer} & $12\,616\,704$ & $12\,616\,704$ \\
\hline
\noalign{\vskip 2pt}
\multicolumn{6}{l}{\emph{\textbf{Summary}}} \\
\noalign{\vskip 2pt}
Class-token readout & $(B, n_{\rm tok}, 512)$ & $(B, 128)$ & Position $0$; \texttt{LN}, $\texttt{L}_{512\!\to\!128}$ & $66\,688$ & $66\,688$ \\
Attn. pool readout & $(B, n_{\rm tok}, 512)$ & $(B, 128)$ & $\SBIe$ only: $8$ learned queries cross-attend the whole sequence ($8$ heads); \texttt{LN}, $\texttt{L}_{4096\!\to\!128}$; added to the class-token readout & -- & $1\,587\,328$ \\
Projection & $(B, 128)$ & $(B, 128)$ & \texttt{LN}, $\texttt{L}_{128\!\to\!128}$, \texttt{GELU}, $\texttt{L}_{128\!\to\!128}$ & $33\,280$ & $33\,280$ \\
Start-time append & $(B, 128)$, $(B, 2)$ & $(B, 130)$ & $[\,\mathbf{z},\, \sin 2\pi t_0,\, \cos 2\pi t_0\,]$ $\to$ flow context & -- & -- \\
Handedness head & $(B, 130)$ & $(B, 1)$ & $\SBIl$ only: $\texttt{L}_{130\!\to\!1}$ on the flow context; logit of $\mathrm{sgn}(\sin\beta\cos\iota)$ for the auxiliary loss of Eq.~\eqref{eq:handaux}; training only & $131$ & -- \\
\hline
\noalign{\vskip 2pt}
\multicolumn{6}{l}{\emph{\textbf{Normalizing flow}} ($\boldsymbol{\theta} \to \mathcal{N}(\mathbf{0}, \mathbf{I}_\rm{D})$} \\
\noalign{\vskip 2pt}
Handedness fold & $(B, 9)$ & $(B, 9)$ & $\SBIl$ only: $\cos\iota \to \cos\iota\,\mathrm{sgn}(\sin\beta)$, Eq.~\eqref{eq:handfold} & -- & -- \\
Standardization & $(B, D)$ & $(B, D)$ & Fixed per-parameter affine z-score & -- & -- \\
$8 \times$Spline coupling & $(B, D)$, $(B, 130)$ & $(B, D)$ & Rational-quadratic spline, $32$ bins; conditioner: residual net ($2 \times 128$) on the context & $1\,376\,604$ & $1\,079\,412$ \\
$8 \times$LU linear & $(B, D)$ & $(B, D)$ & Invertible linear layer after each coupling & $720$ & $96$ \\
Base distribution & $(B, D)$ & -- & $\mathcal{N}(\mathbf{0}, \mathbf{I}_D)$ & -- & -- \\
\noalign{\vskip 2pt}
\hline
\noalign{\vskip 2pt}
\multicolumn{4}{l}{\textbf{Total}} & $14\,992\,687$ & $16\,511\,444$ \\
\hline
\end{tabular}
\caption{Architecture of the two networks. $B$ is the batch size; TD and FD denote the time- and frequency-domain views (Sec.~\ref{sec:target}), with $N$ samples and $L$ bins per channel, patch size $P$, and $n_{\rm TD}$, $n_{\rm FD}$ tokens ($N = 946$, $L = 261$, $P = 8$, $n_{\rm TD} = 119$, $n_{\rm FD} = 33$, $n_{\rm tok} = 153$, $\rm{D} = 9$ for $\SBIl$; $N = 7049$, $L = 1765$, $P = 16$, $n_{\rm TD} = 441$, $n_{\rm FD} = 111$, $n_{\rm tok} = 553$, $\rm{D} = 3$ for $\SBIe$). \texttt{CLS} is the learned class token; \texttt{GELU} the Gaussian error linear unit; \texttt{LN} layer normalization; $\texttt{L}_{x\to y}$ a linear layer mapping $x$ to $y$ neurons; and LU linear an invertible linear layer whose weights are parametrized by their lower--upper decomposition. The two networks share the model width ($512$), the encoder and the flow construction. $\SBIe$ adds an attention-pool readout, initialized at zero so that it leaves the class-token readout unchanged at first (Sec.~\ref{sec:training}), has no handedness fold or head, and its flow acts on three parameters.}
\label{table:network}
\end{table*}
 
\begin{table*}[ht]
\centering
\begin{tabular}{l c c >{\raggedright\arraybackslash}p{7.5cm} r}
\hline
\textbf{Operation} & \textbf{Input shape} & \textbf{Output shape} & \textbf{Details} & \textbf{Params} \\
\hline
\noalign{\vskip 2pt}
\multicolumn{5}{l}{\emph{\textbf{Attention} (MHSA) sub-layer}}\\
\noalign{\vskip 2pt}
\texttt{LN} & $(B, n_{\rm tok}, 512)$ & $(B, n_{\rm tok}, 512)$ & Per-token normalization, learned gain and bias & $1\,024$ \\
\noalign{\vskip 2pt}
Q, K, V projection & $(B, n_{\rm tok}, 512)$ & $3 \times (B, n_{\rm tok}, 512)$ & $\texttt{L}_{512 \to 3 \times 512}$ & $787\,968$ \\
\noalign{\vskip 2pt}
8-Heads & $3 \times (B, n_{\rm tok}, 512)$ & $3 \times (B, 8, n_{\rm tok}, 64)$ & $8$ heads of dimension $d_h = 64$ & -- \\
\noalign{\vskip 2pt}
Attention & $(B, 8, n_{\rm tok}, 64)$ & $(B, 8, n_{\rm tok}, n_{\rm tok})$ & $\mathcal{A} = \mathrm{softmax}\bigl(QK^{\mathsf T}/\sqrt{d_h}\bigr)$; dense, no mask & -- \\
\noalign{\vskip 2pt}
Weighted sum & $(B, 8, n_{\rm tok}, n_{\rm tok})$ & $(B, 8, n_{\rm tok}, 64)$ & $\mathcal{A}\,V$, per head & -- \\
\noalign{\vskip 2pt}
Output projection & $(B, 8, n_{\rm tok}, 64)$ & $(B, n_{\rm tok}, 512)$ & Concatenate the $8$ heads; $\texttt{L}_{512 \to 512}$ & $262\,656$ \\
\noalign{\vskip 2pt}
Residual add & $(B, n_{\rm tok}, 512)$ & $(B, n_{\rm tok}, 512)$ & Added to sub-layer input & -- \\
\hline
\noalign{\vskip 2pt}
\multicolumn{5}{l}{\emph{\textbf{Feed-forward} sub-layer}}\\
\noalign{\vskip 2pt}
\texttt{LN} & $(B, n_{\rm tok}, 512)$ & $(B, n_{\rm tok}, 512)$ & Per-token normalization, learned gain and bias & $1\,024$ \\
\noalign{\vskip 2pt}
Expansion & $(B, n_{\rm tok}, 512)$ & $(B, n_{\rm tok}, 1024)$ & $\texttt{L}_{512 \to 1024}$, \texttt{GELU}; for each token separately & $525\,312$ \\
\noalign{\vskip 2pt}
Contraction & $(B, n_{\rm tok}, 1024)$ & $(B, n_{\rm tok}, 512)$ & $\texttt{L}_{1024 \to 512}$ & $524\,800$ \\
\noalign{\vskip 2pt}
Residual add & $(B, n_{\rm tok}, 512)$ & $(B, n_{\rm tok}, 512)$ & Added to sub-layer input & -- \\
\hline
\noalign{\vskip 2pt}
\multicolumn{4}{l}{\textbf{Total} one block} & $2\,102\,784$ \\
\noalign{\vskip 2pt}
\multicolumn{4}{l}{\textbf{Total} six-block encoder} & $12\,616\,704$ \\
\hline
\end{tabular}
\caption{One pre-layer-normalization transformer block of the encoder; both networks stack six such blocks. $B$ is the batch size, $n_{\rm tok}$ the sequence length (class token $+$ TD $+$ FD tokens: $153$ for $\SBIl$, $553$ for $\SBIe$) and $512$ the model width. MHSA is multi-head self-attention, \texttt{LN} layer normalization, and $\texttt{L}_{x\to y}$ a linear layer mapping $x$ to $y$ neurons. The parameter count does not depend on the sequence length, so it is the same for both networks.}
\label{table:transformer}
\end{table*}

\section{Precision across the prior}
\label{app:res-population}

\begin{figure*}[t]
  \centering
  \includegraphics[width=\textwidth]{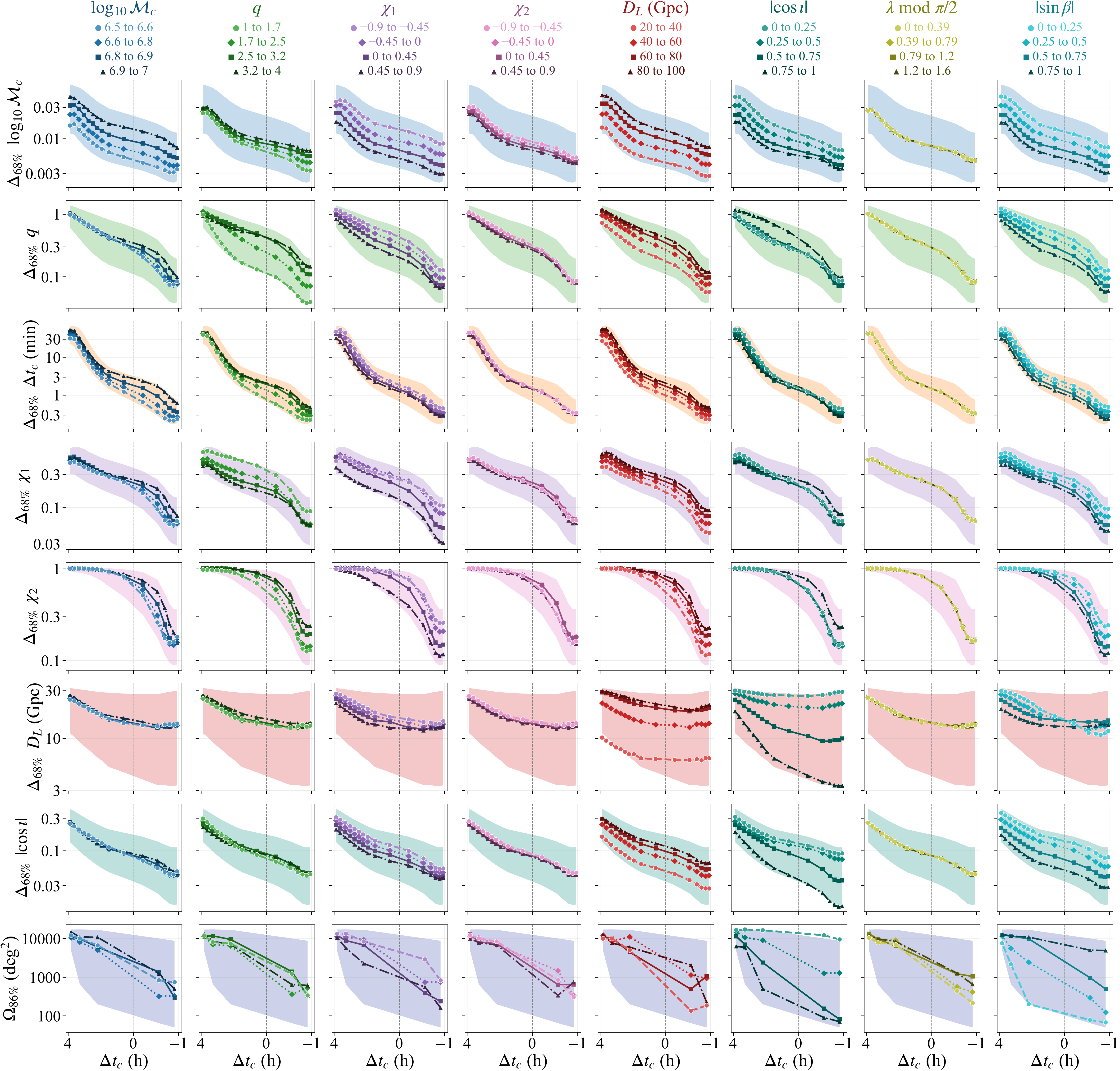}
  \caption{Population precision of $\SBIl$ against $\Delta t_c$. Rows 1--7: the width $\Delta_{68\%}$ (bounds of the highest 68.3\% probability density) of each inferred quantity in the network posterior, from $1.1\times10^{6}$ signals drawn from the $\SBIl$ prior, each with an independent noise realization. Row 8: the sky area $\Omega_{86\%}$ (corresponding to the highest 86.5\% probability density region) of the importance-sampled posterior, for the $\sim1000$ signals of the importance-sampling study (Sec~\ref{sec:res-likelihood}). Columns: the population split into quartiles of the true parameter lying within the listed range. Lines with markers give the median width of each quartile, from the lowest quartile (lightest, dashed, circles) to the highest (darkest, dash-dot, triangles); the shaded band encloses the 16th--84th percentiles of all signals. $\lambda$ is folded onto one quarter turn ($\lambda \bmod \pi/2$) and $\cos\iota$, $\sin\beta$ enter through their absolute values. The y-axes are on a log scale.}
  \label{fig:population}
\end{figure*}

Our amortized network makes it possible to study posterior precision across a population that would be computationally prohibitive with likelihood-based sampling. In Fig.\ref{fig:population}, we analyze $1.1\times10^{6}$ signals drawn from the joint prior. Each signal has an independent noise realization, coalescence time relative to the window, and observation epoch $t_0$. The full study took about five GPU hours. Importance sampling the same population would require about $10$ GPU-years at $282\,\mathrm{s}$ per window (Sec.\ref{sec:res-cost}), while a full MCMC analysis would be substantially more expensive.

Figure~\ref{fig:population} shows how the $\SBIl$ uncertainty in each inferred parameter depends on the true source parameters. Each row gives the highest probability density's $68\%$ width of one inferred quantity, or the $86.5\%$ sky area, as a function of $\Delta t_c$. Each column divides the population into quartiles of the labeled true parameter. The curves show the median uncertainty within each quartile. A cell therefore shows whether the true parameter labeling that column affects the precision of the quantity in that row. The shaded band gives the spread over the full prior. The first seven rows use the network posterior. The final row uses the IS-reweighted sky posteriors of the $\sim1000$ signals from Sec.~\ref{sec:res-likelihood}, since sky area is the quantity most strongly changed by reweighting. We also checked the first seven rows on the IS sample and found the same ordering between quartiles. The trends discussed below are therefore not introduced by the network approximation.

\emph{Time evolution: $\Delta t_c$.} Almost every inferred quantity improves substantially as merger approaches. Between $\Delta t_c=+4$ and $-1\,\mathrm{h}$, the widths fall by factors of $5$ to $120$, with $D_L$ as the main exception. The coalescence time improves most strongly before merger, followed by chirp mass, mass ratio, and inclination. The secondary spin behaves differently and improves mainly once the merger enters the window. Before merger, the phase primarily constrains the effective spin $\chi_{\rm eff} = (m_1\chi_1+m_2\chi_2)/(m_1 + m_2)$, so the corresponding degeneracy can leave the marginal $\chi_2$ posterior as broad as the prior. The final cycles, merger, and ringdown provide the additional information needed to break this degeneracy.

Distance shows the opposite behavior. For most subpopulations, its median uncertainty improves by only about a factor of two before merger and then changes little. The additional SNR from merger and ringdown therefore does not translate into a comparable improvement in $D_L$. Instead, the distance is limited by its degeneracy with inclination, polarization, and latitude. This degeneracy weakens toward face-on orientations, which is why those sources continue to improve after merger.

\emph{Signal-to-noise ratio: $D_L$ and $|\sin\beta|$.} The distance and latitude columns order nearly every row in the direction expected from SNR. From the nearest to the farthest distance quartile, the uncertainties in chirp mass, mass ratio, coalescence time, and spins grow by factors of $2$--$3$. This is comparable to the inverse SNR ratio of about $3$ between the quartile midpoints. The distance uncertainty again behaves differently because it reaches the geometric degeneracy described above.

Latitude produces a similar SNR trend through the antenna response. High-latitude sources with $|\sin\beta|\to1$ have a larger rms response than sources near the constellation plane. Apart from the sky area, the posterior widths decrease by about a factor of two from the in-plane to the highest-latitude quartile. This is consistent with the corresponding rms-response ratio of about $2.4$.

\emph{Cycles in the window: $\log_{10}\mathcal{M}_c$.} Chirp mass mainly controls the quantities measured through phase evolution. Heavier binaries complete fewer cycles in the analysis window and merge at lower frequencies, while the SNR is nearly independent of mass across this prior. From the lightest to the heaviest quartile, the uncertainty at merger grows by a factor of about three in both chirp mass and coalescence time. The mass-ratio and spin widths increase more modestly, by about $20$--$30\%$. Distance and inclination show little change. The dominant effect of chirp mass is therefore the number of informative phase cycles accumulated in the window.

\emph{Subdominant harmonics: $q$, $\chi_1$, $|\cos\iota|$.} These parameters affect different inferred quantities in different directions. Much of the structure follows the relative amplitudes of the $(2,1)$, $(3,3)$, and $(4,4)$ modes with respect to the dominant $(2,2)$ mode. The subdominant modes become stronger for more unequal masses and for larger $\sin\iota$.

Mass ratio also changes how the two spins are measured. More unequal masses improve the primary-spin measurement but worsen the secondary-spin measurement because the heavier black hole contributes more strongly to the spin dynamics. The most equal-mass quartile has the narrowest $q$ posterior. Odd-$m$ harmonics vanish at $q=1$, so their absence strongly constrains departures from equal mass.

The primary spin produces a broader effect. More aligned values of $\chi_1$ improve the precision of most quantities around merger. At fixed time before merger, an aligned-spin binary is at a higher frequency because its inspiral proceeds more slowly. More of its observable cycles therefore lie where the acceleration noise is lower, and the system also merges at a higher frequency.

Inclination introduces a competition between SNR and harmonic content. Face-on sources have the best distance measurements but the worst constraints on mass ratio and spins. Edge-on systems show the opposite behavior. The distance uncertainty improves by a factor of about eight from the most edge-on to the most face-on quartile and spans almost the full population band. Face-on signals are circularly polarized, which removes part of the degeneracy between amplitude, polarization, latitude, and distance.\footnote{For a linearly polarized signal, each channel measures one amplitude, $F_+^{A,E}(\lambda,\beta,\psi),h_+/D_L$, for four unknowns. For a circularly polarized signal, $\psi$ becomes a common phase that can be absorbed into $\phi_c$. The two channel amplitudes and their relative phase then constrain $(\lambda,\beta,D_L)$ up to the discrete sky modes.} This is also why the distance of face-on systems continues to improve after merger while that of more inclined systems saturates.

The mass-ratio and spin widths show more of a threshold than a continuous inclination trend. The three quartiles with $|\cos\iota|<0.75$ nearly overlap. Only the most face-on quartile becomes noticeably broader, by about a factor of $1.5$ in $q$ and $\chi_2$. In this regime the subdominant harmonics are weak and $\sin\iota$ itself is poorly constrained, producing a degeneracy between $q$ and inclination through the harmonic amplitude ratios.

\emph{Weak or no trend: $\lambda$ and $\chi_2$.} Longitude has little effect on the precision of any inferred quantity, as expected from the rotational symmetry of the $(A,E)$ response. The secondary-spin column shows a weaker version of the ordering produced by the primary spin. Both spins enter the inspiral phase mainly through $\chi_{\rm eff}$, but the secondary is weighted by the lighter mass. For the typical $q\simeq2$--$3$ systems in this prior, that weight is only about one third.

\emph{Sky area: $\Omega_{86.5\%}$.} Sky localization depends much more strongly on source geometry than the other inferred quantities. Inclination and latitude dominate the variation across the prior, while distance, masses, and spins have smaller effects. In the late-time subsample, $34\%$ of sources are localized to better than $100\,\mathrm{deg}^2$ and $58\%$ to better than $1000\,\mathrm{deg}^2$. Appendix~\ref{app:res-sky} examines these population trends in detail and uses representative sky maps to show how the individual modes change.
\begin{figure*}[t]
  \centering
  \includegraphics[width=\textwidth]{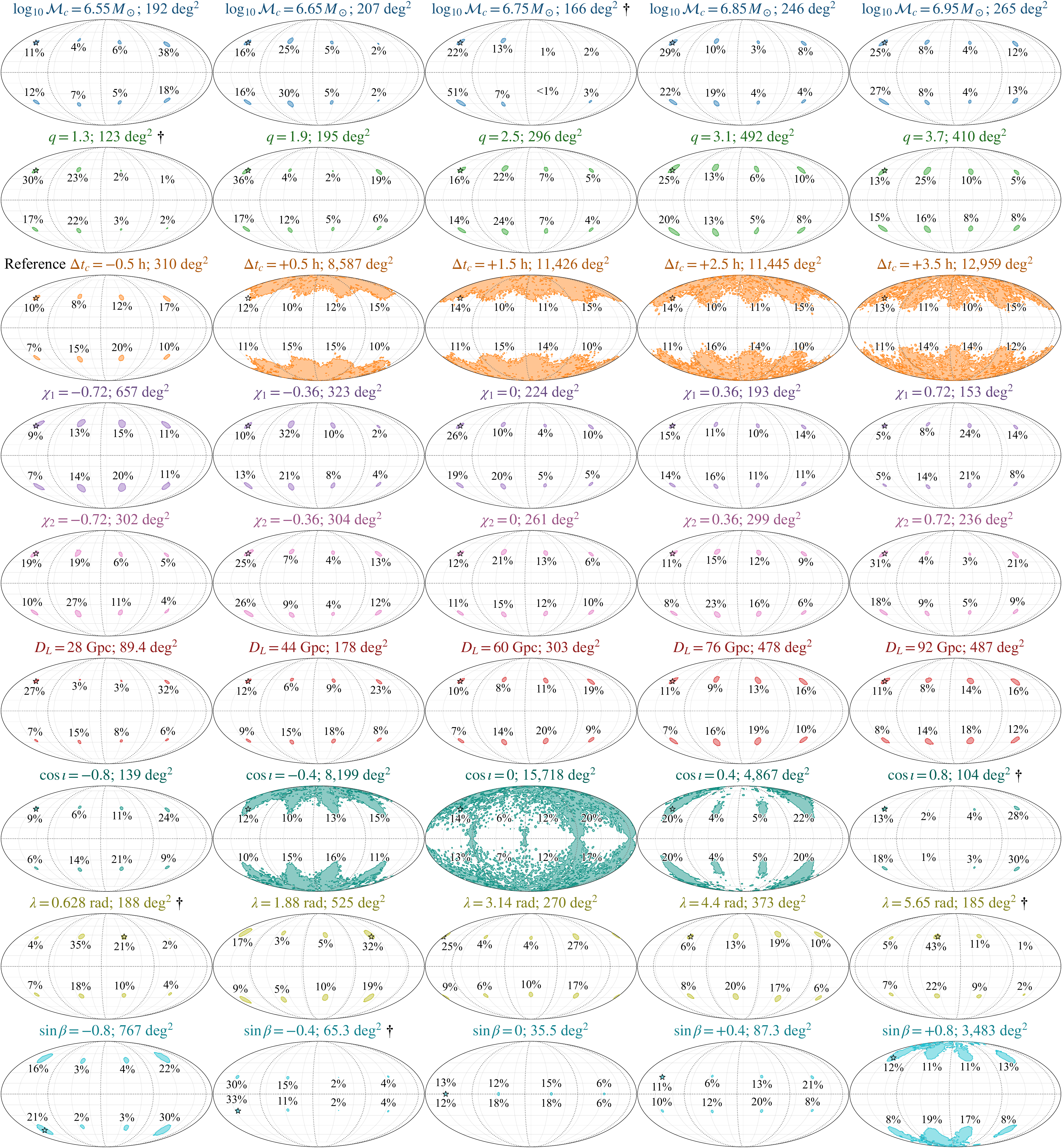}
  \caption{Sky localization of $\SBIl$ after importance sampling for one-parameter variations of the \emph{reference} signal ($\log_{10}(\mathcal{M}_c/M_\odot)=6.836$, $q=2.19$, $D_L=61.1$\,Gpc, $\cos\iota=-0.64$, $\lambda=3.75$\,rad, $\sin\beta=0.67$, $\chi_1=-0.13$, $\chi_2=0.50$ at $\Delta t_c=-0.5$~h). Each row varies one parameter over five values across the $\SBIl$ prior, all others held at their \emph{reference} values. All signals share one noise realization. Panels show the full sky (Mollweide, LISA frame at merger) and shaded pixels form the $86.5\,\%$ ($2\sigma$) region, and titles give its total area $\Omega_{86.5\%}$ over all patches. The star marks the true position. Grey dotted lines divide each sky into eight octants around the true posterior modes. The number in each octant shows the percentage of importance-sampled $\SBIl$ posterior samples that fall in it. A dagger after a panel title marks sources with an octant that holds less than $0.6\,\mathrm{deg}^2$ (three map pixels) of the $\Omega_{86.5\,\%}$ region.  Effective sample sizes range from 52 to $2.7\times10^{4}$, corresponding to an IS efficiency of 0.05 to 27\%.}
  \label{fig:skymaps}
\end{figure*}

\section{Sky localization}
\label{app:res-sky}

Figure~\ref{fig:skymaps} varies one parameter of the \emph{reference} signal at a time and shows the importance-sampled sky posterior $30\,\mathrm{min}$ after merger. We use these examples together with the population study of Appendix~\ref{app:res-population} to understand what controls the total sky area and the structure of its modes. The \emph{reference} signal is localized to $\Omega_{86.5\%}=310\,\mathrm{deg}^2$. This area is divided among eight modes of about $40\,\mathrm{deg}^2$ each. Four modes lie at longitudes separated by a quarter turn, and each has a reflected counterpart across the constellation plane. The individual modes contain between $7$ and $20\%$ of the posterior samples. Before merger, the same source remains much less localized at $8,600$--$13,000\,\mathrm{deg}^2$. Its posterior then forms two broad polar caps whose edges lie near the true latitude and its reflection. The inspiral excludes low latitudes, while the merger provides the additional information that selects the allowed longitudes.

Inclination and latitude have by far the largest effect on localization. In the population, the fraction of sources localized to better than $100\,\mathrm{deg}^2$ in windows ending more than $30\,\mathrm{min}$ after merger rises from $2$ to $62\%$ across the inclination quartiles. Across the latitude quartiles it falls from $69$ to $0\%$. Figure~\ref{fig:skymaps} shows the corresponding changes in the posterior geometry. As $\cos\iota\to0$, the sky area grows by about two orders of magnitude. The handedness $\mathrm{sgn}(\sin\beta\cos\iota)$ becomes difficult to measure and the discrete modes broaden into bands. At $\cos\iota=0$, the posterior covers about $40\%$ of the sky. Latitude produces an equally strong change. In the constellation plane, with $\sin\beta=0$, each reflected pair coincides and the posterior reduces to four patches with a total area of $36\,\mathrm{deg}^2$. At $|\sin\beta|=0.4$, the eight modes remain separate but the total area is only $65$--$90\,\mathrm{deg}^2$. At $|\sin\beta|=0.8$, it grows to $(0.8$--$3.5)\times10^3\,\mathrm{deg}^2$. The population spans the same range. In-plane sources are also localized earlier because their longitude, up to the quarter-turn ambiguity, is constrained throughout the inspiral by the relative amplitudes in the $A$ and $E$ channels. For the in-plane copy of the \emph{reference} signal, the sky area is about $200\,\mathrm{deg}^2$ at $1.5\,\mathrm{h}$ before merger and about $100\,\mathrm{deg}^2$ half an hour before merger (Fig.~\ref{fig:countdown}). Areas below about $7\,\mathrm{deg}^2$ per mode are limited by the smoothing procedure.

The remaining parameters have smaller effects. In the one-parameter scan, most change the size of the eight modes without moving their locations. Longitude is the exception because changing $\lambda$ rotates the full pattern with the source. The total area grows with distance and mass ratio and decreases as the primary spin becomes more aligned. Chirp mass and secondary spin have weaker effects in the scan. The area remains between $170$ and $520\,\mathrm{deg}^2$ in these cases. The population study confirms the same broad trends when inclination and latitude are held fixed. Across the prior, the area grows by a factor of $2.6\pm0.4$ with distance, $1.6\pm0.2$ with chirp mass, and $1.4\pm0.2$ with mass ratio. It falls by a factor of $3.2\pm0.4$ from anti-aligned to aligned primary spin. The secondary spin and longitude have no measurable effect. These changes are small compared with the orders-of-magnitude variation produced by inclination and latitude. This shows that post-merger localization is controlled mainly by geometric degeneracies rather than by SNR alone.

The octant labels in Fig.~\ref{fig:skymaps} show how the posterior mass is distributed among the eight sky modes. The modes are not equally weighted. Across the scan, individual weights range from below $1\%$ to $52\%$. These differences are physical rather than a network artifact. Across the population, the true mode contains $22\%$ of the samples on average, consistent with the mean posterior probability of $23\%$ assigned to it. The reflection degeneracy remains nearly exact. On average, $51\%$ of the posterior mass lies in the hemisphere containing the true source. The quarter-turn degeneracy is only approximate. The true longitude pair contains $41\%$ of the posterior mass, compared with $25\%$ for an exact fourfold degeneracy. This degeneracy is broken more strongly for louder systems and for sources that merge at higher frequencies. Closer and more face-on systems therefore show a larger imbalance between the longitude modes. Mass and spin have a similar effect through their influence on the remnant ringdown frequency. The distance row illustrates the SNR dependence. At $28\,\mathrm{Gpc}$, the most and least populated modes differ by a factor of $12$. At $92\,\mathrm{Gpc}$, the difference is only a factor of $2$.

A longitude pair drops out of the $86.5\%$ region once its posterior mass falls below about $2\%$. The daggers in Fig.~\ref{fig:skymaps} mark these cases, where the removed octants contain only $0.4$--$1.7\%$ of the samples. Across the population, this occurs in $39\%$ of compact posteriors with $\Omega_{86.5\%}\le1000\,\mathrm{deg}^2$. It is more common for sources with higher ringdown frequencies, smaller distances, or more face-on orientations. The $q=1.3$ and $\cos\iota=0.8$ panels are examples. Not every removed mode follows one of these population trends. Three of the remaining cases vary longitude or latitude, which do not control mode removal in the population study. The fourth has $\log_{10}\mathcal{M}_c=6.75$ and lies near the middle of its row, while the lighter systems in the same row retain all eight modes. The apparent number of modes can also decrease for geometric reasons. In the constellation plane, each reflected pair coincides and eight modes become four patches. Near the pole, the four longitude modes lie on a small circle and merge into one polar patch in each hemisphere. Broad posteriors behave differently again. Before merger, or for poorly localized sources, the octants divide a continuous region rather than distinct modes and their masses are nearly equal. About half of the compact population posteriors retain eight separate patches. For a typical source, post-merger localization is therefore determined more by the size and relative weight of the patches than by their number.

\bibliographystyle{utphys}
\bibliography{refs}

@article{Yi:2025pxe,
    author = "Yi, Sophia and Iacovelli, Francesco and Marsat, Sylvain and Wadekar, Digvijay and Berti, Emanuele",
    title = "{Systematic biases in parameter estimation on LISA binaries: The effect of excluding higher harmonics for non-spinning binaries}",
    eprint = "2502.12237",
    archivePrefix = "arXiv",
    primaryClass = "gr-qc",
    doi = "10.1103/3gs3-gmb4",
    journal = "Phys. Rev. D",
    volume = "112",
    number = "12",
    pages = "124063",
    year = "2025"
}

@article{Yi:2026ucv,
    author = "Yi, Sophia and Iacovelli, Francesco and Berti, Emanuele and Chandramouli, Rohit S. and Marsat, Sylvain and Wadekar, Digvijay and Yunes, Nicol{\'a}s",
    title = "{Systematic biases in parameter estimation on LISA binaries. II. The effect of excluding higher harmonics for spin-aligned, high-mass binaries}",
    eprint = "2602.09088",
    archivePrefix = "arXiv",
    primaryClass = "gr-qc",
    doi = "10.1103/k14f-vqnx",
    journal = "Phys. Rev. D",
    volume = "114",
    number = "2",
    pages = "024017",
    year = "2026"
}

@article{Pitte:2023ltw,
    author = "Pitte, Chantal and Baghi, Quentin and Marsat, Sylvain and Besan{\c{c}}on, Marc and Petiteau, Antoine",
    title = "{Detectability of higher harmonics with LISA}",
    eprint = "2304.03142",
    archivePrefix = "arXiv",
    primaryClass = "gr-qc",
    doi = "10.1103/PhysRevD.108.044053",
    journal = "Phys. Rev. D",
    volume = "108",
    number = "4",
    pages = "044053",
    year = "2023"
}

@article{alessandro_santini_2026_19390393,
    author = "Santini, Alessandro and Bandopadhyay, Diganta and Gair, Jonathan",
    title = "{Phentax: A JAX implementation of the IMRPhenomTHM waveform model}",
    eprint = "2611.08596",
    archivePrefix = "arXiv",
    primaryClass = "gr-qc",
    month = "11",
    year = "2026"
}

@article{Estelles:2020twz,
    author = "Estell{\'e}s, H{\'e}ctor and Husa, Sascha and Colleoni, Marta and Keitel, David and Mateu-Lucena, Maite and Garc{\'\i}a-Quir{\'o}s, Cecilio and Ramos-Buades, Antoni and Borchers, Angela",
    title = "{Time-domain phenomenological model of gravitational-wave subdominant harmonics for quasicircular nonprecessing binary black hole coalescences}",
    eprint = "2012.11923",
    archivePrefix = "arXiv",
    primaryClass = "gr-qc",
    doi = "10.1103/PhysRevD.105.084039",
    journal = "Phys. Rev. D",
    volume = "105",
    number = "8",
    pages = "084039",
    year = "2022"
}

@article{garcíaquirós2025gpuacceleratedlisaparameterestimation,
      title={GPU-accelerated LISA parameter estimation with full time domain response},
      author={Cecilio García-Quirós and Shubhanshu Tiwari and Stanislav Babak},
      year={2025},
      eprint={2501.08261},
      archivePrefix={arXiv},
      primaryClass={gr-qc},
      url={https://arxiv.org/abs/2501.08261},
    journal = {Phys. Rev. D},
    volume = {112},
    pages = {064017},
    doi = {10.1103/79kn-53nt}
}

@article{Agrawal:2026ohy,
    author = "Agrawal, Rishav and Allietta, Cl{\'e}mence and Burke, Ollie and Chua, Alvin J. K.",
    title = "{Untangling Extreme Mass Ratio Inspirals and Massive Black Hole Binaries in LISA Data}",
    eprint = "2610.27642",
    archivePrefix = "arXiv",
    primaryClass = "gr-qc",
    month = "10",
    year = "2026"
}

@inproceedings{SplineFlows_2019,
    author = {Durkan, Conor and Bekasov, Artur and Murray, Iain and Papamakarios, George},
    title = {Neural Spline Flows},
    booktitle = {Advances in Neural Information Processing Systems},
    volume = {32},
    year = {2019},
    eprint = {1906.04032},
    archivePrefix = {arXiv}
}

@article{mangiagli2020,
  author        = {Mangiagli, Alberto and Klein, Antoine and Bonetti, Matteo and Katz, Michael L. and Sesana, Alberto and Volonteri, Marta and Colpi, Monica and Marsat, Sylvain and Babak, Stanislav},
  title         = {Observing the inspiral of coalescing massive black hole binaries with {LISA} in the era of multimessenger astrophysics},
  journal       = {Phys. Rev. D},
  volume        = {102},
  pages         = {084056},
  year          = {2020},
  doi           = {10.1103/PhysRevD.102.084056},
  eprint        = {2006.12513},
  archivePrefix = {arXiv},
  primaryClass  = {astro-ph.HE}
}

@article{marsat2021,
  author        = {Marsat, Sylvain and Baker, John G. and Dal Canton, Tito},
  title         = {Exploring the {Bayesian} parameter estimation of binary black holes with {LISA}},
  journal       = {Phys. Rev. D},
  volume        = {103},
  pages         = {083011},
  year          = {2021},
  doi           = {10.1103/PhysRevD.103.083011},
  eprint        = {2003.00357},
  archivePrefix = {arXiv},
  primaryClass  = {gr-qc}
}

@ARTICLE{Attention-is-all-you-need_2017,
       author = {{Vaswani}, Ashish and {Shazeer}, Noam and {Parmar}, Niki and {Uszkoreit}, Jakob and {Jones}, Llion and {Gomez}, Aidan N. and {Kaiser}, Lukasz and {Polosukhin}, Illia},
        title = "{Attention Is All You Need}",
      journal = {arXiv e-prints},
         year = 2017,
        month = jun,
          eid = {arXiv:1706.03762},
        pages = {arXiv:1706.03762},
          doi = {10.48550/arXiv.1706.03762},
archivePrefix = {arXiv},
       eprint = {1706.03762},
 primaryClass = {cs.CL},
       adsurl = {https://ui.adsabs.harvard.edu/abs/2017arXiv170603762V}
}

@article{Korol+2022,
    author = {Korol, Valeriya and Hallakoun, Na’ama and Toonen, Silvia and Karnesis, Nikolaos},
    title = {Observationally driven Galactic double white dwarf population for LISA},
    journal = {Monthly Notices of the Royal Astronomical Society},
    volume = {511},
    number = {4},
    pages = {5936-5947},
    year = {2022},
    month = {02},
    issn = {0035-8711},
    doi = {10.1093/mnras/stac415}
}

@techreport{LISA2018ScienceReq,
  author       = {{LISA Science Study Team}},
  title        = {{LISA Science Requirements Document}},
  institution  = {European Space Agency},
  year         = {2018},
  type         = {Technical Report},
  note         = {Available from \url{www.cosmos.esa.int/documents/678316/1700384/SciRD.pdf}}
}

@article{Alvey:2023npw,
    author = "Alvey, James and Bhardwaj, Uddipta and Domcke, Valerie and Pieroni, Mauro and Weniger, Christoph",
    title = "{Simulation-based inference for stochastic gravitational wave background data analysis}",
    eprint = "2309.07954",
    archivePrefix = "arXiv",
    primaryClass = "gr-qc",
    reportNumber = "CERN-TH-2023-167",
    doi = "10.1103/PhysRevD.109.083008",
    journal = "Phys. Rev. D",
    volume = "109",
    number = "8",
    pages = "083008",
    year = "2024"
}

@article{SBIfrontier:Kranmer+2020,
    author  = {Kyle Cranmer and Johann Brehmer and Gilles Louppe},
    title   = {The frontier of simulation-based inference},
    journal = {Proceedings of the National Academy of Sciences},
    volume  = {117},
    number  = {48},
    pages   = {30055--30062},
    year    = {2020},
    doi     = {10.1073/pnas.1912789117}
}

@article{LISA:2017pwj,
    author = "Amaro-Seoane, Pau and others",
    collaboration = "LISA",
    title = "{Laser Interferometer Space Antenna}",
    eprint = "1702.00786",
    archivePrefix = "arXiv",
    primaryClass = "astro-ph.IM",
    month = "2",
    year = "2017"
}

@article{Barausse:2023yrx,
    author = "Barausse, Enrico and Dey, Kallol and Crisostomi, Marco and Panayada, Akshay and Marsat, Sylvain and Basak, Soumen",
    title = "{Implications of the pulsar timing array detections for massive black hole mergers in the LISA band}",
    eprint = "2307.12245",
    archivePrefix = "arXiv",
    primaryClass = "astro-ph.GA",
    doi = "10.1103/PhysRevD.108.103034",
    journal = "Phys. Rev. D",
    volume = "108",
    number = "10",
    pages = "103034",
    year = "2023"
}

@article{Dey:2021dem,
    author = "Dey, Kallol and Karnesis, Nikolaos and Toubiana, Alexandre and Barausse, Enrico and Korsakova, Natalia and Baghi, Quentin and Basak, Soumen",
    title = "{Effect of data gaps on the detectability and parameter estimation of massive black hole binaries with LISA}",
    eprint = "2104.12646",
    archivePrefix = "arXiv",
    primaryClass = "gr-qc",
    doi = "10.1103/PhysRevD.104.044035",
    journal = "Phys. Rev. D",
    volume = "104",
    number = "4",
    pages = "044035",
    year = "2021"
}

@article{Klein:2015hvg,
    author = "Klein, Antoine and others",
    title = "{Science with the space-based interferometer eLISA: Supermassive black hole binaries}",
    eprint = "1511.05581",
    archivePrefix = "arXiv",
    primaryClass = "gr-qc",
    doi = "10.1103/PhysRevD.93.024003",
    journal = "Phys. Rev. D",
    volume = "93",
    number = "2",
    pages = "024003",
    year = "2016"
}

@article{LISA:2024hlh,
    author = "Colpi, Monica and others",
    collaboration = "LISA",
    title = "{LISA Definition Study Report}",
    eprint = "2402.07571",
    archivePrefix = "arXiv",
    primaryClass = "astro-ph.CO",
    month = "2",
    year = "2024"
}

@article{Dax:2021tsq,
    author = {Dax, Maximilian and Green, Stephen R. and Gair, Jonathan and Macke, Jakob H. and Buonanno, Alessandra and Sch\"olkopf, Bernhard},
    title = "{Real-Time Gravitational Wave Science with Neural Posterior Estimation}",
    eprint = "2106.12594",
    archivePrefix = "arXiv",
    primaryClass = "gr-qc",
    reportNumber = "LIGO-P2100223",
    doi = "10.1103/PhysRevLett.127.241103",
    journal = "Phys. Rev. Lett.",
    volume = "127",
    number = "24",
    pages = "241103",
    year = "2021"
}

@inproceedings{Dax:2021myb,
    author = {Dax, Maximilian and Green, Stephen R. and Gair, Jonathan and Deistler, Michael and Sch\"olkopf, Bernhard and Macke, Jakob H.},
    title = "{Group equivariant neural posterior estimation}",
    eprint = "2111.13139",
    archivePrefix = "arXiv",
    primaryClass = "cs.LG",
    month = "11",
    year = {2022},
    booktitle = {International Conference on Learning Representations}
}

@ARTICLE{Berti:2005ys,
       author = {{Berti}, Emanuele and {Cardoso}, Vitor and {Will}, Clifford M.},
        title = "{Gravitational-wave spectroscopy of massive black holes with the space interferometer LISA}",
      journal = {\prd},
         year = 2006,
        month = mar,
       volume = {73},
       number = {6},
          eid = {064030},
        pages = {064030},
          doi = {10.1103/PhysRevD.73.064030},
archivePrefix = {arXiv},
       eprint = {gr-qc/0512160},
 primaryClass = {gr-qc},
       adsurl = {https://ui.adsabs.harvard.edu/abs/2006PhRvD..73f4030B}
}

@article{Srinivasan2025,
  author  = {Srinivasan, Rahul and Barausse, Enrico and Korsakova, Natalia and Trotta, Roberto},
  title   = {Simulation-based population inference of {LISA}'s {G}alactic binaries: Bypassing the global fit},
  journal = {Phys. Rev. D}, volume = {112}, pages = {103043}, year = {2025},
  doi     = {10.1103/shym-w46f}, eprint = {2506.22543}, archivePrefix = {arXiv}}

@article{Littenberg2023,
  author  = {Littenberg, Tyson B. and Cornish, Neil J.},
  title   = {Prototype global analysis of {LISA} data with multiple source types},
  journal = {Phys. Rev. D}, volume = {107}, pages = {063004}, year = {2023},
  doi     = {10.1103/PhysRevD.107.063004}, eprint = {2301.03673}, archivePrefix = {arXiv}}

@article{Strub2024,
  author  = {Strub, Stefan H. and Ferraioli, Luigi and Schmelzbach, C{\'e}dric and St{\"a}hler, Simon C. and Giardini, Domenico},
  title   = {Global analysis of {LISA} data with {G}alactic binaries and massive black hole binaries},
  journal = {Phys. Rev. D}, volume = {110}, pages = {024005}, year = {2024},
  doi     = {10.1103/PhysRevD.110.024005}, eprint = {2403.15318}, archivePrefix = {arXiv}}

@article{Katz2025,
  author  = {Katz, Michael L. and Karnesis, Nikolaos and Korsakova, Natalia and Gair, Jonathan R. and Stergioulas, Nikolaos},
  title   = {Efficient {GPU}-accelerated multisource global fit pipeline for {LISA} data analysis},
  journal = {Phys. Rev. D}, volume = {111}, pages = {024060}, year = {2025},
  doi     = {10.1103/PhysRevD.111.024060}, eprint = {2405.04690}, archivePrefix = {arXiv}}

@article{mangiagli2022,
  author  = {Mangiagli, Alberto and Caprini, Chiara and Volonteri, Marta and Marsat, Sylvain and Vergani, Susanna and Tamanini, Nicola and Inchausp{\'e}, Henri},
  title   = {Massive black hole binaries in {LISA}: Multimessenger prospects and electromagnetic counterparts},
  journal = {Phys. Rev. D}, volume = {106}, pages = {103017}, year = {2022},
  doi     = {10.1103/PhysRevD.106.103017}, eprint = {2207.10678}, archivePrefix = {arXiv}}

@article{Chua2021,
  author  = {Chua, Alvin J. K. and Katz, Michael L. and Warburton, Niels and Hughes, Scott A.},
  title   = {Rapid generation of fully relativistic extreme-mass-ratio-inspiral waveform templates for {LISA} data analysis},
  journal = {Phys. Rev. Lett.}, volume = {126}, pages = {051102}, year = {2021},
  doi     = {10.1103/PhysRevLett.126.051102}, eprint = {2008.06071}, archivePrefix = {arXiv}}

@article{Katz2021,
  author  = {Katz, Michael L. and Chua, Alvin J. K. and Speri, Lorenzo and Warburton, Niels and Hughes, Scott A.},
  title   = {Fast extreme-mass-ratio-inspiral waveforms: New tools for millihertz gravitational-wave data analysis},
  journal = {Phys. Rev. D}, volume = {104}, pages = {064047}, year = {2021},
  doi     = {10.1103/PhysRevD.104.064047}, eprint = {2104.04582}, archivePrefix = {arXiv}}

@article{ChapmanBird2025,
  author  = {Chapman-Bird, Christian E. A. and Speri, Lorenzo and Nasipak, Zachary and Burke, Ollie and Katz, Michael L. and Santini, Alessandro and Kejriwal, Shubham and Lynch, Philip and Mathews, Josh and Khalvati, Hassan and Thompson, Jonathan E. and Isoyama, Soichiro and Hughes, Scott A. and Warburton, Niels and Chua, Alvin J. K. and Pigou, Maxime},
  title   = {The Fast and the Frame-Dragging: Efficient waveforms for asymmetric-mass eccentric equatorial inspirals into rapidly-spinning black holes},
  journal = {Phys. Rev. D}, volume = {112}, pages = {104023}, year = {2025},
  doi     = {10.1103/scbp-75pf}, eprint = {2506.09470}, archivePrefix = {arXiv}}

@misc{Mojito,
  author       = {{LISA Distributed Data Processing Center}},
  title        = {Mojito: {LISA} Common Dataset 1},
  howpublished = {\url{https://mojito-e66317.io.esa.int/}},
  year         = {2025},
  note         = {Accessed 2026-10-26}}

@article{GWTC4,
  author        = {{The LIGO Scientific Collaboration} and {the Virgo Collaboration} and {the KAGRA Collaboration}},
  title         = {{GWTC-4.0}: Updating the Gravitational-Wave Transient Catalog with Observations from the First Part of the Fourth {LIGO}-{Virgo}-{KAGRA} Observing Run},
  journal       = {Astrophys. J. Lett.},
  volume        = {1004},
  pages         = {L22},
  year          = {2026},
  doi           = {10.3847/2041-8213/ae2c74},
  eprint        = {2508.18082},
  archivePrefix = {arXiv}}

@article{Bogdanovic2022,
  author        = {Bogdanovi{\'c}, Tamara and Miller, M. Coleman and Blecha, Laura},
  title         = {Electromagnetic counterparts to massive black-hole mergers},
  journal       = {Living Rev. Relativ.},
  volume        = {25},
  pages         = {3},
  year          = {2022},
  doi           = {10.1007/s41114-022-00037-8},
  eprint        = {2109.03262},
  archivePrefix = {arXiv}}

@article{Tang2018,
  author        = {Tang, Yike and Haiman, Zolt{\'a}n and MacFadyen, Andrew},
  title         = {The late inspiral of supermassive black hole binaries with circumbinary gas discs in the {LISA} band},
  journal       = {Mon. Not. R. Astron. Soc.},
  volume        = {476},
  pages         = {2249--2257},
  year          = {2018},
  doi           = {10.1093/mnras/sty423},
  eprint        = {1801.02266},
  archivePrefix = {arXiv}}

@article{dAscoli2018,
  author        = {d'Ascoli, St{\'e}phane and Noble, Scott C. and Bowen, Dennis B. and Campanelli, Manuela and Krolik, Julian H. and Mewes, Vassilios},
  title         = {Electromagnetic emission from supermassive binary black holes approaching merger},
  journal       = {Astrophys. J.},
  volume        = {865},
  pages         = {140},
  year          = {2018},
  doi           = {10.3847/1538-4357/aad8b4},
  eprint        = {1806.05697},
  archivePrefix = {arXiv}}

@article{Gutierrez2022,
  author        = {Guti{\'e}rrez, Eduardo M. and Combi, Luciano and Noble, Scott C. and Campanelli, Manuela and Krolik, Julian H. and L{\'o}pez Armengol, Federico G. and Garc{\'i}a, Federico},
  title         = {Electromagnetic signatures from supermassive binary black holes approaching merger},
  journal       = {Astrophys. J.},
  volume        = {928},
  pages         = {137},
  year          = {2022},
  doi           = {10.3847/1538-4357/ac56de},
  eprint        = {2112.09773},
  archivePrefix = {arXiv}}

@article{Haiman2017,
  author        = {Haiman, Zolt{\'a}n},
  title         = {Electromagnetic chirp of a compact binary black hole: A phase template for the gravitational wave inspiral},
  journal       = {Phys. Rev. D},
  volume        = {96},
  pages         = {023004},
  year          = {2017},
  doi           = {10.1103/PhysRevD.96.023004},
  eprint        = {1705.06765},
  archivePrefix = {arXiv}}

@article{DalCanton2019,
  author        = {Dal Canton, Tito and Mangiagli, Alberto and Noble, Scott C. and Schnittman, Jeremy and Ptak, Andrew and Klein, Antoine and Sesana, Alberto and Camp, Jordan},
  title         = {Detectability of modulated {X}-rays from {LISA}'s supermassive black hole mergers},
  journal       = {Astrophys. J.},
  volume        = {886},
  pages         = {146},
  year          = {2019},
  doi           = {10.3847/1538-4357/ab505a},
  eprint        = {1902.01538},
  archivePrefix = {arXiv}}

@article{Krauth2023,
  author        = {Krauth, Luke Major and Davelaar, Jordy and Haiman, Zolt{\'a}n and Westernacher-Schneider, John Ryan and Zrake, Jonathan and MacFadyen, Andrew},
  title         = {Disappearing thermal {X}-ray emission as a tell-tale signature of merging massive black hole binaries},
  journal       = {Mon. Not. R. Astron. Soc.},
  volume        = {526},
  pages         = {5441--5454},
  year          = {2023},
  doi           = {10.1093/mnras/stad3095},
  eprint        = {2304.02575},
  archivePrefix = {arXiv}}

@article{Dittmann2023,
  author        = {Dittmann, Alexander J. and Ryan, Geoffrey and Miller, M. Coleman},
  title         = {The decoupling of binaries from their circumbinary disks},
  journal       = {Astrophys. J. Lett.},
  volume        = {949},
  pages         = {L30},
  year          = {2023},
  doi           = {10.3847/2041-8213/acd183},
  eprint        = {2303.16204},
  archivePrefix = {arXiv}}

@article{Milosavljevic2005,
  author        = {Milosavljevi{\'c}, Milo{\v s} and Phinney, E. Sterl},
  title         = {The afterglow of massive black hole coalescence},
  journal       = {Astrophys. J. Lett.},
  volume        = {622},
  pages         = {L93--L96},
  year          = {2005},
  doi           = {10.1086/429618},
  eprint        = {astro-ph/0410343},
  archivePrefix = {arXiv}}

@article{Tamanini2016,
  author        = {Tamanini, Nicola and Caprini, Chiara and Barausse, Enrico and Sesana, Alberto and Klein, Antoine and Petiteau, Antoine},
  title         = {Science with the space-based interferometer {eLISA}. {III}: Probing the expansion of the {U}niverse using gravitational wave standard sirens},
  journal       = {J. Cosmol. Astropart. Phys.},
  volume        = {2016},
  number        = {04},
  pages         = {002},
  year          = {2016},
  doi           = {10.1088/1475-7516/2016/04/002},
  eprint        = {1601.07112},
  archivePrefix = {arXiv}}

@article{Lops2023,
  author        = {Lops, Gaia and Izquierdo-Villalba, David and Colpi, Monica and Bonoli, Silvia and Sesana, Alberto and Mangiagli, Alberto},
  title         = {Galaxy fields of {LISA} massive black hole mergers in a simulated {U}niverse},
  journal       = {Mon. Not. R. Astron. Soc.},
  volume        = {519},
  pages         = {5962--5986},
  year          = {2023},
  doi           = {10.1093/mnras/stad058},
  eprint        = {2207.10683},
  archivePrefix = {arXiv}}

@article{XinHaiman2021,
  author        = {Xin, Chengcheng and Haiman, Zolt{\'a}n},
  title         = {Ultra-short-period massive black hole binary candidates in {LSST} as {LISA} `verification binaries'},
  journal       = {Mon. Not. R. Astron. Soc.},
  volume        = {506},
  pages         = {2408--2417},
  year          = {2021},
  doi           = {10.1093/mnras/stab1856},
  eprint        = {2105.00005},
  archivePrefix = {arXiv}}

@article{Kocsis2008,
  author        = {Kocsis, Bence and Haiman, Zolt{\'a}n and Menou, Kristen},
  title         = {Premerger localization of gravitational wave standard sirens with {LISA}: Triggered search for an electromagnetic counterpart},
  journal       = {Astrophys. J.},
  volume        = {684},
  pages         = {870--887},
  year          = {2008},
  doi           = {10.1086/590230},
  eprint        = {0712.1144},
  archivePrefix = {arXiv}}

@article{LangHughes2008,
  author        = {Lang, Ryan N. and Hughes, Scott A.},
  title         = {Localizing coalescing massive black hole binaries with gravitational waves},
  journal       = {Astrophys. J.},
  volume        = {677},
  pages         = {1184--1200},
  year          = {2008},
  doi           = {10.1086/528953},
  eprint        = {0710.3795},
  archivePrefix = {arXiv}}

@article{LangHughes2009,
  author        = {Lang, Ryan N. and Hughes, Scott A.},
  title         = {Advanced localization of massive black hole coalescences with {LISA}},
  journal       = {Class. Quantum Grav.},
  volume        = {26},
  pages         = {094035},
  year          = {2009},
  doi           = {10.1088/0264-9381/26/9/094035},
  eprint        = {0810.5125},
  archivePrefix = {arXiv}}

@article{cornish2022,
  author        = {Cornish, Neil J.},
  title         = {Low latency detection of massive black hole binaries},
  journal       = {Phys. Rev. D},
  volume        = {105},
  pages         = {044007},
  year          = {2022},
  doi           = {10.1103/PhysRevD.105.044007},
  eprint        = {2110.06238},
  archivePrefix = {arXiv}}

@article{Weaving2024,
  author        = {Weaving, Connor R. and Nuttall, Laura K. and Harry, Ian W. and Wu, Shichao and Nitz, Alexander},
  title         = {Adapting the {PyCBC} pipeline to find and infer the properties of gravitational waves from massive black hole binaries in {LISA}},
  journal       = {Class. Quantum Grav.},
  volume        = {41},
  pages         = {025006},
  year          = {2024},
  doi           = {10.1088/1361-6382/ad134d},
  eprint        = {2306.16429},
  archivePrefix = {arXiv}}

@article{cabourndavies2025,
  author        = {Cabourn Davies, Gareth and Harry, Ian and Williams, Michael J. and Bandopadhyay, Diganta and Barack, Leor and Bayle, Jean-Baptiste and Hoy, Charlie and Klein, Antoine and Middleton, Hannah and Moore, Christopher J. and Nuttall, Laura and Pratten, Geraint and Vecchio, Alberto and Woan, Graham},
  title         = {Premerger observation and characterization of massive black hole binaries},
  journal       = {Phys. Rev. D},
  volume        = {111},
  pages         = {043045},
  year          = {2025},
  doi           = {10.1103/PhysRevD.111.043045},
  eprint        = {2411.07020},
  archivePrefix = {arXiv}}

@article{Deng:2025qhx,
    author = "Deng, Senwen and Babak, Stanislav and Marsat, Sylvain",
    title = "{Fast detection and reconstruction of merging massive black hole binary signals}",
    eprint = "2504.11322",
    archivePrefix = "arXiv",
    primaryClass = "gr-qc",
    doi = "10.1103/jyr7-fcgp",
    journal = "Phys. Rev. D",
    volume = "112",
    number = "4",
    pages = "043010",
    year = "2025"
}

@article{nobili2026,
  author        = {Nobili, Francesco and Bellotti, Malvina and Buscicchio, Riccardo and Dotti, Massimo and Lupi, Alessandro},
  title         = {Fast premerger detection of massive black-hole binaries in {LISA} based on time-frequency excess power},
  journal       = {Phys. Rev. D},
  volume        = {114},
  pages         = {063001},
  year          = {2026},
  doi           = {10.1103/l85g-ds5n},
  eprint        = {2602.16792},
  archivePrefix = {arXiv},
  primaryClass  = {astro-ph.IM}
}

@article{spadaro2026,
  author        = {Spadaro, Alice and Gair, Jonathan and Gerosa, Davide and Green, Stephen R. and Buscicchio, Riccardo and Gupte, Nihar and Tenorio, Rodrigo and Clyne, Samuel and P{\"u}rrer, Michael and Korsakova, Natalia},
  title         = {Accurate and efficient simulation-based inference for massive black-hole binaries with {LISA}},
  journal       = {Phys. Rev. D},
  volume        = {114},
  pages         = {023038},
  year          = {2026},
  doi           = {10.1103/mmyl-wdgq},
  eprint        = {2603.20431},
  archivePrefix = {arXiv},
  primaryClass  = {astro-ph.HE}
}

@article{zhang2026,
  author        = {Zhang, Xue-Ting and Gair, Jonathan and Messenger, Chris and Korsakova, Natalia and Hu, Yi-Ming and Chen, Hong-Yu},
  title         = {Pre-localization of Massive Black Hole Binaries in the Millihertz Band},
  year          = {2026},
  eprint        = {2604.24330},
  archivePrefix = {arXiv},
  primaryClass  = {gr-qc}
}

@article{cabourndavies2026,
  author        = {Cabourn Davies, Gareth and Harry, Ian},
  title         = {Inpainting to search for premerger massive black hole binaries in realistic {LISA} datasets},
  journal       = {Phys. Rev. D},
  year          = {2026},
  doi           = {10.1103/kksx-247v},
  eprint        = {2605.13738},
  archivePrefix = {arXiv},
  note          = {in press}}

@article{katz2020,
  author        = {Katz, Michael L. and Marsat, Sylvain and Chua, Alvin J. K. and Babak, Stanislav and Larson, Shane L.},
  title         = {{GPU}-accelerated massive black hole binary parameter estimation with {LISA}},
  journal       = {Phys. Rev. D},
  volume        = {102},
  pages         = {023033},
  year          = {2020},
  doi           = {10.1103/PhysRevD.102.023033},
  eprint        = {2005.01827},
  archivePrefix = {arXiv}}

@article{Dax2023,
  title = {Neural Importance Sampling for Rapid and Reliable Gravitational-Wave Inference},
  author = {Dax, Maximilian and Green, Stephen R. and Gair, Jonathan and P{\"u}rrer, Michael and Wildberger, Jonas and Macke, Jakob H. and Buonanno, Alessandra and Sch{\"o}lkopf, Bernhard},
  journal = {Physical Review Letters},
  volume = {130},
  pages = {171403},
  year = {2023},
  doi = {10.1103/PhysRevLett.130.171403}
}

@article{Prince2002,
  author        = {Prince, Thomas A. and Tinto, Massimo and Larson, Shane L. and Armstrong, J. W.},
  title         = {The {LISA} optimal sensitivity},
  journal       = {Phys. Rev. D},
  volume        = {66},
  pages         = {122002},
  year          = {2002},
  doi           = {10.1103/PhysRevD.66.122002},
  eprint        = {gr-qc/0209039},
  archivePrefix = {arXiv}}

@ARTICLE{Geemoo_Deng2025,
       author = {{Deng}, Senwen and {Babak}, Stanislav and {Le Jeune}, Maude and {Marsat}, Sylvain and {Plagnol}, {\'E}ric and {Sartirana}, Andrea},
        title = "{Modular global-fit pipeline for LISA data analysis}",
      journal = {\prd},
         year = 2025,
        month = may,
       volume = {111},
       number = {10},
          eid = {103014},
        pages = {103014},
          doi = {10.1103/PhysRevD.111.103014},
archivePrefix = {arXiv},
       eprint = {2501.10277},
 primaryClass = {gr-qc},
       adsurl = {https://ui.adsabs.harvard.edu/abs/2025PhRvD.111j3014D}
}

@article{Pratten2023,
  author        = {Pratten, Geraint and Klein, Antoine and Moore, Christopher J. and Middleton, Hannah and Steinle, Nathan and Schmidt, Patricia and Vecchio, Alberto},
  title         = {{LISA} science performance in observations of short-lived signals from massive black hole binary coalescences},
  journal       = {Phys. Rev. D},
  volume        = {107},
  pages         = {123026},
  year          = {2023},
  doi           = {10.1103/PhysRevD.107.123026},
  eprint        = {2212.02572},
  archivePrefix = {arXiv}}

@article{sharma2024,
  author        = {Sharma, Abhishek and Sengupta, Anand S. and Mukherjee, Suvodip},
  title         = {Accelerated parameter estimation of supermassive black hole binaries in {LISA} using a meshfree approximation},
  journal       = {Phys. Rev. D},
  volume        = {111},
  pages         = {042009},
  year          = {2025},
  doi           = {10.1103/PhysRevD.111.042009},
  eprint        = {2409.14288},
  archivePrefix = {arXiv}}

@article{Du2024,
  author        = {Du, Minghui and Liang, Bo and Wang, He and Xu, Peng and Luo, Ziren and Wu, Yueliang},
  title         = {Advancing space-based gravitational wave astronomy: Rapid parameter estimation via normalizing flows},
  journal       = {Sci. China Phys. Mech. Astron.},
  volume        = {67},
  pages         = {230412},
  year          = {2024},
  doi           = {10.1007/s11433-023-2270-7},
  eprint        = {2308.05510},
  archivePrefix = {arXiv}}

@article{cnf2024,
  author        = {Liang, Bo and Du, Minghui and Wang, He and Xu, Yuxiang and Liu, Chang and Wei, Xiaotong and Xu, Peng and Qiang, Li-e and Luo, Ziren},
  title         = {Rapid parameter estimation for merging massive black hole binaries using continuous normalizing flows},
  journal       = {Mach. Learn.: Sci. Technol.},
  volume        = {5},
  pages         = {045040},
  year          = {2024},
  doi           = {10.1088/2632-2153/ad8da9},
  eprint        = {2407.07125},
  archivePrefix = {arXiv}}

@article{snl2024,
  author        = {V{\'i}lchez, Iv{\'a}n Mart{\'i}n and Sopuerta, Carlos F.},
  title         = {Efficient massive black hole binary parameter estimation for {LISA} using sequential neural likelihood},
  journal       = {J. Cosmol. Astropart. Phys.},
  volume        = {2025},
  number        = {04},
  pages         = {022},
  year          = {2025},
  doi           = {10.1088/1475-7516/2025/04/022},
  eprint        = {2406.00565},
  archivePrefix = {arXiv}}

@software{Bayle,
  author       = {Bayle, Jean-Baptiste and Hees, Aur{\'e}lien and Lilley, Marc and Le Poncin-Lafitte, Christophe and Martens, Waldemar and Joffre, Eric},
  title        = {{LISA} Orbits},
  version      = {3.0.0},
  publisher    = {Zenodo},
  year         = {2025},
  doi          = {10.5281/zenodo.17296456},
  url          = {https://gitlab.in2p3.fr/lisa-simulation/orbits}}

@book{Maggiore,
  author       = {Maggiore, Michele},
  title        = {Gravitational Waves. {Vol.} 1: Theory and Experiments},
  publisher    = {Oxford University Press},
  address      = {Oxford},
  year         = {2007},
  doi          = {10.1093/acprof:oso/9780198570745.001.0001}}

@article{PeterMatthews,
  author       = {Peters, P. C. and Mathews, J.},
  title        = {Gravitational radiation from point masses in a {K}eplerian orbit},
  journal      = {Phys. Rev.},
  volume       = {131},
  pages        = {435--440},
  year         = {1963},
  doi          = {10.1103/PhysRev.131.435}}

@article{Buonanno2008,
  author       = {Buonanno, Alessandra and Kidder, Lawrence E. and Lehner, Luis},
  title        = {Estimating the final spin of a binary black hole coalescence},
  journal      = {Phys. Rev. D},
  volume       = {77},
  pages        = {026004},
  year         = {2008},
  doi          = {10.1103/PhysRevD.77.026004},
  eprint       = {0709.3839},
  archivePrefix = {arXiv}}

@inproceedings{Loshchilov2019,
  author       = {Loshchilov, Ilya and Hutter, Frank},
  title        = {Decoupled weight decay regularization},
  booktitle    = {International Conference on Learning Representations (ICLR)},
  year         = {2019},
  eprint       = {1711.05101},
  archivePrefix = {arXiv}}

@misc{Goyal2017,
  author       = {Goyal, Priya and Doll{\'a}r, Piotr and Girshick, Ross and Noordhuis, Pieter and Wesolowski, Lukasz and Kyrola, Aapo and Tulloch, Andrew and Jia, Yangqing and He, Kaiming},
  title        = {Accurate, large minibatch {SGD}: Training {ImageNet} in 1 hour},
  year         = {2017},
  eprint       = {1706.02677},
  archivePrefix = {arXiv},
  primaryClass = {cs.CV},
    note = {Published by Facebook AI Research (Data @ Scale), 2017}
}

@inproceedings{Loshchilov2017,
  author       = {Loshchilov, Ilya and Hutter, Frank},
  title        = {{SGDR}: Stochastic gradient descent with warm restarts},
  booktitle    = {International Conference on Learning Representations (ICLR)},
  year         = {2017},
  eprint       = {1608.03983},
  archivePrefix = {arXiv}}

@misc{Sangria,
  author       = {Le Jeune, Maude and Babak, Stanislav},
  title        = {{LISA} Data Challenge {Sangria} ({LDC2a})},
  year         = {2022},
  publisher    = {Zenodo},
  version      = {v2},
  doi          = {10.5281/zenodo.7132178},
  note         = {Dataset}}

@article{Samajdar:2021egv,
    author = "Samajdar, Anuradha and Janquart, Justin and Van Den Broeck, Chris and Dietrich, Tim",
    title = "{Biases in parameter estimation from overlapping gravitational-wave signals in the third-generation detector era}",
    eprint = "2102.07544",
    archivePrefix = "arXiv",
    primaryClass = "gr-qc",
    doi = "10.1103/PhysRevD.104.044003",
    journal = "Phys. Rev. D",
    volume = "104",
    number = "4",
    pages = "044003",
    year = "2021"
}

@article{Pizzati:2021apa,
    author = "Pizzati, Elia and Sachdev, Surabhi and Gupta, Anuradha and Sathyaprakash, Bangalore",
    title = "{Toward inference of overlapping gravitational-wave signals}",
    eprint = "2102.07692",
    archivePrefix = "arXiv",
    primaryClass = "gr-qc",
    doi = "10.1103/PhysRevD.105.104016",
    journal = "Phys. Rev. D",
    volume = "105",
    number = "10",
    pages = "104016",
    year = "2022"
}

@ARTICLE{2024MNRAS.533.1907B,
       author = {{Bhowmick}, Aklant K. and {Blecha}, Laura and {Torrey}, Paul and {Somerville}, Rachel S. and {Kelley}, Luke Zoltan and {Vogelsberger}, Mark and {Weinberger}, Rainer and {Hernquist}, Lars and {Sivasankaran}, Aneesh},
        title = "{Growth of high-redshift supermassive black holes from heavy seeds in the BRAHMA cosmological simulations: implications of overmassive black holes}",
      journal = {\mnras},
         year = 2024,
        month = sep,
       volume = {533},
       number = {2},
        pages = {1907-1926},
          doi = {10.1093/mnras/stae1819},
archivePrefix = {arXiv},
       eprint = {2406.14658},
 primaryClass = {astro-ph.GA},
       adsurl = {https://ui.adsabs.harvard.edu/abs/2024MNRAS.533.1907B}
}

@article{Armitage:2002uu,
    author = "Armitage, Philip J. and Natarajan, Priyamvada",
    title = "{Accretion during the merger of supermassive black holes}",
    eprint = "astro-ph/0201318",
    archivePrefix = "arXiv",
    doi = "10.1086/339770",
    journal = "Astrophys. J. Lett.",
    volume = "567",
    pages = "L9--L12",
    year = "2002"
}

@article{Palenzuela:2010nf,
    author = "Palenzuela, Carlos and Lehner, Luis and Liebling, Steven L.",
    title = "{Dual Jets from Binary Black Holes}",
    eprint = "1005.1067",
    archivePrefix = "arXiv",
    primaryClass = "astro-ph.HE",
    reportNumber = "PI-OTHER-170",
    doi = "10.1126/science.1191766",
    journal = "Science",
    volume = "329",
    pages = "927",
    year = "2010"
}

@article{Moesta:2011bn,
    author = "Moesta, Philipp and Alic, Daniela and Rezzolla, Luciano and Zanotti, Olindo and Palenzuela, Carlos",
    title = "{On the detectability of dual jets from binary black holes}",
    eprint = "1109.1177",
    archivePrefix = "arXiv",
    primaryClass = "gr-qc",
    doi = "10.1088/2041-8205/749/2/L32",
    journal = "Astrophys. J. Lett.",
    volume = "749",
    pages = "L32",
    year = "2012"
}

@article{Giacomazzo:2012iv,
    author = "Giacomazzo, Bruno and Baker, John G. and Miller, M. Coleman and Reynolds, Christopher S. and van Meter, James R.",
    title = "{General Relativistic Simulations of Magnetized Plasmas around Merging Supermassive Black Holes}",
    eprint = "1203.6108",
    archivePrefix = "arXiv",
    primaryClass = "astro-ph.HE",
    doi = "10.1088/2041-8205/752/1/L15",
    journal = "Astrophys. J. Lett.",
    volume = "752",
    pages = "L15",
    year = "2012"
}

@article{Kaplan:2011mz,
    author = "Kaplan, D. L. and O'Shaughnessy, R. and Sesana, A. and Volonteri, M.",
    title = "{Blindly Detecting Merging Supermassive Black Holes with Radio Surveys}",
    eprint = "1105.3653",
    archivePrefix = "arXiv",
    primaryClass = "astro-ph.HE",
    doi = "10.1088/2041-8205/734/2/L37",
    journal = "Astrophys. J. Lett.",
    volume = "734",
    pages = "L37",
    year = "2011"
}

@article{OShaughnessy:2011nwl,
    author = "O'Shaughnessy, R. and Kaplan, D. L. and Sesana, A. and Kamble, A.",
    title = "{Blindly detecting orbital modulations of jets from merging supermassive black holes}",
    eprint = "1109.1050",
    archivePrefix = "arXiv",
    primaryClass = "astro-ph.CO",
    doi = "10.1088/0004-637X/743/2/136",
    journal = "Astrophys. J.",
    volume = "743",
    pages = "136",
    year = "2011"
}

@article{Laghi:2026ujb,
    author = "Laghi, Danny and Tamanini, Nicola and Sesana, Alberto and Gair, Jonathan and Barausse, Enrico and Caprini, Chiara and Del Pozzo, Walter and Mangiagli, Alberto and Marsat, Sylvain",
    title = "{Reducing cosmological degeneracies by combining multiple classes of LISA gravitational-wave standard sirens}",
    eprint = "2603.23612",
    archivePrefix = "arXiv",
    primaryClass = "astro-ph.CO",
    month = "3",
    year = "2026"
}

@article{Barausse:2012fy,
    author = "Barausse, Enrico",
    title = "{The evolution of massive black holes and their spins in their galactic hosts}",
    eprint = "1201.5888",
    archivePrefix = "arXiv",
    primaryClass = "astro-ph.CO",
    doi = "10.1111/j.1365-2966.2012.21057.x",
    journal = "Mon. Not. Roy. Astron. Soc.",
    volume = "423",
    pages = "2533--2557",
    year = "2012"
}

@article{Sesana:2014bea,
    author = "Sesana, A. and Barausse, E. and Dotti, M. and Rossi, E. M.",
    title = "{Linking the spin evolution of massive black holes to galaxy kinematics}",
    eprint = "1402.7088",
    archivePrefix = "arXiv",
    primaryClass = "astro-ph.CO",
    doi = "10.1088/0004-637X/794/2/104",
    journal = "Astrophys. J.",
    volume = "794",
    pages = "104",
    year = "2014"
}

@article{Antonini:2015sza,
    author = "Antonini, Fabio and Barausse, Enrico and Silk, Joseph",
    title = "{The Coevolution of Nuclear Star Clusters, Massive Black Holes, and their Host Galaxies}",
    eprint = "1506.02050",
    archivePrefix = "arXiv",
    primaryClass = "astro-ph.GA",
    doi = "10.1088/0004-637X/812/1/72",
    journal = "Astrophys. J.",
    volume = "812",
    number = "1",
    pages = "72",
    year = "2015"
}

@article{Bonetti:2018tpf,
    author = "Bonetti, Matteo and Sesana, Alberto and Haardt, Francesco and Barausse, Enrico and Colpi, Monica",
    title = "{Post-Newtonian evolution of massive black hole triplets in galactic nuclei {\textendash} IV. Implications for LISA}",
    eprint = "1812.01011",
    archivePrefix = "arXiv",
    primaryClass = "astro-ph.GA",
    doi = "10.1093/mnras/stz903",
    journal = "Mon. Not. Roy. Astron. Soc.",
    volume = "486",
    number = "3",
    pages = "4044--4060",
    year = "2019"
}

@article{Barausse:2020mdt,
    author = "Barausse, Enrico and Dvorkin, Irina and Tremmel, Michael and Volonteri, Marta and Bonetti, Matteo",
    title = "{Massive Black Hole Merger Rates: The Effect of Kiloparsec Separation Wandering and Supernova Feedback}",
    eprint = "2006.03065",
    archivePrefix = "arXiv",
    primaryClass = "astro-ph.GA",
    doi = "10.3847/1538-4357/abba7f",
    journal = "Astrophys. J.",
    volume = "904",
    number = "1",
    pages = "16",
    year = "2020"
}

@article{Houba:2025dnr,
    author = "Houba, Niklas and Giarda, Giovanni and Speri, Lorenzo",
    title = "{SlotFlow: Amortized Trans-Dimensional Inference with Slot-Based Normalizing Flows}",
    eprint = "2511.23228",
    archivePrefix = "arXiv",
    primaryClass = "astro-ph.IM",
    month = "11",
    year = "2025"
}

@misc{LISAX,
    author       = {Pitte, Chantal and Bayle, Jean-Baptiste},
    title        = {{lisax}: A fast and accurate {LISA} response in {JAX}},
    year         = {2026},
    note         = {Manuscript in preparation},
    url          = {https://github.com/ChanPitte/lisax}
}

\end{document}